\documentclass[11pt,a4paper,twoside]{article}
\usepackage{amsmath,amsfonts,amsthm,amssymb,color,cite,graphics}
\newcommand{\be}{\begin{eqnarray}}
\newcommand{\ee}{\end{eqnarray}}
\newcommand{\bez}{\begin{eqnarray*}}
\newcommand{\eez}{\end{eqnarray*}}
\newcommand{\pa}{\partial}
\newcommand{\la}{\lambda}

\newcommand{\cA}{\mathcal{A}}
\newcommand{\cB}{\mathcal{B}}
\newcommand{\cC}{\mathcal{C}}
\newcommand{\cD}{\mathcal{D}}
\newcommand{\cN}{\mathcal{N}}
\newcommand{\cP}{\mathcal{P}}
\newcommand{\cS}{\mathcal{S}}
\newcommand{\cT}{\mathcal{T}}
\newcommand{\cU}{\mathcal{U}}
\newcommand{\cV}{\mathcal{V}}

\newcommand{\bA}{\mathbf{A}}
\newcommand{\bB}{\mathbf{B}}
\newcommand{\bC}{\mathbf{C}}
\newcommand{\bG}{\mathbf{G}}
\newcommand{\bGamma}{\mathbf{\Gamma}}
\newcommand{\bI}{\mathbf{I}}
\newcommand{\bJ}{\mathbf{J}}
\newcommand{\bK}{\mathbf{K}}
\newcommand{\bM}{\mathbf{M}}

\newcommand{\bP}{\mathbf{P}}
\newcommand{\bPhi}{\mathbf{\Phi}}
\newcommand{\bPi}{\mathbf{\Pi}}
\newcommand{\bPsi}{\mathbf{\Psi}}
\newcommand{\bQ}{\mathbf{Q}}
\newcommand{\bR}{\mathbf{R}}
\newcommand{\bS}{\mathbf{S}}
\newcommand{\bSigma}{\mathbf{\Sigma}}
\newcommand{\bT}{\mathbf{T}}

\newcommand{\bU}{\mathbf{U}}
\newcommand{\bV}{\mathbf{V}}
\newcommand{\bW}{\mathbf{W}}
\newcommand{\bX}{\mathbf{X}}
\newcommand{\bXi}{\mathbf{\Xi}}
\newcommand{\bY}{\mathbf{Y}}
\newcommand{\bZ}{\mathbf{Z}}

\newcommand{\bbC}{\mathbb{C}}
\newcommand{\bbN}{\mathbb{N}}

\newcommand{\bbR}{\mathbb{R}}
\newcommand{\bbS}{\mathbb{S}}
\newcommand{\bbT}{\mathbb{T}}
\newcommand{\bbZ}{\mathbb{Z}}

\newcommand{\txi}{\tilde{\xi}}

\newcommand{\tbP}{\tilde{\mathbf{P}}}
\newcommand{\tbR}{\tilde{\mathbf{R}}}
\newcommand{\tbU}{\tilde{\mathbf{U}}}
\newcommand{\tbV}{\tilde{\mathbf{V}}}
\newcommand{\tbX}{\tilde{\mathbf{X}}}

\renewcommand{\d}{\mathrm{d}}

\newcommand{\imag}{\mathrm{i}}
\renewcommand{\Re}{\mathrm{Re}}

\newcommand{\tminus}{\mbox{\tiny ($-$)}}
\newcommand{\tplus}{\mbox{\tiny ($+$)}}

\theoremstyle{plain}
\newtheorem{theorem}{Theorem}[section]
\newtheorem{lemma}[theorem]{Lemma}
\newtheorem{proposition}[theorem]{Proposition}
\newtheorem{corollary}[theorem]{Corollary}

\theoremstyle{definition}
\newtheorem{definition}[theorem]{Definition}
\newtheorem{remark}[theorem]{Remark}
\newtheorem{example}[theorem]{Example}

\renewcommand{\theequation} {\arabic{section}.\arabic{equation}}

\title{\huge Solutions of matrix NLS systems and their discretisations: 
             A unified treatment}

\author{ {\scshape Aristophanes Dimakis} \\
 Department of Financial and Management Engineering, \\
 University of the Aegean, 41, Kountourioti Str., GR-82100 Chios, Greece \\
 E-mail: \emph{dimakis@aegean.gr}
             \and
 {\scshape Folkert M\"uller-Hoissen} \\
 Max-Planck-Institute for Dynamics and Self-Organization \\
 Bunsenstrasse 10, D-37073 G\"ottingen, Germany \\
 E-mail: \emph{folkert.mueller-hoissen@ds.mpg.de} 
       }

\date{}

\begin{document}

\maketitle

\begin{abstract}
Using a bidifferential graded algebra approach to ``integrable'' partial differential 
or difference equations, a unified treatment of continuous, semi-discrete 
(Ablowitz-Ladik) and fully discrete matrix NLS systems is presented. 
These equations originate from a universal equation within this framework, 
by specifying a representation of the bidifferential graded algebra and imposing 
a reduction. By application of a general result, corresponding families of exact 
solutions are obtained that in particular comprise the matrix soliton solutions 
in the focusing NLS case. The solutions are parametrised in terms of 
constant matrix data subject to a Sylvester equation (which previously 
appeared as a \emph{rank condition} in the integrable systems literature). 
These data exhibit a certain redundancy, which we diminish to a large extent. 

More precisely, we first consider more general AKNS-type systems from which two different 
matrix NLS systems emerge via reductions. In the continuous case, the familiar Hermitian 
conjugation reduction leads to a continuous matrix (including vector) NLS equation, 
but it is well-known that this does not work as well in the discrete cases. 
On the other hand there is a complex conjugation reduction, which apparently has not 
been studied previously. It leads to \emph{square} matrix NLS systems, but works  
in all three cases (continuous, semi- and fully-discrete). 
A large part of this work is devoted to an exploration of the corresponding solutions, 
in particular regularity and asymptotic behaviour of matrix soliton solutions. 
\end{abstract}

\small
\tableofcontents
\normalsize

\section{Introduction}
\setcounter{equation}{0}
Using the framework of bidifferential calculi (or bidifferential graded algebras), a unification of 
integrability aspects and solution generating techniques for a wide class of integrable models 
has been achieved \cite{DMH08bidiff} (see also the references cited therein). 
In particular, there is a fairly simple and universal method to construct large families of exact 
solutions from solutions of a linear system of equations. The formalism treats continuous and 
discrete systems on an equal level. This suggests a corresponding study of the well-known nonlinear 
Schr\"odinger (NLS) equation (see e.g. \cite{Fadd+Takh87,APT04book}), its 
semi-discretisation, the integrable discrete NLS or Ablowitz-Ladik (AL) equation 
\cite{Ablo+Ladi76JMP,Ablo+Ladi76SAM,Ablo+Ladi77,Ablo77}, and a full discretisation
\cite{Ablo+Ladi76SAM}. 
The literature on mathematical aspects of the continuous NLS equation 
and its applications in science is meanwhile impossible to summarise. A rather incomplete list 
of references dealing with mathematical aspects of the AL equation is  \cite{Chiu+Ladik77,Levi+Ragn78,Kako+Mugi79,NQC83,QNCL84,Ahma+Chow87,Herb+Ablo89,Geng89,ZTOF91,Scha+Bish91,Veks+Kono92,CKVV92,MRT94,CBG-J94,CBG-J95,MEKL95,PBL95,BLP96,Pogr+Prati96,Suri97,KMZ97,Garn01,Vani01,AAU02,MAA03,SAA03,APT04book,Rour04,DMR04,Nenc05,Simon05,Simon07,Maru+Ohta06,CCX06,Peli06,KRSSS06,TCLS07,Dokt+Leble07,Gekh+Nenc09,Tsuch09}, 
and 
\cite{Schil89,Doli+Sant95,Zeng+Rauc95,Veks95,Veks96,Veks98,Veks98univ,Veks99,Veks02,Veks06,Sada03,GDC03,Erco+Loza06,ZNBC06,Geng+Dai07,GDZ07,GHMT07,GHMT08II,GHMT08,GHMT08SAM,Fan+Yang09,Mich09,GHMT10} 
more generally for its hierarchy. 
Physical applications of the AL equation can be found e.g. in 
\cite{Vakh+Gaid86,CKKS93,IIKS93,Kivs+Camp93,Vakh+Vakh95,AAPMLTW96,Kund00}. 
In this work, we consider more generally \emph{matrix} versions of the (continuous and discrete) NLS equations and 
present corresponding solutions. 
Matrix generalisations, including vector versions, of the scalar continuous NLS equation have 
been studied in particular in 
\cite{Manak74,Kuli+Skly81,Ford+Kuli83,Harn+Wiss92,Radha+Laks95,RLH97,TUW98,FMMW00,EEK00,Koda+Mikh01,Park+Shin00,Park+Shin02,Han+Shin04,KTA04,Tsuch04,APT04DPDE,APT04IP,APT04book,IUW07,Warr+Elgi07,Warr07,Dega+Lomb07,Demo+Mee08,Schie08,Tern+Uhle09BT,Sakh09}. A brief summary of the physical 
relevance can be found in \cite{APT04book}. In particular, a (symmetric) $2 \times 2$ matrix NLS equation 
turned out to be of relevance for the description of a special Bose-Einstein condensate (with atoms 
in a spin 1 state)  \cite{IMW04PRL,IMW04,LLMML05,IMW06,Wada+Tsuc06,UIW06,UIW07,Ieda+Wada07,Kuro+Wada07,DRK07,KAGG07,AGGK08,DWY08,KSKLW09}. 
Semi-discrete matrix NLS equations appeared in
\cite{Gerd+Ivan82,Tsuc+Wada98,TUW99,AOT99,Ohta00,Ding00,Malo+Yang02,APT03,APT04book,APT06,Hiro06,Cont+Chow08,Maru+Prin08,TNRFK09,Tsuch10}, 
a full discretisation has been elaborated in \cite{Tsuc00RMP} and a dispersionless limit studied in \cite{Maru+Prin08}. 

In section~\ref{sec:framework} we introduce the general framework. 
Section~\ref{sec:NLS} then presents bidifferential calculi for the NLS system and  
its discrete versions. Section~\ref{sec:NLS_sol} derives exact solutions, by application 
of the method of section~\ref{sec:framework}. The solutions are parametrised by  
matrix data that have to solve a Sylvester equation.\footnote{See e.g. 
\cite{Bhat+Rose97,Dym07,Horn+John91} for results on the Sylvester equation. 
A vast literature deals with its solutions and applications.}  
In section~\ref{sec:decomp}, these results are translated into a decomposed form, where 
the NLS systems attain a more familiar form. 

Section~\ref{sec:cc-reduction} deals with a complex conjugation 
reduction which applies to the continuous as well as the discrete NLS systems. 
Quite surprisingly, this has apparently not been studied previously for matrix NLS equations. 
Our analysis in particular addresses regularity and asymptotic behaviour of matrix soliton solutions. 
Section~\ref{sec:HermRed} treats the more familiar Hermitian conjugation reduction which, 
however, only works for the continuous NLS system, at least without severe restrictions or 
introduction of non-locality (cf. appendix~D). 
Section~\ref{sec:concl} contains some concluding remarks. 

Some supplementary material has been shifted to appendices. 
Appendix~A briefly treats the considerably simpler example of the matrix KdV equation, in particular to 
provide the reader with a quick access to the methods used in this work, but also in order to 
stress the similarities even in details of the calculation. 
Appendix~B establishes a relation between the semi-discrete matrix NLS (AL) system and the 
matrix modified KdV equation. 
Appendix~C demonstrates that our fully discrete matrix NLS system reduces in the scalar case 
to the corresponding equation in \cite{Ablo+Ladi76SAM}. 
Appendix~D presents a non-local reduction of the semi-discrete NLS system.
Appendix~E recovers Lax pairs for the three NLS systems, starting with the respective 
bidifferential calculus. 
Finally, Appendix~F contains some remarks on the continuum limit of the fully discrete matrix 
NLS system and the solutions obtained in this work.

\section{The framework}
\label{sec:framework}
\setcounter{equation}{0}
\begin{definition}
A (complex) \emph{graded algebra} is an algebra\footnote{All algebras in this work will be
(assumed to be) associative. }
$\Omega$ that has a direct sum decomposition
\be
     \Omega = \bigoplus_{r \geq 0} \Omega^r
\ee
into complex vector spaces $\Omega^r$, with the property
\be
     \Omega^r \, \Omega^s \subseteq \Omega^{r+s} \; .
\ee
Furthermore, $\cA := \Omega^0$ is a subalgebra and, for $r>0$, $\Omega^r$ is an $\cA$-bimodule. 
\end{definition}

\begin{definition}
A \emph{bidifferential graded algebra} or \emph{bidifferential calculus} is a graded algebra
$\Omega$ equipped with two graded derivations
$\d, \bar{\d} : \Omega \rightarrow \Omega$ of degree one (hence $\d \Omega^r \subseteq \Omega^{r+1}$,
$\bar{\d} \Omega^r \subseteq \Omega^{r+1}$), with the properties
\be
    \d \circ \d = 0 \, , \quad
    \bar{\d} \circ \bar{\d} = 0 \, , \quad
    \d \circ \bar{\d} + \bar{\d} \circ \d = 0 \, , \label{bidiff}
\ee
and the graded Leibniz rule
\be
    \d(\chi \, \chi') = (\d \chi) \, \chi' + (-1)^r \, \chi \, \d \chi' \,  , \qquad
    \bar{\d}(\chi \, \chi') = (\bar{\d}\chi) \, \chi + (-1)^r \, \chi \, \bar{\d} \chi' \, ,
    \label{Leibniz}
\ee
for all $\chi \in \Omega^r$ and $\chi' \in \Omega$.
\end{definition}

For an algebra $\cA$, a corresponding graded algebra is given by
\be
    \Omega = \cA \otimes \bigwedge(\bbC^2) \, ,    \label{Omega=AotimesC^2}
\ee
where $\bigwedge(\bbC^2)$ denotes the exterior algebra of $\bbC^2$.\footnote{More generally,
one may consider the exterior algebra of $\bbC^k$, $k>1$. }
In this case it is sufficient to define the maps $\d, \bar{\d}$ on $\cA$. They extend
in an obvious way to $\Omega$ such that the Leibniz rule holds. Elements of $\bigwedge(\bbC^2)$
are treated as ``constants''. 
This structure underlies all our examples. In this work, $\zeta_1,\zeta_2$ will denote a basis of $\bigwedge^1(\bbC^2)$. 
\vskip.1cm

Given a unital algebra $\cB$ and a 
bidifferential graded algebra $(\Omega,\d, \bar{\d})$ with $\Omega^0 = \cB$,
the actions of $\d$ and $\bar{\d}$ extend componentwise to matrices over $\cB$. 
A corresponding example is provided in Appendix~A, the matrix KdV equation. 
By application of results stated below we recover in a strikingly simple way a class of 
KdV solutions derived previously via inverse scattering theory (see in particular \cite{Akto+Mee06}).
A treatment of NLS-type equations requires a generalisation, however. Here $\d$ and $\bar{\d}$ 
involve a structure that depends on the size of the matrix on which they act. 
\vskip.1cm

The space of all matrices over $\cB$, with size greater or equal to that of
$n_0 \times n_0$ matrices,\footnote{In some cases it may turn out to be 
necessary to further restrict the values of $n$ and $n'$. In typical examples we can set $n_0 = 1$, 
but the NLS case requires $n_0=2$ (see section~\ref{sec:NLS}).} 
\be
    \mathrm{Mat}_{n_0}(\cB) := \bigoplus_{n',n \geq n_0} \mathrm{Mat}(n',n,\cB) \, ,
\ee
has the structure of a complex algebra with the usual matrix product extended trivially by setting
$A B=0$ whenever the dimensions do not match. 
Extending $\cA = \mathrm{Mat}_{n_0}(\cB)$ to a graded algebra (\ref{Omega=AotimesC^2}), we shall 
require that $\d$ and $\bar{\d}$ respect the size of matrices in the sense that 
$\d( \mathrm{Mat}(n',n,\cB) \otimes \bigwedge(\bbC^2) ) \subseteq 
\mathrm{Mat}(n',n,\cB) \otimes \bigwedge(\bbC^2))$, and correspondingly for $\bar{\d}$. 
The following result originated in \cite{DMH08disp} and appeared in \cite{DMH08bidiff} in 
a more general form.

\begin{theorem}
\label{theorem:main}
Let $(\Omega, \d, \bar{\d})$ be a bidifferential graded algebra with\footnote{More general 
choices of $\Omega$ are actually allowed.} 
$\Omega = \cA \otimes \bigwedge(\bbC^2)$ and $\cA = \mathrm{Mat}_{n_0}(\cB)$, 
for some $n_0 \in \bbN$. 
For fixed $n,n' \geq n_0$, let $\bX \in \mathrm{Mat}(n,n,\cB)$
and $\bY \in \mathrm{Mat}(n',n,\cB)$ be solutions of
\be
   \bar{\d} \bX = (\d \bX) \, \bP  \, , \qquad
      \bar{\d} \bY = (\d \bY) \, \bP \, ,        \label{XY_linsys}
\ee
and\footnote{(\ref{RX+QY=XP}) has a form analogous to that of a Sylvester equation. The 
ingredients of the latter are just matrices over $\bbC$, however. Nevertheless, we shall see that in 
our examples (\ref{RX+QY=XP}) actually results in an ordinary Sylvester equation.} 
\be
    \bR \, \bX - \bX \, \bP = -\bQ \, \bY    \label{RX+QY=XP} \, ,
\ee
with $\d$- and $\bar{\d}$-constant matrices $\bP,\bR \in \mathrm{Mat}(n,n,\cB)$, and
\be
    \bQ = \tbV \, Q \, \tbU \, ,  \label{bQ=bVQbU}
\ee
where $\tbU \in \mathrm{Mat}(m',n',\cB)$, $\tbV \in \mathrm{Mat}(n,m,\cB)$ and 
$Q  \in \mathrm{Mat}(m,m',\cB)$ are also constants of the calculus. 
If $\bX$ is invertible, then
\be
     \phi := \tbU \bY \bX^{-1} \tbV  \in \mathrm{Mat}(m',m,\cB)    \label{phi=UPhiV}
\ee
satisfies
\be
   \bar{\d} \phi = (\d \phi) \, Q \, \phi + \d \theta \qquad \mbox{with} \quad
   \theta = \tbU \bY \bX^{-1} \bR \tbV \; .    \label{bdphi}
\ee
Furthermore, (by application of $\d$) $\phi$ solves 
\be
     \bar{\d} \d \phi = \d \phi \, Q \, \d \phi \; .   \label{univeqQ}
\ee
\end{theorem}
\noindent
\textit{Proof:} As a consequence of (\ref{XY_linsys}) and (\ref{RX+QY=XP}), we find
(cf. \cite{DMH08bidiff}) that
\bez
     \bPhi := \bY \bX^{-1} \in \mathrm{Mat}(n',n,\cB)
\eez
solves
\bez
      \bar{\d} \bPhi = (\d \bPhi) \, \bQ \, \bPhi + \d(\bPhi \, \bR) \; .
\eez
Multiplying from the left by $\tbU $ and from the right by
$\tbV$, resolving $\bQ$ according to (\ref{bQ=bVQbU}), and
using the graded Leibniz rule for $\d$ and $\bar{\d}$, we obtain (\ref{bdphi}).
\hfill  $\square$
\vskip.2cm

The theorem reveals an integrability feature of the nonlinear equation (\ref{univeqQ}).
 From solutions of a system of $n \times n$, respectively $n' \times n$, matrix linear equations,
it generates a large class of solutions of the $m' \times m$ matrix equation (\ref{univeqQ}).
Since $n,n'$ can be arbitrarily large, in this way we obtain infinite families of solutions
of (\ref{univeq}), where the members depend on a number of parameters that increases with
$n$ and $n'$.
In this work we will restrict our considerations to the case where
\be
     m' = m \, , \qquad Q = I_m  \; .
\ee
Accordingly, our central equation will be
\be
     \bar{\d} \d \phi = \d \phi \; \d \phi  \qquad \mbox{where} \quad
     \phi \in \mathrm{Mat}(m,m,\cB) \; .                              \label{univeq}
\ee

\begin{remark}
Concerning the general case, if $Q$ admits a factorisation 
\bez
    Q = A \, \left( \begin{array}{cc} I_r & 0_{r \times (m'-r)} \\ 
                    0_{(m-r) \times r} & 0_{(m-r) \times (m'-r)} \end{array} \right) \, B
\eez
into $\d$- and $\bar{\d}$-constant matrices, where $A$ and $B$ are invertible,
then (2.12) is equivalent to a (inhomogeneous) \emph{linear extension} 
(cf. \cite{Kupe01}) of (\ref{univeq}) (where $m$ has to be replaced by $r$), 
after the redefinition $B \phi A \mapsto \phi$. This holds in particular if 
$Q$ is a constant complex matrix (Smith form decomposition, see e.g. \cite{Bern09}), and 
if all such matrices are $\d$- and $\bar{\d}$-constant. Our methods can also be applied to 
such a linear extension. 
In case of the bidifferential calculi in section~\ref{sec:NLS}, $\d$- and 
$\bar{\d}$-constancy of a matrix places additional restrictions on a constant matrix, 
so that (2.12) is \emph{not} in general equivalent to a linear extension of (\ref{univeq}). 
All this opens additional possibilities that will be disregarded in this work, however. 
A matrix $Q$ different from an identity matrix effects a deformation of the ordinary 
matrix product, which indeed offers interesting applications \cite{DMH06DNLS}.
\hfill $\square$
\end{remark}

Many (e.g. in the sense of the inverse scattering method) integrable partial differential or
difference equations can be recovered from the universal equation (\ref{univeq}), by choosing 
a suitable bidifferential graded algebra. 
At first sight it seems that (\ref{univeq}) can only produce equations 
with quadratic nonlinearity. But $\phi$ need not directly correspond to the independent variable 
of the equation of final interest. Indeed, the main examples in this work, the NLS systems, 
possess a cubic nonlinearity. 
The (anti-) self-dual Yang-Mills equation (more precisely, a gauge reduced version of it), 
which is well-known as a source of many integrable partial differential equations, is a special 
case of (\ref{univeq}) \cite{DMH08bidiff}. Whereas the KP equation and its hierarchy do not 
quite fit into the self-dual Yang-Mills framework, they do fit well into our generalised 
framework. The generalisation (\ref{univeq}) covers much more, 
however, and in particular integrable partial \emph{difference} equations, i.e. discretised 
versions of continuous integrable equations. 
\vskip.2cm

In all examples that will be considered in this work, the general solution of the linear equations 
(\ref{XY_linsys}) can be expressed in the form
\be
    \bX = \bA_0 + \bA_1 \, \hat{\bXi} \, , \qquad
    \bY = \bB_0 + \bB_1 \, \hat{\bXi} \, ,    \label{X,Y_univ}
\ee
where the matrices $\bA_0,\bA_1,\bB_0,\bB_1$ are $\d$- and $\bar{\d}$-constant, and $\hat{\bXi}$ is 
not. Assuming that $\bA_0$ is invertible, we can rewrite the solution $\phi$ determined by 
the theorem as
\be
    \phi = \tbU \, (\bB_0 + \bB_1 \, \hat{\bXi})(\bI- \hat{\bK} \hat{\bXi})^{-1} \bV 
         = \hat{\bU} \, \hat{\bXi} \, (\bI- \hat{\bK} \hat{\bXi})^{-1} \bV + \tbU \bB_0 \bV  
              \label{phi_univ0}
\ee
in terms of the $\d$- and $\bar{\d}$-constant matrices
\be
    \hat{\bK} = -\bA_0^{-1} \bA_1   \, ,  \qquad
    \hat{\bU} = \tbU \, ( \bB_1 + \bB_0 \hat{\bK} ) \, ,  \qquad
    \bV = \bA_0^{-1} \tbV \; .   \label{KUV_univ}
\ee
The last summand in (\ref{phi_univ0}) is $\d$- and $\bar{\d}$-constant. Since such a term 
can always be added to any solution of (\ref{univeq}), we shall neglect it. 
Next we insert (\ref{X,Y_univ}) in (\ref{RX+QY=XP}) and separate the $\d$- and $\bar{\d}$-constant 
summands from those that depend on $\hat{\bXi}$. 
Assuming that both parts have to be satisfied separately (which is true in our 
examples), the first part becomes $\bR = (\bA_0 \bP - \tbV \tbU \bB_0) \, \bA_0^{-1}$. 
Eliminating $\bR$ from the second, then yields 
\be
      [ \bP , \hat{\bK} \hat{\bXi} ] = \bV \hat{\bU} \hat{\bXi} \; .  \label{constr_univ}
\ee
We summarise our results. 

\begin{corollary}
\label{cor:univ_sol}
Let $(\Omega = \cA \otimes \bigwedge(\bbC^2), \d, \bar{\d})$ be a bidifferential graded 
algebra with $\cA = \mathrm{Mat}_{n_0}(\cB)$ for some $n_0 \in \bbN$.
Let $\hat{\bXi} \in \mathrm{Mat}(n,n,\cB)$ be any solution of 
\be
   \bar{\d} \hat{\bXi} = (\d \hat{\bXi}) \, \bP 
\ee
and (\ref{constr_univ}), 
with $\d$- and $\bar{\d}$-constant matrices $\bP,\hat{\bK} \in \mathrm{Mat}(n,n,\cB)$, 
$\bV \in \mathrm{Mat}(n,m,\cB)$ and $\hat{\bU} \in \mathrm{Mat}(m,n,\cB)$.
Then 
\be
   \phi = \hat{\bU} \, \hat{\bXi} \, (\bI- \hat{\bK} \hat{\bXi})^{-1} \bV  \label{phi_univ}
\ee
solves (\ref{univeq}), provided that $\bI- \hat{\bK} \hat{\bXi}$ is invertible.
\hfill $\square$
\end{corollary}

Having determined $\hat{\bXi}$ (which should not be $\d$- and $\bar{\d}$-constant) in 
a concrete example, the remaining task is thus to find, 
for given $\d$- and $\bar{\d}$-constant matrices $\hat{\bU}$ and $\bV$, a 
$\d$- and $\bar{\d}$-constant matrix $\hat{\bK}$ such that (\ref{constr_univ}) holds.

\begin{remark}
The construction of exact solutions according to theorem~\ref{theorem:main} hardly exhausts the 
set of all solutions of (\ref{univeq}) (which is (\ref{univeqQ}) with $Q=I_m$). In examples like 
matrix KdV (see appendix~A) and the matrix NLS systems treated in the main part of this work, 
the latter equation is equivalent to $\bar{\d} \phi = (\d \phi) \, \phi + \d \theta $ 
with an arbitrary matrix function $\theta$. But (\ref{bdphi}) chooses $\theta$ from a 
certain family of matrix functions, which clearly means a restriction to a subclass of 
solutions. On the other hand, it should be noted that the expression for $\theta$ in 
(\ref{bdphi}) involves matrices of arbitrarily large size and, with suitable technical 
assumptions, we may allow them to be infinitely large, corresponding to operators on 
a Banach space (see also \cite{March88,Schie04,GHS06,GHS09}). It is by far not evident what kind 
of solutions can be reached in this way. 
\hfill $\square$
\end{remark}

\begin{remark}
Theorem~\ref{theorem:main} can be generalised by replacing (\ref{XY_linsys}) with 
the weaker equations
\be
   \bar{\d} \bX = (\d \bX) \, \bP + \bX \, \bPsi  \, , \qquad
      \bar{\d} \bY = (\d \bY) \, \bP + \bY \, \bPsi \, ,   \label{gen_XY_linsys}
\ee
with any $\bPsi \in \Omega^1$ of size $n \times n$, 
and dropping the requirement that the $n \times n$ matrix $\bP$ is 
$\d$- and $\bar{\d}$-constant. $\bP$ and $\bPsi$ are constrained as a 
consequence of the above system, however. Indeed, acting with $\bar{\d}$ on 
(\ref{RX+QY=XP}), using the Leibniz rule, (\ref{gen_XY_linsys}) and (\ref{RX+QY=XP}) again, we obtain 
\be
    \bar{\d} \bP = (\d \bP) \, \bP + \bP \, \bPsi - \bPsi \, \bP \; . \label{bard(bP)}
\ee
Acting with $\bar{\d}$ on (\ref{gen_XY_linsys}), using the Leibniz rule and $\bar{\d} \d = - \d \bar{\d}$, 
again (\ref{gen_XY_linsys}) and then (\ref{bard(bP)}), leads to 
\be
    \bar{\d} \bPsi = (\d \bPsi) \, \bP - \bPsi^2 \; . \label{bard(bPsi)}
\ee 
Setting 
\be
   \bPsi = \bG^{-1} \, \left( \bar{\d} \bG - (\d \bG) \, \bP \right) \, ,  \label{pure_gauge}
\ee
with an invertible $n \times n$ matrix function $\bG$, then 
$0 = \bar{\d}^2 \bG = \bar{\d} ( G \bPsi + (\d \bG) \bP) 
   = \bG \, ( \bar{\d} \bPsi - (\d \bPsi) \, \bP + \bPsi^2 )$, 
by use of (\ref{bard(bP)}), shows that (\ref{bard(bPsi)}) is satisfied. In terms of $\bP' = \bG \bP \bG^{-1}$, 
(\ref{bard(bP)}) then reads 
\be
    \bar{\d} \bP' = (\d \bP') \, \bP' \; . \label{P-equation}
\ee
If for any solution $(\bP,\bPsi)$ of (\ref{bard(bP)}) and (\ref{bard(bPsi)}) the equation 
(\ref{pure_gauge}) has a solution $\bG$, it follows that $\bPsi$ can be set to zero without 
restriction of generality. A restriction seems to remain nevertheless in theorem~\ref{theorem:main}, 
since there we solved the non-linear equation (\ref{P-equation}) 
trivially by requiring that $\bP'$ (renamed to $\bP$) is $\d$- and $\bar{\d}$-constant. 
In case of other solutions of (\ref{P-equation}), the linear equations (\ref{XY_linsys}) involve 
variable coefficients and also the constraint (\ref{RX+QY=XP}) becomes considerably more complicated. 
We will therefore not consider this generalisation further in this work. 
\hfill $\square$
\end{remark}

\section{NLS systems}
\label{sec:NLS}
\setcounter{equation}{0}

\subsection{A class of bidifferential calculi} 
Let $\cA = \mathrm{Mat}_2(\cB)$ 
and $\Omega = \cA \otimes \bigwedge(\bbC^2)$. For every $n \geq 2$ we choose a matrix $J_n$ 
with the properties
\be
        J_n^2 = I_n \, , \qquad J_n \neq \pm I_n \, , 
\ee
where $I_n$ denotes the $n \times n$ identity matrix. Then  
\be
    e_2(f) := \frac{1}{2} [J,f] := \frac{1}{2} (J_{n'} f - f J_n)  
    \qquad \mbox{for} \quad  f \in \mathrm{Mat}(n',n,\cB) 
\ee
defines a derivation of $\cA$. Setting  
\be
 \d f = e_1(f) \, \zeta_1 + e_2(f) \, \zeta_2 
                 \, , \qquad
 \bar{\d} f = \bar{e}_1(f) \, \zeta_1 + \bar{e}_2(f) \, \zeta_2 \, ,
                          \label{class_bidiff}
\ee
then determines a bidifferential calculus on $\Omega$ 
if $e_1,\bar{e}_1,\bar{e}_2$ are commuting derivations of $\cA$ and 
also commute with $e_2$. In the following, $e_1,\bar{e}_1,\bar{e}_2$ will be related to 
partial derivative or partial difference operators. In order to be $\d$- and $\bar{\d}$-constant, 
a matrix must then not depend on the respective variables. 
We shall see that, with appropriate choices, (\ref{univeq}) reproduces a continuous, 
semi-discrete, respectively fully discrete NLS system.

\subsection{Continuous NLS} 
Let $\cB$ be the space of smooth complex functions on $\bbR^2$. In (\ref{class_bidiff}) 
we choose\footnote{Despite of our notation, in some important applications (notably in optics) 
$x$ plays the role of a time coordinate and $t$ that of a physical space coordinate.}
\be
    e_1 = \pa_x \, , \qquad
    \bar{e}_1 = - \imag \, \pa_t \, , \qquad
    \bar{e}_2 = \pa_x \, ,                       \label{e's_NLS}
\ee
where $\pa_x$ is the partial derivative with respect to $x$, and $\imag = \sqrt{-1}$.
(\ref{univeq}) takes the form 
\be
   - \frac{\imag}{2} [ J, \phi_t ] = \phi_{xx} + \frac{1}{2} [ \phi_x , [J,\phi]] \, ,
                      \label{NLS1}
\ee
where $J = J_m$. Writing
\be
    \phi = J \, \varphi \, ,  \label{phi=Jvarphi}    
\ee
and decomposing the $m \times m$ matrix $\varphi$ as follows into the sum of an ``even'' 
and an ``odd'' part,
\be
    \varphi = \cD + \cU \, , \qquad 
    \cD = \frac{1}{2} (\varphi + J \varphi J) \, , \quad
    \cU = \frac{1}{2} (\varphi - J \varphi J) \, ,   
\ee
we have 
\be
      J \cD = \cD J \, , \qquad  
      J \cU = - \cU J \; .    \label{JcD,JcU}
\ee
Hence (\ref{NLS1}) splits into 
\be
  \imag J \, \cU_t + \cU_{xx} + \cD_x \, \cU 
      + \cU \cD_x = 0          \label{3dNLSa}
\ee
and $\cD_{xx} = -(\cU^2)_x$. The latter can be integrated to
\be
    \cD_x = -\cU^2 + \cC \, ,           \label{3dNLSb}
\ee
where $\cC$ is an even matrix independent of $x$. This can be used to eliminate $\cD$ in (\ref{3dNLSa}). 
We conclude that, for the chosen bidifferential calculus, (\ref{univeq}) is equivalent to
\be
   \imag J \, \cU_t + \cU_{xx} - 2 \, \cU^3 
   + \cC \cU + \cU \cC = 0 \, , \qquad 
   \cD_x = - \cU^2 + \cC \; .   \label{NLS_sys_C}
\ee
Without restriction of generality, we can set\footnote{For  
$\cU' = \cT \cU \cT^{-1}$, where $\cT_t = - \imag \, J \cC \cT$, and 
$\cD' = \cD - \cC \, x$, (\ref{NLS_sys}) holds without the terms involving $\cC$. 
The modification of the NLS equations in sections \ref{sec:cc-reduction} and \ref{sec:HermRed}, 
caused by a nonvanishing $\cC$, in the scalar case reproduces a ``background term''
which turned out to be convenient in order to study finite density type solutions 
(see e.g. (1.5) in \cite{Fadd+Takh87}). } 
$\cC=0$, so that
\be
   \imag J \, \cU_t + \cU_{xx} - 2 \, \cU^3 = 0 \, , \qquad 
   \cD_x = - \cU^2 \; .   \label{NLS_sys}
\ee
This will be called ($m \times m$ matrix) continuous \emph{NLS system}. 

The $\zeta_2$-part of (\ref{bdphi}) (with $Q=I_m$) is
\be
   \varphi_x = \frac{1}{2} \, [J,\varphi] \, J \, \varphi + \frac{1}{2} \, J \, [J,\theta] \, ,
\ee
which implies $\cD_x = - \cU^2$. As a consequence, the solutions generated 
via theorem~\ref{theorem:main} actually have vanishing $\cC$.\footnote{We actually lost 
the freedom of the choice of an even matrix $\cC$ in the step from (\ref{bdphi}) to 
(\ref{univeqQ}), which would still allow the addition of a $\d$-closed 1-form 
that is not $\d$-exact, namely $J \cC \, \zeta_2$ in the case under consideration. }

\subsection{Semi-discrete NLS}
Let $\cB_0$ be the algebra of functions on $\bbZ \times \bbR$, smooth in the second variable. 
We extend it to an algebra $\cB$ by adjoining the shift operator $\bbS$ with respect to the discrete 
variable (the discretised coordinate $x$). In (\ref{class_bidiff}) we choose
\be
    e_1(f) = [\bbS,f] \, , \qquad
    \bar{e}_1(f) = - \imag \, \dot{f} \, , \qquad
    \bar{e}_2(f) = - [\bbS^{-1},f] \, ,                       \label{e's_sdNLS}
\ee
where $\dot{f}= df/dt$. 
In the following we use the notation $f^+ = \bbS f \bbS^{-1}$, $f^- = \bbS^{-1} f \bbS$. 
Setting 
\be
    \phi = J \varphi \bbS^{-1} \, ,   \label{sd_NLS_phi->varphi}
\ee
(\ref{univeq}) becomes
\be
      \frac{\imag}{2} [J,\dot{\varphi}] + \varphi^+ -2 \, \varphi + \varphi^- 
    + \frac{1}{2} (\varphi^+ - \varphi) J[J,\varphi] 
    - \frac{1}{2} [J,\varphi] J (\varphi-\varphi^-) = 0 \, , 
\ee
where $\varphi$ can be restricted to be an element of $\mathrm{Mat}(m,m,\cB_0)$. 
With the decomposition $\varphi = \cD + \cU$ as used above, we find 
\be
    \cD^+ - \cD = - \cU^+ \cU  + \cC \, ,        \label{sd_NLS_D} 
\ee
where $\cC$ is even and does not depend on the discrete variable, and
\be
  \imag \, J \dot{\cU} + \cU^+ - 2 \, \cU +\cU^- - \cU^+ \cU^2 - \cU^2 \cU^- 
               + \{ \cC , \cU \} = 0 \; .  \label{sd_NLS_U}
\ee
With the above choice of bidifferential calculus, (\ref{univeq}) is thus 
equivalent to (\ref{sd_NLS_D}) and (\ref{sd_NLS_U}). 
Again, $\cC$ can be eliminated by redefinitions, and we have
\be
  \imag \, J \, \dot{\cU} + \cU^+ - 2 \, \cU + \cU^- - \cU^+ \cU^2 - \cU^2 \cU^- = 0 
         \, , \qquad
  \cD^+ - \cD = - \cU^+ \cU  \; .  \label{sd_NLS_sys}
\ee
These two equations constitute a semi-discretisation of (\ref{NLS_sys}) and will be 
called ($m \times m$ matrix) \emph{semi-discrete NLS} system. 
By inspection of (\ref{bdphi}), the solutions generated 
via theorem~\ref{theorem:main} have vanishing $\cC$.

\subsection{Fully discrete NLS}
Here we make the following choice in (\ref{class_bidiff}), 
\be
    e_1(f) = [\bbT \bbS,f] \, , \qquad
    \bar{e}_1(f) = - \imag \, [\bbT,f] \, , \qquad
    \bar{e}_2(f) = - [\bbS^{-1},f] \, ,                       \label{e's_fdNLS}
\ee
where $\bbT$ is the shift operator in discrete ``time''. 
Here $\cB$ is the algebra of functions $\cB_0$ on $\bbZ \times \bbZ$, extended by 
adjoining $\bbS$ and $\bbT$. 
In the following, we also set $f_+ = \bbT f \bbT^{-1}$ and $f_- = \bbT^{-1} f \bbT$. Writing  
\be
    \phi = J \varphi \bbS^{-1} \, , 
\ee
(\ref{univeq}) reads
\be
     \frac{\imag}{2} [J,\varphi_+ -\varphi]
    + (\varphi^+-\varphi)_+ -(\varphi-\varphi^-)
    + \frac{1}{2} (\varphi^+_+ - \varphi) J [J,\varphi_+] 
    - \frac{1}{2} [J,\varphi] J (\varphi_+ - \varphi^-) = 0 \, ,  \label{fd_NLS_varphi_eq}
\ee
and $\varphi$ can be restricted to be an element of $\mathrm{Mat}(m,m,\cB_0)$. 
With the decomposition $\varphi = \cD + \cU$, we obtain
\be
  && \imag \, J \, (\cU_+-\cU) + (\cU^+ - \cU)_+ - (\cU- \cU^-) - \cU^+_+ \, \cU_+^2 - \cU^2 \, \cU^- 
     + (\cD_+ - \cD) \, \cU_+ + \cU \, (\cD_+ - \cD) = 0 \, , \nonumber \\
  &&  \cD^+ - \cD = - \cU^+ \cU \, ,   \label{fd_NLS_sys}
\ee
which will be called ($m \times m$ matrix) \emph{fully discrete NLS system}. We again dropped terms 
involving an even matrix $\cC$ that does not depend on the discrete variable $x$. 
Again, inspection of (\ref{bdphi}) shows 
that solutions generated via theorem~\ref{theorem:main} have in fact vanishing $\cC$ and 
are therefore solutions of (\ref{fd_NLS_sys}).

\subsection{Normal form of the matrix \textnormal{\textit{J}} }
By an application of the Jordan decomposition theorem, $J$ is related by a similarity
transformation to a diagonal matrix with diagonal entries $\pm 1$. Since we can apply 
this similarity transformation to the above equations, without restriction of generality we can choose 
\be
       J = \left( \begin{array}{cc} I_{m_1} & 0 \\ 0 & - I_{m_2} \end{array} \right)
                     \label{J_normalform}
\ee
with any choice $m_1,m_2 \in \bbN$ such that $m_1 + m_2 = m$.
Note that, for $m>2$, the bidifferential calculus 
depends on the splitting of $m$ into two summands. This means that there are actually 
different bidifferential calculi on $\mathrm{Mat}(m,m,\cB)$, and different choices correspond 
to different NLS systems.

\subsection{Complex conjugation reduction}
Let $J$ satisfy
\be
        J^\ast = J \, ,
\ee
where ${}^\ast$ denotes complex conjugation, and let $\Gamma$ be an invertible constant 
$m \times m$ matrix with the properties
\be
    J \Gamma = - \Gamma J \, , \quad 
    \Gamma^\ast = \Gamma^{-1} \; .  \label{ccred_Gamma_conds}
\ee
Then the transformation
\be
    \cU \mapsto \epsilon \, \Gamma^{-1} \cU^\ast \Gamma 
    \quad \mbox{with} \quad \epsilon = \pm 1 \, , \qquad
    \cD \mapsto \Gamma^{-1} \cD^\ast \Gamma \, ,   \label{NLS_ast_sym}
\ee
leaves all three NLS systems invariant, including (\ref{JcD,JcU}) 
(which applies to all three cases). Hence a consistent reduction is given by 
\be
    \cU^\ast = \epsilon \, \Gamma \cU \Gamma^{-1} \, , \qquad
    \cD^\ast = \Gamma \cD \Gamma^{-1} \; .    \label{ccred_U,D}
\ee 
Transforming $J$ to its normal form (\ref{J_normalform}), the first of (\ref{ccred_Gamma_conds}) 
restricts $\Gamma$ to off-block-diagonal form. The second of conditions (\ref{ccred_Gamma_conds}) 
then requires that $m$ is even and 
\be
      m_1 = m_2 \; .
\ee
We will actually choose
\be
    \Gamma = \left( \begin{array}{cc} 0 & I_{m_1} \\ I_{m_1} & 0 \end{array} \right) \; .
                     \label{ccred_Gamma_m1=m2} 
\ee

\subsection{Hermitian conjugation reduction}
The \emph{continuous} NLS system with 
\be
       J^\dagger = J \, ,
\ee
where ${}^\dagger$ denotes Hermitian conjugation (i.e. transposition combined 
with complex conjugation), is invariant under 
\be
    \cU \mapsto \epsilon \, \cU^\dagger 
      \quad \mbox{with} \quad \epsilon = \pm 1 \, , \qquad
    \cD \mapsto \cD^\dagger \; .   \label{NLS_dag_sym}
\ee 
Hence 
\be
   \cU^\dagger = \epsilon \, \cU \, , \qquad
   \cD^\dagger = \cD            \label{NLS_dag_red}
\ee 
is a consistent reduction. This does \emph{not} work as well for the matrix 
discrete NLS systems, since ${}^\dagger$ is an anti-involution (whereas ${}^\ast$ 
is an involution). Applying it to the nonlinear term in the first of equations 
(\ref{sd_NLS_sys}), we have
$(\cU^+ \, \cU^2 + \cU^2 \, \cU^-)^\dagger 
   = \cU^{\dagger 2} \, \cU^{\dagger +}  + \cU^{\dagger -} \, \cU^{\dagger 2}$. 
Clearly the matrices in this expression will not commute in general. 
In order to achieve that the equations obtained from (\ref{sd_NLS_sys}) 
by Hermitian conjugation become equivalent to the original equations, we can either \\
(1) require that $\cU^2$ commutes with $\cU^{\pm}$, while keeping (\ref{NLS_dag_red}), or \\
(2) replace (\ref{NLS_dag_red}) by $\cU(x,t)^\dagger = \epsilon \, \cU(-x,t)$ and 
$\cD(x,t)^\dagger = - \cD(-x,t)$ (see also \cite{Gerd+Ivan82}). 
\\
We will not discuss the first case further in this work, but only refer to 
\cite{TUW99,Tsuc00RMP,APT03,APT04DPDE,APT04book,APT06}, see also remark~\ref{rem:Hred_comm_constr}.
The second case leads (in a decomposed form in the sense of section~\ref{sec:decomp}) in general 
to a \emph{non-local} equation, see appendix~D.

\section{Solutions of the matrix NLS systems} 
\label{sec:NLS_sol}
\setcounter{equation}{0}
A central result of this section is expressed in the next proposition. 
In the following, $\bI = I_n$ denotes the $n \times n$ identity matrix.

\begin{proposition} 
\label{prop:NLSsol}
\be
    \cU = \bU \, ( \bXi^{-1} - \bK \bXi \bK )^{-1} \, \bV   \, , \qquad
    \cD = \bU \, \bXi \bK ( \bXi^{-1} - \bK \bXi \bK )^{-1} \, \bV    \label{cU,cD_sol}
\ee
solves\footnote{More precisely, here we assume that the inverses appearing in these 
formulae exist. This will also be done in the formulation of results that originate 
from this proposition. A considerable part of this work deals with this regularity problem, 
however. }
the continuous, semi-discrete, respectively fully discrete $m \times m$ matrix NLS system 
(involving the $m \times m$ matrix $J$ with $J^2 = I_m$) if 
\renewcommand{\arraystretch}{1.5}
\begin{center}
\begin{tabular}{c|c|c|c|c|c|c} 
        & $\bJ \neq \pm \bI$ & $\bS$ & $\bU$ & $\bV$ & $\bK$  \\
\hline
 size   & $n \times n$ & $n \times n$ & $m \times n$ & $n \times m$ & $n \times n$
\end{tabular}
\end{center}
are constant matrices such that
\be
     \bJ^2 = \bI \, , \quad 
   \bJ \bS = \bS \bJ \, , \quad
     J \bU = - \bU \bJ \, , \quad 
   \bJ \bV = \bV J \, , \quad
   \bJ \bK = - \bK \bJ \, ,     \label{JSUVK_rels}
\ee
and 
\begin{center}
\begin{tabular}{c|c|c} 
 NLS  & Sylvester equation & $\bXi$   \\
\hline
 continuous  &  $\bS \bK + \bK \bS = \bV \bU$ & $e^{-x \, \bS - \imag \, t \, \bS^2 \bJ}$  \\
\hline 
 semi-discrete & $\bS^{-1} \bK - \bK \bS = \bV \bU$ & $\bS^x \, e^{- \imag \, t \, (\bS+\bS^{-1}-2\bI) \, \bJ}$ \\
\hline
 fully discrete & $\bS^{-1} \bK - \bK \bS = \bV \bU$ & $\bS^x \, [( \bI + \imag \, \bJ \, (\bI-\bS^{-1}) )
                ( \bI - \imag \, \bJ \, (\bI-\bS))^{-1}]^t$ 
\end{tabular}
\end{center}
where $x \in \bbZ$ in the semi- and fully discrete NLS case, $t \in \bbZ$ in the fully discrete NLS case, 
and $x,t \in \bbR$ otherwise. Of course, we need $\bS$ invertible in the discrete cases. 
\hfill $\square$
\end{proposition}
\vskip.2cm

The matrices $\bS, \bU, \bV, \bK$ are subject to a \emph{Sylvester equation}.
Given matrix data $(\bJ, \bS, \bU, \bV)$, the problem  
is to find a solution $\bK$ of the Sylvester equation. 
Let $\sigma(\bS)$ denote the spectrum of $\bS$, i.e. the set of its eigenvalues. 
A sufficient condition for the existence of a solution of the respective Sylvester equation is 
\be
     \begin{array}{r} \sigma(\bS) \cap \sigma(-\bS) = \emptyset \\
                      \sigma(\bS) \cap \sigma(\bS^{-1}) = \emptyset
     \end{array} \quad \mbox{for} \quad
     \begin{array}{l} \mbox{continuous NLS} \\ \mbox{discrete NLS} \end{array} 
         \label{Sylvester_spec_cond}
\ee
(see e.g. theorem~4.4.6 in \cite{Horn+John91}). Moreover, the solution is then unique, 
and as a consequence $\bK$ automatically satisfies $\bJ \bK = - \bK \bJ$. 
\vskip.1cm

There is a certain redundancy in the matrix data, since different matrix data can determine the 
same solution, but we will narrow this down to a considerable extent (section~\ref{subsec:transf}, 
see also \cite{ABDM09sym}). 
\vskip.1cm

We note that the NLS equations can determine $\cD$ only up to addition of an arbitrary 
$m \times m$ matrix $\cD_0$ that does not depend on $x$ (and $\cD_0$ has to commute with $J$). 
We shall therefore understand 
the expression for $\cD$ in (\ref{cU,cD_sol}) always modulo addition of an arbitrary 
constant matrix commuting with $J$. To put it another way, we simply drop any constant 
summand that arises in an expression for $\cD$, and we actually did this in order to 
arrive at the expression in (\ref{cU,cD_sol}). 
\vskip.1cm

In section~\ref{subsec:matrixNLSsol_proof} we derive the results formulated in
proposition~\ref{prop:NLSsol}. 
Some symmetries of the solution formulae in proposition~\ref{prop:NLSsol}
are revealed in subsection~\ref{subsec:transf}. Subsection~\ref{subsec:NLS_super}
formulates a superposition rule on the level of matrix data. 
Furthermore, section~\ref{subsec:redsol} specialises to the complex conjugation, 
respectively Hermitian conjugation reductions. 

\begin{remark}
The results formulated in proposition 4.1 will be derived below by application 
of corollary~\ref{cor:univ_sol}. The latter actually allows to replace $\bXi$ in 
proposition~\ref{prop:NLSsol} by $\bC \bXi$ with any constant 
matrix $\bC$ that commutes with $\bJ$ and $\bS$.
But the line of arguments that led us from theorem~2.3 to corollary 2.4 in fact 
shows that such a matrix can be eliminated by redefinitions of the matrix variables 
entering the solution formula. 
Conversely, in special cases (see e.g. section 6.2) it will be possible and 
convenient to shift all the freedom in the choice of $\bU$ and $\bV$ into 
such a matrix $\bC$. In other cases it will be convenient to move at least part of 
the freedom from $\bU$ and $\bV$ to a matrix $\bC$.
Although the corresponding reparametrisation transformation (see section 4.2.3) 
requires some restrictions that result in an invertible $\bC$ (commuting with $\bS$), 
the aforementioned ensures us that this restriction for $\bC$ is not really 
required in order to obtain an NLS solution. 
\hfill $\square$
\end{remark}

\subsection{Proof of proposition~\ref{prop:NLSsol}}
\label{subsec:matrixNLSsol_proof}
In the following we start with the bidifferential calculi 
on $\Omega = \mathrm{Mat}_2(\cB) \otimes \bigwedge(\bbC^2)$ associated 
in section~\ref{sec:NLS} with the continuous NLS, semi-discrete NLS, respectively fully 
discrete NLS system. Then we apply theorem~\ref{theorem:main} in the form specified 
in corollary~\ref{cor:univ_sol}. 
For fixed $n$ we write $\bJ$ for $J_n$.
The requirement that the matrices $\bP,\hat{\bU}, \bV$ have to be $\d$- and $\bar{\d}$-constant 
now means that they must not depend on the independent (continuous, 
respectively discrete) variables, and they have to satisfy 
\be
     [\bJ,\bP] = 0 \, , \qquad
     J \, \hat{\bU} = \hat{\bU} \, \bJ  \, , \qquad 
     \bJ \, \bV = \bV \, J \; .          \label{JU=UJ,JV=VJ}  
\ee

Any $n \times n$ matrix $\bA$ decomposes into the sum of an ``even'' and an ``odd'' part,  
\be
     \bA = \bA_e + \bA_o  \qquad \mbox{where} \quad \bJ \bA_e = \bA_e \bJ \, , \quad
                    \bJ \bA_o = - \bA_o \bJ  \; . 
\ee
Such a matrix is $\d$- and $\bar{\d}$-constant iff it is even and constant 
(i.e. independent of $x$ and $t$). 
A constant $n \times n$ matrix $\bSigma$ that satisfies
\be
    \bSigma^2 = \bI \, , \qquad \bJ \bSigma = - \bSigma \bJ \, , 
\ee
can be used to convert odd into even matrices and vice versa.

\paragraph{Continuous NLS.}
We write 
\be
     \bP = \bJ \bS \, ,
\ee
with a new object
$\bS$. Then $\bar{\d} \bX = (\d \bX) \, \bP = (\d \bX) \, \bJ \, \bS$ 
becomes
\be
    - \imag \, \bX_t = \bX_x \bS \bJ \, , \qquad
    \bX_x = \frac{1}{2}[\bJ,\bX] \bJ \bS \; .  \label{NLS_X_eqs}
\ee
The general solution is
\be
    \bX = \bA_0 + \bA_1 \, \bSigma \, \bXi 
    \qquad \mbox{with} \qquad 
    \bXi = e^{- x \, \bS  - \imag \, t \, \bS^2 \bJ} \; . 
\ee
Here $\bA_0,\bA_1$
are constant even matrices (hence 
$\d$- and $\bar{\d}$-constant). Assuming that $\bA_0$ is invertible,  
application of theorem~\ref{theorem:main} reduces to an application of 
corollary~\ref{cor:univ_sol}. 
In terms of the redefined matrices
\be
   \bK = \hat{\bK} \bSigma \, , \qquad \bU = \hat{\bU} \bSigma \, , \label{K->hatK,U->hatU}
\ee
and with $\hat{\bXi} =  \bSigma \bXi$, (\ref{constr_univ}) takes the form 
\be
    \bS \bK + \bK \bS = \bV \bU  \, ,   \label{NLS_rank}
\ee
Note that $\bJ \bV = \bV J$, but $J \bU = - \bU \bJ$. 
(\ref{phi_univ}) and (\ref{phi=Jvarphi}) lead to
\be
   \varphi = \bU \, \bXi \, (\bI-\bK \bXi)^{-1} \bV  
           = \bU \, \bXi \, (\bI + \bK \bXi)( \bI - (\bK \bXi)^2 )^{-1} \bV \; .
\ee
The last expression is convenient in order to split $\varphi$ into an even and an odd part. 
In this way we obtain (\ref{cU,cD_sol}).

\paragraph{Semi-discrete NLS.}
Writing
\be
   \bX = \bbS \tbX \, , \qquad
   \bP = \bJ \bbS^{-1} \tbP \, ,   \label{sd_bX,bP}
\ee 
we find that $\bar{\d} \bX = (\d \bX) \, \bP = (\d \bX) \, \bJ \, \bbS^{-1} \tbP$ 
is equivalent to
\be
   -\imag \, \dot{\tbX} = (\tbX^+ - \tbX) \bJ \tbP \, , \qquad
    \tbX - \tbX^- = \frac{1}{2} [\bJ,\tbX] \bJ \tbP \, , 
\ee
with the general solution
\be
  \tbX = \bA_0 + \bA_1 \, \bSigma \, \bXi \, , 
\ee
where
\be
    \bXi = \bS^x \, e^{- \imag \, t \, \omega(\bS) \, \bJ} \, , \quad
    \omega(\bS) = \bS + \bS^{-1} - 2 \bI \, , \quad
    \bS = (\bI + \tbP)^{-1} \, ,    \label{bS<->tbP}
\ee
and $x \in \bbZ$. Using 
the same redefinitions (\ref{K->hatK,U->hatU}) as in the continuous NLS case, 
(\ref{constr_univ}) becomes
\be
    \bS^{-1} \bK - \bK \, \bS = \bV \bU \, ,  \label{AL_rank}
\ee
and (\ref{cU,cD_sol}) is obtained in precisely the same way as in the continuous NLS case.

\paragraph{Fully discrete NLS.}
Using again (\ref{sd_bX,bP}), 
we find that $\bar{\d} \bX = (\d \bX) \, \bP$ 
is equivalent to
\be
   - \imag \, (\tbX_+ - \tbX) = (\tbX^+_+ - \tbX) \, \bJ \tbP \, , \qquad
    \tbX - \tbX^- = \frac{1}{2} [\bJ,\tbX] \, \bJ \tbP \; .  \label{dAL_Xeqs}
\ee
Assuming that $\bI - \imag \, \bJ \tbP$ is invertible, it follows 
that the even part of $\tbX$ is constant. If furthermore $\bS = (\bI+\tbP)^{-1}$  
and $[\bI-\imag \, \bJ (\bI -\bS)]^{-1}$ exist, the above system yields the 
following equations for the odd part of $\tbX$, 
\be
   \tbX_o^+ = \tbX_o \, \bS \, , \qquad  
  (\tbX_o)_+ = \tbX_o \, [\bI + \imag \, \bJ (\bI - \bS^{-1})] \, 
               [\bI-\imag \, \bJ (\bI -\bS)]^{-1} \; .
\ee
Hence we obtain the solution 
\be
    \tbX = \bA_0 + \bA_1 \, \bSigma \, \bXi \, , 
\ee
with constant even coefficient matrices and 
\be
  \bXi = \bS^x \, [(\bI+ \imag \, \bJ (\bI-\bS^{-1})] \, [\bI-\imag \, \bJ(\bI-\bS))^{-1}]^t 
         \qquad \qquad  x,t \in \bbZ \; .
\ee
Again, assuming $\bA_0$ invertible and adopting the definitions 
(\ref{K->hatK,U->hatU}), we arrive at (\ref{cU,cD_sol}) and (\ref{AL_rank}).

\subsection{Transformations of the matrix data determining solutions}
\label{subsec:transf}

\subsubsection{Similarity transformations} 
A transformation 
\be
    \bJ \mapsto \bM \bJ \bM^{-1} \, , \quad
    \bK \mapsto \bM \bK \bM^{-1} \, , \quad
    \bS \mapsto \bM \bS \bM^{-1} \, , \quad 
    \bU \mapsto \bU \bM^{-1} \, , \quad 
    \bV \mapsto \bM \bV \, , 
\ee 
where $\bM$ is any constant invertible $n \times n$ matrix, 
obviously leaves $\cU$ and $\cD$ in proposition~\ref{prop:NLSsol} invariant, and 
also the corresponding Sylvester equation. 
It can thus be used to achieve that $\bJ$ takes the form
\be
    \bJ = \left( \begin{array}{cc} I_{n_1} & 0 \\ 0 & -I_{n_2} \end{array} \right) \, , 
              \label{bJ_normalform}
\ee
with $n_1,n_2 \in \bbN$ such that $n_1+n_2 = n$. 
The remaining similarity transformations such that
\be
    [ \bJ , \bM ] = 0    \label{[bJ,bM]=0}
\ee
can then be exploited to diminish the redundancy in the ``parametrisation'' of the solutions 
by the set of matrices $\bS, \bU, \bV$. In particular, it can be used to transform $\bS$ 
to Jordan normal form.

\subsubsection{Reflection symmetries} 
\label{subsubsec:reflection}
If $\bK$ is invertible and $\bPi$ a constant $n \times n$ matrix with
\be
     \bJ \bPi = - \bPi \bJ \, , 
\ee
then the transformation
\be
    \bS \mapsto \bS' \, , \; \;
    \bU \mapsto - \bU \bK^{-1} \bPi^{-1} \, , \; \; 
    \bV \mapsto \bPi \bK^{-1} \bV \, , \; \;
    \bK \mapsto \bPi \bK^{-1} \bPi^{-1} \, ,     \label{bS_reflection}
\ee
with 
\be
   \bS' = \left\{ \begin{array}{l@{\qquad}l}
          -\bPi \bS \bPi^{-1} & \mbox{continuous NLS} \\
          \bPi \bS^{-1} \bPi^{-1} & \mbox{semi- and fully discrete NLS} \end{array} \right.
                 \label{bS'}
\ee
implies
\be
    \bXi \mapsto \bPi \, \bXi^{-1} \, \bPi^{-1} 
\ee
and leaves $\cU$ in proposition~\ref{prop:NLSsol} and also the corresponding Sylvester equation invariant. 
We refer to such a transformation as a \emph{reflection}. Since 
\be
   \cD \mapsto \bU \, (\bXi \bK)^{-1} \, ( \bXi^{-1} - \bK \bXi \bK )^{-1} \, \bV 
               = \cD + \bU \, \bK^{-1} \, \bV    \, ,
\ee
$\cD$ only changes by a constant matrix, which can be dropped as pointed out above. 
Hence a reflection is a symmetry of the solutions given in proposition~\ref{prop:NLSsol} for 
all the NLS systems. 
A symmetry transformation of this kind for continuous NLS solutions already appeared 
in \cite{ABDM09sym}.

\subsubsection{Reparametrisation transformations} 
\label{subsubsec:reparametrisation}
Let $\bA,\bB$ be any constant invertible 
$n \times n$ matrices that commute with $\bJ$ and $\bS$, i.e.
\be
   [\bJ,\bA] = [\bJ,\bB] = [\bS,\bA] = [\bS,\bB] = 0 \; .  \label{reparam_1}
\ee
Then 
\be
   \bS \mapsto \bS \, , \quad 
   \bU \mapsto \bU \bB^{-1} \, , \quad 
   \bV \mapsto \bA^{-1} \bV \, , \quad
   \bK \mapsto \bA^{-1} \bK \bB^{-1} \, , \quad
   \bXi \mapsto \bB \bA \, \bXi \, ,    \label{reparam_2}
\ee
is a \emph{reparametrisation} of the solution (\ref{cU,cD_sol}). 
This is helpful in order to reduce the number of parameters of $\bU$ and $\bV$ 
(see in particular section~\ref{subsec:ccred_solitons}).

\subsection{Superposition of solutions}
\label{subsec:NLS_super}
Let $(\bJ_i,\bS_i, \bU_i, \bV_i)$, $i=1,2$, be data that determine two solutions $(\cU_i, \cD_i)$ 
of one of the three systems (\ref{NLS_sys}), (\ref{sd_NLS_sys}) or (\ref{fd_NLS_sys}). 
Let $\bK_i$, $i=1,2$, be corresponding solutions of the Sylvester equations. Then 
\be
   \bJ = \left( \begin{array}{cc} \bJ_1 & 0 \\ 0 & \bJ_2 \end{array} \right) , \,
   \bS = \left( \begin{array}{cc} \bS_1 & 0 \\ 0 & \bS_2 \end{array} \right) , \,
   \bU = \left( \begin{array}{cc} \bU_1 & \bU_2 \end{array} \right) , \,
   \bV = \left( \begin{array}{c} \bV_1 \\ \bV_2 \end{array} \right) , \,
   \bK = \left( \begin{array}{cc} \bK_1 & \bK_{12} \\ \bK_{21} & \bK_2 \end{array} \right) \;
                       \label{sup_data}
\ee
determine new solutions if $\bK_{12}$ and $\bK_{21}$ satisfy
\be
    \bJ_1 \bK_{12} = - \bK_{12} \bJ_2 \, , \qquad
    \bJ_2 \bK_{21} = - \bK_{21} \bJ_1 \, ,
\ee
and solve the respective equations below. \\
\textbf{Continuous NLS}:
\be
   \bS_1 \bK_{12} + \bK_{12} \bS_2 = \bV_1 \bU_2 \, , \qquad
   \bS_2 \bK_{21} + \bK_{21} \bS_1 = \bV_2 \bU_1 \; .    \label{NLS_bK12,bK21}
\ee
These equations possess unique solutions if $\sigma(\bS_1) \cap \sigma(-\bS_2) = \emptyset$. \\
\textbf{Semi-discrete and fully discrete NLS}:
\be
   \bS_1^{-1} \bK_{12} - \bK_{12} \bS_2 = \bV_1 \bU_2 \, , \qquad
   \bS_2^{-1} \bK_{21} - \bK_{21} \bS_1 = \bV_2 \bU_1  \; .    \label{dNLS_bK12,bK21}
\ee
These equations have unique solutions if $\sigma(\bS_1) \cap \sigma(\bS_2^{-1}) = \emptyset$.
(\ref{Sylvester_spec_cond}) applied to $\bS = \mbox{block-diag}(\bS_1,\bS_2)$ contains the 
abovementioned spectrum conditions and is therefore a sufficient (but not necessary) condition 
for the above superposition principle to generate again a solution. 
Let us list some fairly obvious properties of superpositions. 
\begin{enumerate}
\item The superposition of two solutions is commutative. Note that
\be
    \bM = \left( \begin{array}{cc} 0 & \bI_{n_2} \\ \bI_{n_1} & 0 \end{array} \right)
\ee
is a similarity transformation that exchanges the matrix blocks corresponding to the 
two solutions in the superposed data.
\item Superpositions are associative. 
\item A similarity transformation of the matrix data corresponding to a constituent 
solution extends to a similarity transformation of the superposition, 
hence remains an equivalence transformation. This is simply based on the fact that 
a similarity transformation of the superposed data, with an invertible matrix 
of the form 
\be
    \bM = \left( \begin{array}{cc} \bM_1 & 0 \\ 0 & \bM_2 \end{array} \right) \, , 
\ee
covers arbitrary similarity transformations of the constituent data. 
\item A reparametrisation transformation of a constituent extends to a 
reparametrisation of the superposition, hence remains an equivalence transformation. 
(Choose $\bA,\bB$ block-diagonal in (\ref{reparam_1}) and (\ref{reparam_2}).)
\end{enumerate}

Recall that, for any data $(\bJ,\bS,\bU,\bV)$ that determine a solution, 
via a similarity transformation $\bJ$ can be brought to standard form (\ref{bJ_normalform}) 
and $\bS$ simultaneously to Jordan normal form. 
We call the data $(\bJ,\bS,\bU,\bV)$ \emph{simple} if at least one of the blocks of 
$\bS$, corresponding to a block of $\bJ$, is a single Jordan block. 
For any solution given by data $(\bJ,\bS,\bU,\bV)$ that are not simple, we have 
the block structure in (\ref{sup_data}). These data have to solve the Sylvester equation, 
hence in particular there are matrices $\bK_1,\bK_2$ solving its diagonal blocks. 
It follows that $(\bJ,\bS,\bU,\bV)$ decomposes into data $(\bJ_i,\bS_i, \bU_i, \bV_i)$, 
$i=1,2$, that determine solutions. As a consequence, for any solution obtained from 
non-simple data, the simple subdata also determine solutions. In this sense, 
the solution can always be regarded as a superposition of solutions. 
\vskip.1cm

Though reflection symmetries are not quite compatible with the structure of 
a superposition, we still have the following result (based on arguments in \cite{ABDM09sym}).

\begin{proposition}
\label{prop:reflection_super}
Let (\ref{Sylvester_spec_cond}) hold for $\bS = \mbox{\textnormal{block-diag}}(\bS_1,\bS_2)$. 
A reflection symmetry of (only) one of the two matrix data sets can then be extended in such a way 
that it is also a symmetry of the superposition solution. 
\end{proposition}
\noindent
\textit{Proof:} We consider a reflection symmetry on the second data set. Hence we have to 
assume that $\bK_2$ is invertible. 
Writing $\bZ_i = \bXi_i^{-1} - \bK_i \bXi_i \bK_i$, so that 
$\cU_i = \bU_i \bZ_i^{-1} \bV_i$, we have $\cU = \bU \bZ^{-1} \bV$ with
\bez
 \bZ = \left(\begin{array}{cc}
   \bZ_1 - \bK_{12} \bXi_2 \bK_{21} & - \bK_1 \bXi_1 \bK_{12} - \bK_{12} \bXi_2 \bK_2 \\
   - \bK_2 \bXi_2 \bK_{21} - \bK_{21} \bXi_1 \bK_1 & \bZ_2 - \bK_{21} \bXi_1 \bK_{12}
             \end{array}\right) \; .
\eez
According to the reflection transformation rules (\ref{bS_reflection}), we have
\bez
    \bS' = \left(\begin{array}{cc} \bS_1 & 0 \\ 0 & \bS'_2 \end{array}\right) \, ,
\eez
where $\bS'_2$ is defined for the three NLS cases in (\ref{bS'}). 
The other transformed quantities $\bU_2', \bV_2', \bK_2'$ are given by (\ref{bS_reflection}). 
Corresponding transformation rules for $\bK_{12}$ and $\bK_{21}$ should then be read off 
from (\ref{NLS_bK12,bK21}), respectively (\ref{dNLS_bK12,bK21}). It turns out that also 
$\bU_1$ and $\bV_1$ have to be transformed. In fact, setting
\bez
    \bU' = \bU \bA \, , \; \bV' = \bB \, \bV \quad \mbox{with} \quad 
    \bA = \left(\begin{array}{cc} \bI & 0  \\
                     - \bK_2^{-1} \bK_{21} & -\bK_2^{-1} \bPi^{-1}
                     \end{array}\right) \, , \; 
    \bB = \left(\begin{array}{cc} \bI & -\bK_{12} \bK_2^{-1} \\
                                      0 & \bPi \bK_2^{-1} 
                  \end{array}\right) \, , 
\eez
and 
\bez
  \bK' = \left(\begin{array}{cc} \bK_1 - \bK_{12} \bK_2^{-1}\bK_{21} & -\bK_{12} \bK_2^{-1} \bPi^{-1} \\
         \bPi \bK_2^{-1} \bK_{21} & \bPi \bK_2^{-1} \bPi^{-1}
              \end{array}\right) 
       = \left(\begin{array}{cc} \bK_1 & \bK_{12} \\ 0 & - \bPi \end{array}\right) \, \bA \, ,
\eez 
one obtains $\bZ' = \bB \bZ \bA$ and hence $\cU' = \bU' \, {\bZ'}^{-1} \, \bV' = \cU$. 
Furthermore, we find that
\bez
    \cD' = \cD + \bU \left(\begin{array}{cc} 0 & 0 \\ 0 & \bK_2^{-1} \end{array}\right) \bV \, ,
\eez
i.e. $\cD$ only changes by an irrelevant constant summand.
\hfill $\square$
\vskip.2cm

If we superpose the same data, we obtain nothing new. More generally, we have the 
following result.\footnote{One can still refine this proposition in order to weaken 
the assumptions.}

\begin{proposition}
\label{prop:sup_sim_data}
Let $(\bJ_i,\bS_i,\bU_i,\bV_i)$, $i=1,2$, be two data sets that determine solutions of an  
NLS system according to proposition~\ref{prop:NLSsol} and let (\ref{Sylvester_spec_cond}) 
hold for $\bS = \mbox{\textnormal{block-diag}}(\bS_1,\bS_2)$.
Suppose there are matrices $\bA, \bB$ such that $\bU_2 = \bU_1 \bB$, $\bV_2 = \bA \bV_1$,  
$\bA \bS_1 = \bS_2  \bA$ and $\bS_1 \bB = \bB \, \bS_2$. 
Let further $\bC = \bI + \bB \bA$ be invertible. 
Then the superposed data (\ref{sup_data}) determine a solution that can also be obtained 
from the data $(\bJ_1,\bS_1,\bU_1 \bC,\bV_1)$ (i.e. simply by a redefinition of $\bU_1$). 
\end{proposition}
\noindent
\textit{Proof:} We consider the continuous NLS case only. The discrete cases are treated 
in a similar way. 
The assumptions together with the respective Sylvester equations for the data 
$(\bJ_i,\bS_i,\bU_i,\bV_i)$ imply 
\bez
    \bS_2 \, ( \bK_2 - \bA \, \bK_1 \, \bB) + ( \bK_2 - \bA \, \bK_1 \, \bB) \, \bS_2 = 0 \; .
\eez 
Since the assumption (\ref{Sylvester_spec_cond}) in particular implies 
$\sigma(\bS_2) \cap \sigma(-\bS_2) =\emptyset$, the last equation 
implies $\bK_2 = \bA \, \bK_1 \, \bB$. 
For (\ref{sup_data}) to determine a solution, (\ref{NLS_bK12,bK21}) has to be satisfied, and thus
\bez
   \bS_1 \, ( \bK_{12} - \bK_1 \, \bB) + ( \bK_{12} - \bK_1 \, \bB ) \, \bS_2 = 0 \, , \quad
   \bS_2 \, ( \bK_{21} - \bA \bK_1) + ( \bK_{21} - \bA \bK_1 ) \, \bS_1 = 0 \; .
\eez
Since (\ref{Sylvester_spec_cond}) also implies $\sigma(\bS_1) \cap \sigma(-\bS_2) =\emptyset$, 
these equations have the unique solutions $ \bK_{12} = \bK_1 \, \bB$ and $\bK_{21} = \bA \bK_1$, 
hence
\bez
    \bK = \left( \begin{array}{c} \bI \\ \bA \end{array} \right) \, \bK_1 \, 
          \left( \begin{array}{cc} \bI & \bB \end{array} \right) \; .
\eez
This in turn implies
\bez
   \bW := \bI - \bXi \, \bK \, \bXi \, \bK 
        = \bI - \left( \begin{array}{c} \bI \\ \bA \end{array} \right) \,
                 \bXi_1 \bK_1 \, \bC \,  \bXi_1 \bK_1 \, 
                 \left( \begin{array}{cc} \bI & \bB \end{array} \right) \, ,
\eez
and thus, assuming that $\bC = \bI + \bB \bA$ is invertible and introducing 
$\tilde{\bXi}_1 = \bC \, \bXi_1$, 
\bez
  \bW^{-1} = \bI + \left( \begin{array}{c} \bI \\ \bA \end{array} \right) \,
                   \bC^{-1} \, (\tilde{\bW}_1^{-1} -\bI) \, 
                   \left( \begin{array}{cc} \bI & \bB \end{array} \right) \, ,
\eez
where $\tilde{\bW}_1 = \bI - \tilde{\bXi}_1 \, \bK_1 \, \tilde{\bXi}_1 \, \bK_1$.
With straightforward computations (\ref{cU,cD_sol}) now leads to 
\bez
  \cU = \bU_1 \, \left( \tilde{\bXi}_1^{-1} - \bK_1 \, \tilde{\bXi}_1 \, \bK_1 \right)^{-1} \bV_1 \, , 
          \qquad
  \cD = \bU_1 \, \tilde{\bXi}_1 \, \bK_1 \, \left( \tilde{\bXi}_1^{-1} - \bK_1 \, \tilde{\bXi}_1 
        \, \bK_1 \right)^{-1} \bV_1 \, ,
\eez
from which our assertion is easily deduced. 
\hfill $\square$

\subsection{Imposing reduction conditions on the solutions}
\label{subsec:redsol}

\subsubsection{Complex conjugation reduction} 
Let $\bGamma$ be an $n \times n$ matrix with the properties 
\be
     \bJ \bGamma = - \bGamma \bJ \, , \quad 
     \bGamma^\ast = \bGamma^{-1}  \; .  \label{bGamma_conds}
\ee
Choosing the normal form (\ref{bJ_normalform}) of $\bJ$, this implies that $\bGamma$ 
is off-block-diagonal. 
The last condition in (\ref{bGamma_conds}) requires that $n$ is even and 
\be
      n_1 = n_2 \; . 
\ee
Imposing the following constraints on the matrices $\bS, \bU, \bV$, 
\be
    \bS^\ast = \bGamma \bS \bGamma^{-1} \, , \qquad
    \bU^\ast = \epsilon \, \epsilon' \, \Gamma \bU \bGamma^{-1} \, , \qquad
    \bV^\ast = \epsilon' \, \bGamma \bV \Gamma^{-1} \, ,   \label{ccred_S,U,V}
\ee
with $\epsilon' = \pm 1$ and $\Gamma$ as in section~\ref{sec:NLS}, the Sylvester equations demand 
\be
    \bK^\ast = \epsilon \, \bGamma \bK \bGamma^{-1} \, ,   \label{ccred_K}
\ee
and $\cU$ and $\cD$ given by (\ref{cU,cD_sol}) indeed satisfy the reduction conditions 
(\ref{ccred_U,D}). 

The transformations considered in section~\ref{subsec:transf} should now be restricted 
in such a way that the reduction conditions are preserved. 
The matrix $\bM$ of a similarity transformation (preserving the form of $\bJ$), 
now also has to satisfy $\bM^\ast = \bGamma \bM \bGamma^{-1}$.
For a reflection symmetry, the matrix $\bPi$ has to satisfy 
$\bPi^\ast = \epsilon \, \bGamma \bPi \bGamma^{-1}$. In case of a reparametrisation 
transformation, we have to add the conditions $\bA^\ast = \bGamma \bA \bGamma^{-1}$ and 
$\bB^\ast = \bGamma \bB \bGamma^{-1}$.

\subsubsection{Hermitian conjugation reduction} 
This only applies to the continuous NLS case. 
 From (\ref{cU,cD_sol}) we obtain
\be
  \cU^\dagger = \bV^\dagger \, (\bXi^{\dagger -1} - \bK^\dagger \bXi^\dagger \bK^\dagger)^{-1} 
                \, \bU^\dagger   \, , 
\ee
In order to achieve the reduction condition $\cU^\dagger = \epsilon \, \cU$ with $\epsilon = \pm 1$, 
we set
\be
    \bS^\dagger = \bT \bS \bT^{-1} \, , \qquad
    \bU = \bV^\dagger \, \bT  \, ,   \label{Hred_Sdag,U}
\ee
with an invertible $n \times n$ matrix $\bT$ with the properties
\be
    \bJ \bT = - \bT \bJ \, , \qquad 
    \bT^\dagger = \epsilon \, \bT \; .   \label{Hred_T_conds}
\ee
These conditions again restrict us to the case where 
\be
      n_1 = n_2 \; . 
\ee
The Sylvester equation now reads
\be
      \bS \bK + \bK \bS = \bV \bV^\dagger \bT \, ,  \label{Hred_NLS_rank}
\ee
and is preserved by Hermitian conjugation if
\be
    \bK^\dagger = \epsilon \, \bT \bK \bT^{-1} \; .  \label{Hred_Kdag}
\ee
Then the expression for $\cU$ in (\ref{cU,cD_sol}) indeed satisfies
$\cU^\dagger = \epsilon \, \cU$. Furthermore, as a consequence of the identity 
$\bZ^{-1} \bK \bXi = \bXi \bK \bZ^{-1}$ where $\bZ = \bXi^{-1} - \bK \bXi \bK$, 
the expression for $\cD$ in (\ref{cU,cD_sol}) satisfies $\cD^\dagger = \cD$. 

 For a similarity transformation with matrix $\bM$, which commutes with $\bJ$, we have to require 
$\bT \mapsto (\bM^\dagger)^{-1} \bT \bM^\dagger$. With $\bJ$ given by (\ref{bJ_normalform}) 
(where $n_1=n_2$), this can be used to achieve that 
\be
   \bT = \left(\begin{array}{cc} 0 & I_{n_1} \\ \epsilon \, I_{n_1} & 0 \end{array} \right) \; .
                  \label{Hred_bGamma_choice}
\ee

After having fixed $\bT$, the matrix $\bM$ of a similarity transformation preserving 
the Hermitian reduction conditions has to satisfy $\bM^\dagger = \bT \bM^{-1} \bT^{-1}$. 
For a reflection symmetry, $\bPi^\dagger = - \epsilon \, \bT \bPi^{-1} \bT^{-1}$. 
In case of a reparametrisation, $\bA^\dagger = \bT \bB \bT^{-1}$.

\section{Block-decomposition of the NLS systems and their solutions} 
\label{sec:decomp}
\setcounter{equation}{0}
In this section we express the NLS systems in a more familiar form\footnote{See also 
appendix~E for a derivation of Lax pairs for these NLS systems within the bidifferential 
calculus framework. }
and translate proposition~\ref{prop:NLSsol} correspondingly. 
We choose for $J$ the normal form (\ref{J_normalform}) and write
\be
  \varphi = \left(\begin{array}{cc} p & q \\ \bar{q} & \bar{p} \end{array} \right) \, , \label{varphi_matrix}
\ee
where $p,\bar{p},q,\bar{q}$ are, respectively, $m_1 \times m_1$, $m_2 \times m_2$, $m_1 \times m_2$ 
and $m_2 \times m_1$ matrices. As a consequence of (\ref{JcD,JcU}), 
\be
   \cU = \left(\begin{array}{cc} 0 & q \\ \bar{q} & 0 \end{array} \right) \, , \qquad
   \cD = \left(\begin{array}{cc} p & 0 \\ 0 & \bar{p} \end{array} \right) \; .  
\ee

\paragraph{Continuous NLS system.} 
(\ref{NLS_sys}) becomes the matrix \emph{NLS system} 
(see e.g. \cite{Ford+Kuli83,Ablo+Clar91,Harn+Wiss92,APT04book})
\be
   \imag \, q_t + q_{xx} - 2 \, q \, \bar{q} \, q = 0 \, , \quad 
   - \imag \, \bar{q}_t + \bar{q}_{xx} - 2 \, \bar{q} \, q \, \bar{q} = 0  \, , \label{NLSsystem}
\ee
together with
\be
    p_x = - q \, \bar{q} \, , \quad \bar{p}_x = - \bar{q} \, q  \; .   \label{NLS_p-system}
\ee

\paragraph{Semi-discrete NLS system.} 
(\ref{sd_NLS_sys}) leads to
\be
    p^+ - p = -q^+ \bar{q} \, , \qquad 
    \bar{p}^+ - \bar{p} = -\bar{q}^+ q \, , 
\ee
and the matrix \emph{Ablowitz-Ladik (AL) system} 
\be
   \imag \, \dot{q} + q^+ - 2 q + q^- - q^+ \bar{q} \, q - q \, \bar{q} \, q^- &=& 0 \, , \nonumber \\
   - \imag \, \dot{\bar{q}} + \bar{q}^+ - 2 \bar{q} + \bar{q}^- 
   - \bar{q}^+ q \, \bar{q} - \bar{q} \, q \, \bar{q}^- &=& 0 \; .   \label{ALsystem}
\ee

\paragraph{Fully discrete NLS system.} 
 From (\ref{fd_NLS_sys}) we obtain 
\be
    p^+ - p= -q^+ \bar{q} \, , \qquad
    \bar{p}^+ - \bar{p} = -\bar{q}^+ q \, ,         \label{sdNLS_p_eq}
\ee
(as in the semi-discrete case) and
\be
  \imag \, (q_+ - q) + (q^+ - q)_+ - (q-q^-) - (q^+ \bar{q} q)_+ - q \bar{q} q^- 
  + (p_+ - p) q_+ + q (\bar{p}_+ - \bar{p}) 
         &=& 0  \, ,                   \nonumber \\
  - \imag \, (\bar{q}_+ - \bar{q}) + (\bar{q}^+ - \bar{q})_+ - (\bar{q}-\bar{q}^-) - (\bar{q}^+ q \bar{q})_+ 
  - \bar{q} q \bar{q}^- + (\bar{p}_+ - \bar{p}) \bar{q}_+ + \bar{q} (p_+ - p) 
         &=& 0  \; .          \label{sdNLS_qr_eq}
\ee
Unlike the situation in the continuous NLS and semi-discrete NLS (AL) cases, here the variables 
$p,\bar{p}$ cannot be eliminated from the equations for $q$ and $\bar{q}$ without introduction of non-local 
terms. A reduction of this system appeared in \cite{Ablo+Ladi76SAM} (see also appendix~C). 
Integrable full discretisations of matrix NLS systems appeared in \cite{Tsuc00RMP}.

\paragraph{Solutions.} 
For $\bJ$ we choose the normal form (\ref{bJ_normalform}). 
 From the properties of the matrices appearing in proposition~\ref{prop:NLSsol}, 
we find the following block structures, 
\be
    \bS = \left(\begin{array}{cc} S & 0 \\ 0 & \bar{S} \end{array} \right) \, , \quad
    \bU = \left(\begin{array}{cc} 0 & U \\ \bar{U} & 0 \end{array} \right) \, , \quad
    \bV = \left(\begin{array}{cc} V & 0 \\ 0 & \bar{V} \end{array} \right) \, , \quad
    \bK = \left(\begin{array}{cc} 0 & K \\ \bar{K} & 0 \end{array} \right)     \, ,
                 \label{S,U,V,K_blocks}
\ee 
and $\bXi$ is block-diagonal with diagonal blocks $\Xi,\bar{\Xi}$. 
Now proposition~\ref{prop:NLSsol} translates into the following statement.

\begin{proposition}
\label{prop:sol_NLSsystems}
Let  
\renewcommand{\arraystretch}{1.5}
\begin{center}
\begin{tabular}{c|c|c|c|c|c|c|c|c} 
        & $U$ & $\bar{U}$ & $V$ & $\bar{V}$ & $S$ & $\bar{S}$ & $K$ & $\bar{K}$  \\
\hline
 size   & $m_1 \times n_2$ & $m_2 \times n_1$ & $n_1 \times m_1$ & $n_2 \times m_2$ 
        & $n_1 \times n_1$ & $n_2 \times n_2$ & $n_1 \times n_2$ & $n_2 \times n_1$
\end{tabular}
\end{center}
be constant complex matrices, with $S$ and $\bar{S}$ invertible in the discrete NLS cases, and 
\begin{center}
\begin{tabular}{c|c|c|c|l} 
 NLS  & Sylvester equations & $\Xi$ & $\bar{\Xi}$ &    \\
\hline
 continuous  & \begin{tabular}{c} $S K + K \bar{S} = V U$ \\ 
                                  $\bar{S} \bar{K} + \bar{K} S = \bar{V} \bar{U}$ \end{tabular} 
             & $e^{-x \, S - \imag \, t \, S^2}$ & $e^{-x \, \bar{S} + \imag \, t \, \bar{S}^2}$ 
             &  $x,t \in \bbR$   \\
\hline 
 semi-discrete & \begin{tabular}{c} $S^{-1} K - K \bar{S} = V U$ \\
                                  $\bar{S}^{-1} \bar{K} - \bar{K} S = \bar{V} \bar{U}$
                 \end{tabular} 
             & $S^x \, e^{- \imag \, t \, \omega(S)}$ 
             & $\bar{S}^x \, e^{\imag \, t \, \omega(\bar{S})}$   
             & $x \in \bbZ$, $t \in \bbR$        \\
\hline
 fully discrete& \begin{tabular}{c} $S^{-1} K - K \bar{S} = V U$ \\
                                  $\bar{S}^{-1} \bar{K} - \bar{K} S = \bar{V} \bar{U}$ \end{tabular}
             & $S^x \, \, \Omega(S)^t$ 
             & $\bar{S}^x \, \bar{\Omega}(\bar{S})^t$ 
             & $x,t \in \bbZ$ 
\end{tabular}
\end{center}
where 
\be
    \omega(S) &=& S + S^{-1}-2I_{n_1}  \, , \label{omega}       \\
    \Omega(S) &=& [I_{n_1} + \imag \, (I_{n_1}- S^{-1})][I_{n_1} 
          - \imag \, (I_{n_1}- S)]^{-1} \, , \nonumber \\
    \bar{\Omega}(\bar{S}) &=& [I_{n_2} - \imag \, (I_{n_2} - \bar{S}^{-1})][I_{n_2} 
          + \imag \, (I_{n_2} - \bar{S})]^{-1}   \; .
\ee
Then
\be
  \begin{array}{l@{\quad}l}
  q = U \, \left(\bar{\Xi}^{-1} - \bar{K} \, \Xi K \right)^{-1} \, \bar{V} \, , & 
  \bar{q} = \bar{U} \, \left(\Xi^{-1} - K \bar{\Xi} \bar{K} \right)^{-1} \, V  \, ,  \\
  p = U \bar{\Xi} \bar{K} \, \left(\Xi^{-1} - K \bar{\Xi} \bar{K} \right)^{-1} \, V  \, , & 
  \bar{p} = \bar{U} \Xi K \, \left(\bar{\Xi}^{-1} - \bar{K} \Xi K \right)^{-1} \, \bar{V} \, ,
  \end{array}               \label{q_p_sol}
\ee
solves the respective NLS system.
\hfill $\square$
\end{proposition}

\subsection{Transformations of the matrix data determining solutions}
The transformations described in section~\ref{subsec:transf} decompose correspondingly. \\
\textbf{1.} A similarity transformation satisfies (\ref{[bJ,bM]=0}), with $\bJ$ given 
by (\ref{bJ_normalform}), iff the transformation matrix $\bM$ is block-diagonal, i.e. 
$\bM = \mbox{block-diag}(M,\bar{M})$. Then we obtain 
\be
 &&  S \mapsto M \, S \, M^{-1} \, , \quad 
   \bar{S} \mapsto \bar{M} \, \bar{S} \, \bar{M}^{-1} \, , \quad
   U \mapsto U \, \bar{M}^{-1} \, , \quad 
   \bar{U} \mapsto \bar{U} \, M^{-1} \, , \nonumber \\
 &&  K \mapsto M \, K \, \bar{M}^{-1}  \, , \quad
   \bar{K} \mapsto \bar{M} \, \bar{K} \, M^{-1} \, , \quad
   V \mapsto M \, V  \, , \quad 
   \bar{V} \mapsto \bar{M} \, \bar{V}  \; .
\ee
As a consequence, without restriction of generality we can assume that 
$S$ and $\bar{S}$ have Jordan normal form.  \\
\textbf{2.} The matrix $\bPi$ of a reflection symmetry has to be off-block-diagonal. 
For later use we consider the example
\be
    \bPi = \left(\begin{array}{cc} 0 & \beta \, I_{n_1} \\ \bar{\beta} \, I_{n_1} & 0 \end{array}\right) \, ,
\ee
assuming $n_1 = n_2$ (and thus $n$ even). The transformation then acts as follows, 
\be
    \begin{array}{lclcl}
    K \mapsto \beta \bar{\beta}^{-1} K^{-1} \, , & \qquad &
    U \mapsto -\beta \, U K^{-1} \, , & \qquad &
    V \mapsto \bar{\beta}^{-1} K^{-1} V \, , \\[.2cm]
    \bar{K} \mapsto \beta^{-1} \bar{\beta} \, \bar{K}^{-1} \, , & \qquad &
    \bar{U} \mapsto -\bar{\beta} \, \bar{U} \bar{K}^{-1} \, , & \qquad &
    \bar{V} \mapsto \beta^{-1} \bar{K}^{-1} \bar{V} \, ,
    \end{array}
\ee
together with 
\be 
    S \mapsto - \bar{S} \, , \qquad
    \bar{S} \mapsto -S \, , 
\ee
for continuous NLS, and
\be
    S \mapsto  \bar{S}^{-1} \, , \qquad
    \bar{S} \mapsto S^{-1} 
\ee
for semi-discrete and fully discrete NLS. This implies
\be
    \Xi \mapsto \bar{\Xi}^{-1} \, , \qquad
    \bar{\Xi} \mapsto \Xi^{-1}  \; .
\ee
\textbf{3.} In case of a reparametrisation transformation, we have
$\bA = \mbox{block-diag}(A,\bar{A})$, $\bB = \mbox{block-diag}(B,\bar{B})$, and 
$[S,A]=[S,B] = [\bar{S},\bar{A}] = [\bar{S},\bar{B}]=0$. The transformation is then 
given by
\be
  && S \mapsto S \, , \quad
   \bar{S} \mapsto \bar{S} \, , \quad
   U \mapsto U \bar{B}^{-1} \, , \quad
   \bar{U} \mapsto \bar{U} B^{-1} \, , \quad
   V \mapsto A^{-1} V \, , \quad
   \bar{V} \mapsto \bar{A}^{-1} \bar{V}  \nonumber \\
  &&  K \mapsto A^{-1} K \bar{B}^{-1} \, , \quad
      \bar{K} \mapsto \bar{A}^{-1} \bar{K} B^{-1} \, , \quad
      \Xi \mapsto B A \, \Xi \, , \quad
      \bar{\Xi} \mapsto \bar{B} \bar{A} \, \bar{\Xi} \; .  \label{decomp_repar_trans}
\ee
\textbf{4.} A transformation 
\be
    U \mapsto \cS U \, , \quad 
    \bar{U} \mapsto \bar{\cS} \bar{U} \, , \quad 
    V \mapsto V \cS^{-1} \, , \quad
    \bar{V} \mapsto \bar{V} \bar{\cS}^{-1} \, ,   \label{ext_transf}
\ee 
with invertible constant matrices $\cS, \bar{\cS}$, has the effect 
\be
    q \mapsto \cS q \bar{\cS}^{-1} \, , \quad
    \bar{q} \mapsto \bar{\cS} \bar{q} \cS^{-1} \, , \quad
    p \mapsto \cS p \cS^{-1} \, , \quad
    \bar{p} \mapsto \bar{\cS} \bar{p} \bar{\cS}^{-1} \; .  
\ee
In contrast to the previous transformations, this is not a symmetry 
of the solutions. But it may be regarded as an equivalence transformation of solutions.

\section{The complex conjugation reduction} 
\label{sec:cc-reduction}
\setcounter{equation}{0}
We recall that, for 
\be
           m_1=m_2 \, , 
\ee
all three NLS systems admit the complex conjugation reduction introduced in section~\ref{sec:NLS}. 
In the following we rename $m_1$ to $m$. 
Choosing $\Gamma$ as in (\ref{ccred_Gamma_m1=m2}), the reduction conditions (\ref{ccred_U,D}) 
take the form
\be
    \bar{q} = \epsilon \, q^\ast  \quad \mbox{with} \quad \epsilon = \pm 1 \, , 
    \qquad \bar{p} = p^\ast  \, ,   \label{NLS_ccred}
\ee
Here $q,\bar{q},p,\bar{p}$ are all \emph{square} matrices of size $m \times m$.

\paragraph{Continuous NLS.} 
(\ref{NLS_ccred}) reduces the (continuous) NLS system to the following (square) matrix NLS equations,
\be
  \imag \, q_t + q_{xx} - 2 \, \epsilon \, q q^\ast q = 0 \, , \qquad
  p_x = - \epsilon \, q \, q^\ast \; .             \label{NLS_cc}
\ee
The cases $\epsilon = -1$ and $\epsilon = 1$ are sometimes referred to as \emph{focusing} and \emph{defocusing}, respectively. 

For $m=2$ and \emph{symmetric} $q$, i.e. $q^\intercal = q$ (where $q^\intercal$ is the transpose 
of $q$), so that $q$ has the form
\be
     q = \left( \begin{array}{cc} q_1 & q_0 \\ q_0 & q_{-1} \end{array} \right) \, , 
                 \label{2x2sym_q}
\ee
the matrix NLS equation with $\epsilon=-1$ describes a special case of a spin-1 Bose-Einstein condensate \cite{IMW04PRL,IMW04,LLMML05,IMW06,Wada+Tsuc06,UIW06,UIW07,Ieda+Wada07,Kuro+Wada07,DRK07,KAGG07,AGGK08,DWY08,KSKLW09}. (\ref{NLS_cc}) is then equivalent to the system
\be
 \imag \,q_{1,t}+q_{1,xx} + 2 (|q_1|^2+2|q_0|^2) \, q_1 + 2 q_0^2 \, q_{-1}^\ast &=& 0 
                          \, , \nonumber \\
 \imag \,q_{-1,t}+q_{-1,xx} + 2 (|q_{-1}|^2+2|q_0|^2) \, q_{-1} + 2 q_0^2 \, q_1^\ast &=& 0 
                          \, , \nonumber \\
 \imag \,q_{0,t}+q_{0,xx} + 2 (|q_1|^2+|q_0|^2+|q_{-1}|^2) \, q_0 + 2 q_1 \, q_0^\ast \, q_{-1} &=& 0 
                          \; .  \label{2x2symNLSeqs}
\ee

\paragraph{Semi-discrete NLS.}
In this case (\ref{NLS_ccred}) leads to the (square) matrix \emph{AL equations}
\be
    \imag \, \dot{q} + q^+ - 2 \, q + q^- - \epsilon \, (q^+ q^\ast q + q \, q^\ast q^-) = 0 \, , \qquad
    p^+ - p = - \epsilon \, q^+ q^\ast \; .       \label{sdNLS_cc}
\ee
The scalar version of the first equation first appeared in \cite{Ablo+Ladi76JMP,Ablo+Ladi76SAM,Ablo+Ladi77} 
and has since been the subject of many publications (see the introduction).

\paragraph{Fully discrete NLS.}
Now (\ref{NLS_ccred}) yields
\be
  &&  \imag \, (q_+ - q) + (q^+ - q)_+ - (q - q^-) - \epsilon \, (q^+ q^\ast q)_+
        - \epsilon \, q q^\ast q^- + (p_+ - p) q_+ + q (p_+ - p)^\ast = 0 \, , \nonumber \\
  &&  p^+ - p = - \epsilon \, q^+ q^\ast \; .              \label{fdNLS_cc}
\ee
Here $p$ does not decouple from the equation for $q$. Appendix~C shows that, by elimination 
of $p$ at the price of introducing non-local terms in the first equation, one recovers an 
equation first obtained by Ablowitz and Ladik \cite{Ablo+Ladi76SAM} in the scalar case.

\paragraph{Solutions.} We set $n_1 = n_2$ and rename it to $n$. Furthermore, we choose
\be
     \bGamma = \left(\begin{array}{cc} 0 & I_n \\ I_n & 0 \end{array} \right) \; .
\ee
Using (\ref{S,U,V,K_blocks}), the reduction conditions (\ref{ccred_S,U,V}) and (\ref{ccred_K}) become
\be
    \bar{S} = S^\ast \, , \qquad    
    \bar{U} = \epsilon \epsilon' \, U^\ast \, , \qquad
    \bar{V} = \epsilon' \, V^\ast \, , \qquad 
    \bar{K} = \epsilon \, K^\ast \; .
\ee
Proposition~\ref{prop:sol_NLSsystems} now implies the following result. 

\begin{proposition}
\label{prop:NLS_cc}
Let $S,U,V$ be constant complex matrices of size $n \times n$, $m \times n$ and 
$n \times m$, respectively. In the discrete NLS cases, we assume invertibility of $S$. 
Then 
\be
  q &=& \pm \, U \, \left(\Xi^{\ast -1} - \epsilon \, K^\ast \Xi K \right)^{-1} \, V^\ast \, , 
                       \label{q_NLS_cc}               \\
  p &=& \epsilon \, U \Xi^\ast K^\ast \, 
           (\Xi^{-1} - \epsilon \, K \Xi^\ast K^\ast)^{-1} \, V    \label{p_NLS_cc}
\ee
solves the $m \times m$ matrix continuous NLS equations (\ref{NLS_cc}), the semi-discrete 
NLS equations  (\ref{sdNLS_cc}), respectively the fully discrete NLS equations (\ref{fdNLS_cc}), 
if there is a matrix $K$ solving the following Sylvester equation and if $\Xi$ is given by the 
following expression,
\begin{center}
\renewcommand{\arraystretch}{1.5}
\begin{tabular}{c|c|c@{\qquad}l} 
 NLS  & Sylvester equation & $\Xi$    \\
\hline
 continuous  & $S K + K S^\ast = V U$  
             & $e^{-x \, S - \imag \, t \, S^2}$ 
             & $x,t \in \bbR$   \\
\hline 
 semi-discrete & $S^{-1} K - K S^\ast = V U$ 
             & $S^x \, e^{- \imag \, t \, \omega(S)}$    
             & $x \in \bbZ, \, t \in \bbR$ \\
\hline
 fully discrete& $S^{-1} K - K S^\ast = V U$ 
             & $S^x \, \, \Omega(S)^t$ 
             & $x,t \in \bbZ$
\end{tabular}
\end{center}
where
\be
    \omega(S) = S + S^{-1} -2 I_n  \, , \qquad
    \Omega(S) = [I_n + \imag \, (I_n- S^{-1})][I_n - \imag \, (I_n- S)]^{-1}  \; .
\ee
\hfill $\square$
\end{proposition}

With reference e.g. to theorem~4.4.6 in \cite{Horn+John91}, a sufficient condition 
for the above Sylvester equations to possess a solution $K$, irrespective of the form 
of $U$ and $V$, is
\be
     \begin{array}{r} \sigma(S) \cap \sigma(-S^\ast) = \emptyset \\
                      \sigma(S^\ast) \cap \sigma(S^{-1}) = \emptyset
     \end{array} \quad \mbox{for} \quad
     \begin{array}{l} \mbox{continuous NLS} \\ \mbox{discrete NLS} \end{array} 
         \label{cc_Sylvester_spec_cond}
\ee
(where $\sigma(S)$ denotes the spectrum of the matrix $S$). The solution matrix $K$ 
is then unique. 

Choosing $S$ diagonal, in the scalar focusing case the solutions determined by 
the last proposition contain the familiar solitons. They will be described in a slightly different 
form in section~\ref{subsec:ccred_solitons}.

\begin{remark}
\label{rem:cc_S-transf}
For given data $(S,U,V)$ with an $n \times n$ matrix $S$, let $K$ be a solution of 
the Sylvester equation in the continuous NLS case. Then it follows that $K$ also 
solves the Sylvester equation with $S' = S + \imag \, \beta \, I_n$, $\beta \in \bbR$. 
Hence there is a new NLS solution obtained by replacing $\Xi$ in (\ref{q_NLS_cc}) and (\ref{p_NLS_cc}) by 
\be
    \Xi(x,t;S') 
  = e^{-x \, S' - \imag \, t \, (S')^2}
  = e^{-\imag \, \beta \, (x-\beta \, t)} \, \Xi(x-2\beta t,t;S)      \; .
\ee
This induces the following Galilean transformation of the NLS solution, 
\be
    q'(x,t) = e^{\imag \, \beta \, (x-\beta \, t)} \, q(x-2\beta t,t) \, ,
\ee
which is a symmetry of the NLS equation. 
\hfill $\square$
\end{remark}

\begin{remark}
\label{rem:symmetric}
If $S$ is symmetric, i.e. $S^\intercal = S$, and if $U=V^\dagger$, in the continuous NLS case 
the Sylvester equation becomes the Lyapunov equation\footnote{More generally, the 
Lyapunov equation has the form $S K + K S^\dagger = Q$ with a complex matrix $Q$ (see e.g. 
\cite{Lanc+Tism85,Horn+John91,Dym07,Bhat07}). It plays an important role in control theory 
\cite{Gajic+Qures95}. } 
\be
        S K + K S^\dagger = V V^\dagger \, ,   \label{Lyapunov}
\ee
where $K$ can be chosen Hermitian. Furthermore, the solution $q$ given by (\ref{q_NLS_cc}) is 
symmetric, i.e. $q^\intercal = q$. 
In particular, this yields solutions of the abovementioned spin-1 Bose-Einstein condensate model. 
For later reference (see remark~\ref{rem:symmetric2}), we mention the following result in \cite{Snyd+Zakai70}. \\
(\ref{Lyapunov}) possesses a Hermitian positive definite 
solution $K$ if and only if the following three conditions hold, \\
(i) all eigenvalues of $S$ have non-negative real parts, \\
(ii) the subspace of $\bbC^n$ spanned by all eigenvectors of $S$ corresponding to eigenvalues 
with positive real part coincides with the controllability subspace associated with the 
pair $(S,V)$ (which is the subspace spanned by the columns of the matrices 
$V, SV, S^2 V, \ldots, S^{n-1}V$), \\
(iii) eigenvalues of $S$ with zero real parts are not degenerate (i.e. they are simple 
roots of the minimal polynomial). 
\hfill $\square$
\end{remark}

\subsection{On the structure of the solutions}
\label{subsec:cc_structure}

\subsubsection{Solving the Sylvester equation}
\label{subsubsec:cc_Sylvester}
Without restriction of generality we can assume that $S$ has Jordan 
normal form (since this can be achieved by a similarity transformation). 
The Sylvester equation then has to be satisfied for each Jordan block (and in addition 
we have to solve equations for the off-diagonal blocks of $K$). 
Let us therefore now consider a matrix $S$ consisting of an $r \times r$ Jordan block only, 
hence 
\be
    S = s \, I + \cN  \, , 
\ee
where $I$ is the $r \times r$ identity matrix and the components of $\cN$ are given by 
$\cN_{ij} = \delta_{i,j-1}$. Writing $K = (k_{ij})$ and 
\be
  U = \left( \begin{array}{ccc} \boldsymbol{u}_1 & \cdots & \boldsymbol{u}_r \end{array} \right) 
      \, , \quad
  V = \left( \begin{array}{c} \boldsymbol{v}_1^\intercal \\ \vdots \\ \boldsymbol{v}_r^\intercal 
       \end{array} \right)  \, ,
\ee
with $m$-component column vectors $\boldsymbol{u}_i, \boldsymbol{v}_i$, $i=1,\ldots,r$, 
in the \emph{continuous} NLS case the Sylvester equation takes the form
\be
    (s+s^\ast) \, k_{ij} + k_{i+1,j} + k_{i,j-1} = \boldsymbol{v}_i^\intercal \boldsymbol{u}_j
     \qquad \quad     i,j=1,\ldots,r \, ,  \label{S-Jordan-block_kij-eq}
\ee
where $k_{r+1,j} = k_{i,0} = 0$. Assuming $s \neq - s^\ast$ (in which case we know that the 
Sylvester equation has a unique solution), this implies
\be
    k_{r 1} = \frac{\boldsymbol{v}_r^\intercal \boldsymbol{u}_1}{s+s^\ast} \, , \qquad
 && k_{r j} = \frac{1}{s+s^\ast} \, ( \boldsymbol{v}_r^\intercal \boldsymbol{u}_j - k_{r,j-1} ) 
              \qquad j=2,\ldots,r \, ,    \nonumber \\
 && k_{i 1} = \frac{1}{s+s^\ast} \, ( \boldsymbol{v}_i^\intercal \boldsymbol{u}_1 - k_{i+1,1} ) 
              \qquad i= r-1,\ldots,1 \, ,
\ee
which recursively determines the entries $k_{i 1}, k_{r i}$, $i=1,\ldots,r$. In fact, 
(\ref{S-Jordan-block_kij-eq}) also determines the remaining entries of $K$ recursively. 
For example, for $r=2$ we obtain
\be
   K = \left( \begin{array}{cc} \frac{\boldsymbol{v}_1^\intercal \boldsymbol{u}_1}{s+s^\ast}
       - \frac{\boldsymbol{v}_2^\intercal \boldsymbol{u}_1}{(s+s^\ast)^2} &
       \frac{\boldsymbol{v}_1^\intercal \boldsymbol{u}_2}{s+s^\ast} 
       -\frac{\boldsymbol{v}_1^\intercal \boldsymbol{u}_1 
         + \boldsymbol{v}_2^\intercal \boldsymbol{u}_2 }{(s+s^\ast)^2}
       + 2 \, \frac{\boldsymbol{v}_2^\intercal \boldsymbol{u}_1}{(s+s^\ast)^3} \\
       \frac{\boldsymbol{v}_2^\intercal \boldsymbol{u}_1}{s+s^\ast} &
       \frac{\boldsymbol{v}_2^\intercal \boldsymbol{u}_2}{s+s^\ast}
       - \frac{\boldsymbol{v}_2^\intercal \boldsymbol{u}_1}{(s+s^\ast)^2} \end{array} \right) \; .
\ee
Returning to arbitrary $r$, we observe that the first column of $K$ vanishes if 
$\boldsymbol{u}_1 =0$, and the last row of $K$ vanishes if $\boldsymbol{v}_r =0$. 
Inspection of (\ref{q_NLS_cc}) and 
(\ref{p_NLS_cc}) shows that in these cases the solution is completely determined by the 
correspondingly reduced matrix data. In the case where $\boldsymbol{u}_1 =0$, these  
reduced data are obtained by deleting the first row and first column of $S$, and also 
deleting the first column of $U$ and the first row of $V$. This also holds in case of the 
discrete NLS versions. 
We conclude that, for any solution determined by matrix data $(S,U,V)$, where $U$ and $V$ 
are different from zero and $S$ is a Jordan block subject to the condition 
(\ref{cc_Sylvester_spec_cond}), there are equivalent data $(S',U',V')$ where $S'$ is 
a Jordan block (with the same eigenvalue and of size smaller or equal to that of $S$) 
such that the first column of $U'$ as well as the last row of $V'$ are different from zero. 

More generally, for any non-vanishing solution determined by matrix data $(S,U,V)$, with 
$S$ given in Jordan normal form subject to the condition (\ref{cc_Sylvester_spec_cond}), 
we can assume without restriction of generality that for any Jordan block of $S$ 
the first column of the corresponding part of $U$ as well as the last row of 
the corresponding part of $V$ are different from zero.

\subsubsection{Superpositions} 
Given two data sets $(S_i,U_i,V_i)$, $i=1,2$, that determine NLS solutions, 
with corresponding solutions $K_i$ of the respective Sylvester equations, 
we can superpose them, 
\be
    S = \left( \begin{array}{cc} S_1 & 0 \\ 0 & S_2 \end{array} \right) \, , \quad
    U = \left( \begin{array}{cc} U_1 & U_2 \end{array} \right) \, , \quad
    V = \left( \begin{array}{c} V_1 \\ V_2 \end{array} \right) \, , \quad 
    K = \left( \begin{array}{cc} K_1 & K_{12} \\ K_{21} & K_2 \end{array} \right) \, , 
         \label{cc_superpos}
\ee
where $K_{12}$ and $K_{21}$ have to solve 
\be
    S_1 K_{12} + K_{12} S_2^\ast = V_1 U_2 \, , \quad 
    S_2 K_{21} + K_{21} S_1^\ast = V_2 U_1 \qquad && \mbox{continuous NLS} 
             \label{cc_superpos_K12,K21_cont} \\
    K_{12} - S_1 K_{12} S_2^\ast = V_1 U_2 \, , \quad
    K_{21} - S_2 K_{21} S_1^\ast = V_2 U_1  \qquad && \mbox{discrete NLS} \; . 
             \label{cc_superpos_K12,K21_discr} 
\ee
(\ref{cc_Sylvester_spec_cond}) is a sufficient condition for such a solution 
to exist (which is then unique). Then we have 
\be
  Z = \left( \begin{array}{cc} Z_1 - \epsilon \, K_{12}^\ast \Xi_2 K_{21} & 
                 -\epsilon \, K_1^\ast \Xi_1 K_{12} - \epsilon \, K_{12}^\ast \Xi_2 K_2 \\ 
                -\epsilon \, K_2^\ast \Xi_2 K_{21} - \epsilon \, K_{21}^\ast \Xi_1 K_1 &
                Z_2 - \epsilon \, K_{21}^\ast \Xi_1 K_{12} \end{array} \right) \, ,
            \label{cc_Z_superpos}
\ee
where $Z_i = \Xi_i^{\ast -1} - \epsilon \, K_i^\ast \Xi_i K_i$. 
Its inverse, needed to evaluate $q$ and $p$ more explicitly, can be expressed in terms of 
Schur complements.

Conversely, given solution data $(S,U,V)$, where $S$ is block-diagonal with at least two 
blocks, then these data can be understood as a superposition (see also 
section~\ref{subsec:NLS_super}). 

The following result is simply proposition~\ref{prop:sup_sim_data} adapted to the 
case under consideration. It will be helpful in the sequel. 

\begin{proposition}
\label{prop:cc_sup_sim_data}
Let $(S_i,U_i,V_i)$, $i=1,2$, be two data sets that determine solutions of an  
NLS system according to proposition~\ref{prop:NLS_cc} and let (\ref{cc_Sylvester_spec_cond}) 
hold for $S = \mbox{\textnormal{block-diag}}(S_1,S_2)$.
Suppose there are matrices $A, B$ such that $U_2 = U_1 B$, $V_2 = A V_1$,  
$A S_1 = S_2  A$ and $S_1^\ast B = B \, S_2^\ast$. 
Let further $C = I + B^\ast A$ be invertible. 
Then the superposed data (\ref{cc_superpos}) determine a solution that can 
equivalently be obtained from the data $(S_1,U_1 C,V_1)$ (i.e. simply by a redefinition of $U_1$). 
\end{proposition}

\subsubsection{The use of reflection symmetries} 
The following result has its origin in \cite{ABDM09sym}, where only the (focusing) continuous 
NLS equation was considered, however. We show that for a solution determined 
by matrix data via proposition~\ref{prop:NLS_cc}, in general equivalent matrix data 
$(S,U,V)$ exist such that all eigenvalues $s$ of $S$ with $\Re(s) \neq 0$ in 
the continuous NLS case, respectively $|s| \neq 1$ in the discrete cases, satisfy \\
(i) $\Re(s) > 0$ in the continuous NLS case, \\
(ii) $|s| < 1$ in the semi- and fully discrete NLS case. \\
In case (i) the matrix $S$ is called \emph{positive stable} \cite{Horn+John91}. 
In case (ii) it is sometimes called stable with respect to the unit circle \cite{Lanc+Tism85}. 

Suppose $S_2$ is a Jordan block that does not satisfy condition (i), respectively (ii). 
Writing 
\be
    S = \left(\begin{array}{cc} S_1 & 0 \\ 0 & S_2 \end{array}\right) \, , \quad
    U = \left(\begin{array}{cc} U_1 & U_2 \end{array}\right) \, , \quad
    V = \left(\begin{array}{c} V_1 \\ V_2 \end{array}\right) \, , 
\ee 
presents the solution data in the form of a superposition, and one has to solve 
(\ref{cc_superpos_K12,K21_cont}), respectively (\ref{cc_superpos_K12,K21_discr}), 
in order to construct the solution. 
The idea is now to apply a reflection symmetry and to use the analogue of proposition~\ref{prop:reflection_super}. 
A reflection symmetry of the form considered in section~\ref{subsubsec:reflection}
preserves the reduction conditions if 
\be 
        \bar{\beta} = \epsilon \, \beta^\ast \; .
\ee 
Let us choose $\beta = \epsilon$ (so that $\bar{\beta}=1$). The reflection symmetry is
then given by 
\be
    U_2 \mapsto - \epsilon \, U_2 K_2^{-1} \, , \quad
    V_2 \mapsto K_2^{-1} V_2 \, , \quad
    K_2 \mapsto - \epsilon \, U_2 K_2^{-1} \, , 
\ee
together with 
\be
    S_2 \mapsto S_2' = \left\{ \begin{array}{l@{\quad}l} - S_2^\ast & \mbox{continuous NLS} \\
                              S_2^{\ast -1} & \mbox{semi- and fully discrete NLS} 
                           \end{array} \right.  \, .  \label{cc_S_2-reflection}
\ee
This is a reflection in an obvious sense. As in the proof of proposition~\ref{prop:reflection_super}, 
we can construct an extension to matrix data that determine the same superposition solution. 
For the above reflection, this is achieved by
\be
    S \mapsto \left(\begin{array}{cc} S_1 & 0 \\ 0 & S'_2 \end{array}\right) \, , 
\ee
\be
    U \mapsto U \left(\begin{array}{cc} I & 0 \\ 
          -K_2^{-1}K_{21} & -\epsilon\,K_2^{-1} \end{array}\right) \, , \qquad
    V \mapsto \left(\begin{array}{cc} I & - K_{12} K_2^{-1} \\ 
                   0 & K_2^{-1} \end{array}\right)V  \, ,
\ee
and
\be
    K \mapsto \left(\begin{array}{cc} K_1-K_{12}K_2^{-1}K_{21} & -\epsilon \, K_{12}K_2^{-1} \\
    K_2^{-1} K_{21} & \epsilon \, K_2^{-1} \end{array}\right) \; .
       \label{cc_reflection_K}
\ee
The only assumption needed for this to work is that the Sylvester equation with data 
$(S_2,U_2,V_2)$ has an \emph{invertible} solution $K_2$. 

In case of a Jordan block, the result of the reflection (\ref{cc_S_2-reflection}) has no longer 
the standard Jordan form. 
But with the help of a similarity transformation
one can restore the usual Jordan structure (without changing the eigenvalues).

\subsection{Solutions of the \emph{scalar} continuous and discrete NLS equations}
\label{subsec:ccred_solitons} 
In this subsection we concentrate on the \emph{scalar} NLS case, i.e. $m=1$. Without 
restriction of generality, we can assume that the $n \times n$ matrix $S$ has Jordan normal form.  

\begin{lemma} 
\label{lemma:Toeplitz}
\hspace{1cm} \\
(1) Let $S$ be a single $k \times k$ Jordan block. For any $k$-component column vector $V$,  
there is a $k \times k$ matrix $A$ that commutes with $S$ and satisfies
\be
     V = A \, V_0 \qquad \mbox{where} \qquad 
     V_0^\intercal = (0,\ldots,0,1) \; .  
\ee 
$A$ can be chosen invertible if the last component of $V$ is different from zero. 

Furthermore, for any $k$-component row vector $U$, there is a $k \times k$ matrix $B$ 
that commutes with $S$ and satisfies
\be
     U = U_0 \, B^\ast \qquad \mbox{where} \qquad 
     U_0 = (1,0,\ldots,0) \; .  
\ee 
$B$ can be chosen invertible if the first component of $U$ is different from zero. \\
(2) Let $S = \mbox{block-diag}(S_1,\ldots,S_r)$ with Jordan blocks $S_i$. 
For any correspondingly decomposed $n$-component vector 
$V^\intercal = (V_1^\intercal, \ldots, V_r^\intercal)$,  
there is an $n \times n$ matrix $A$ that commutes with $S$ and satisfies
\be
    V = A \, V_0  \qquad \mbox{with} \qquad 
    V_0^\intercal = (V_{0,1}^\intercal, \ldots, V_{0,r}^\intercal) \, ,
\ee
where each vector $V_{0,i}$ (of same size as $V_i$) has a $1$ in the last entry 
and zeros otherwise.
$A$ can be chosen invertible if the last component of each $V_i$ is different from zero. 

For any correspondingly decomposed $n$-component row vector 
$U = (U_1, \ldots, U_r)$ there is an $n \times n$ matrix $B$ that commutes with $S$ 
and satisfies
\be
    U = U_0 \, B^\ast  \qquad \mbox{with} \qquad 
    U_0 = (U_{0,1}, \ldots, U_{0,r}) \, ,
\ee
where each row vector $U_{0,i}$ (of same size as $U_i$) has a $1$ in the first entry.
$B$ can be chosen invertible if the first component of each $U_i$ is different from zero.
\end{lemma}
\noindent
\textit{Proof:} 
(1) $A$ commutes with the Jordan block $S$ if it is an upper-triangular Toeplitz matrix. Hence
\bez
    S = \left(\begin{array}{cccccc} s &      1 & 0      & \cdots & 0       \\ 
                                    0 & \ddots & \ddots & \ddots & \vdots  \\
                               \vdots & \ddots & \ddots & \ddots & 0       \\
                               \vdots &        & \ddots & \ddots & 1       \\
                                    0 & \cdots & \cdots & 0      & s 
               \end{array}\right) \, , \quad
    A = \left(\begin{array}{cccccc} a_1 &      a_2 & a_3 & \cdots & a_k       \\ 
                                    0 & \ddots & \ddots  & \ddots & \vdots    \\
                               \vdots & \ddots & \ddots  & \ddots & a_3       \\
                               \vdots &        & \ddots  & \ddots & a_2       \\
                                    0 & \cdots & \cdots  & 0      & a_1 
               \end{array}\right) \, , 
\eez 
with arbitrary constants $a_1,\ldots,a_k$. 
It is now immediately verified that, for any given vector $V$, there is a matrix $A$ of 
the above form such that $A V_0 = V$ holds. $A$ is invertible iff $a_1 \neq 0$, and 
$a_1$ coincides with the last component of $V$. Let $u_1,\ldots,u_k$ be the components 
of $U$ and $B$ the Toeplitz matrix built with the complex conjugates of these constants. 
Then $U = U_0 B^\ast$. Clearly, $B$ is invertible iff $u_1 \neq 0$. 
(2) immediately follows from (1).
\hfill $\square$
\vskip.2cm

As a consequence of the lemma, we can always achieve that 
\be
    U = U_0 \, B^\ast \, , \qquad  V = A \, V_0 \, , 
\ee
with matrices $A,B$ commuting with $S$. As shown in section~\ref{subsubsec:cc_Sylvester}, 
the condition in the above lemma achieving that $A$ and $B$ are invertible is no restriction 
of generality, 
provided that $S$ satisfies the spectrum condition (\ref{cc_Sylvester_spec_cond}). 
The latter implies that $S$ is invertible. By a reparametrisation transformation\footnote{The 
complex conjugation reduction requires $\bar{A}=A^\ast$ and $\bar{B}=B^\ast$ 
in (\ref{decomp_repar_trans}). } 
(and redefinitions of $K$ and $V$ in the discrete cases) the above family of solutions 
can then be expressed as 
\be
 q = U_0 \, (I - \epsilon \, \Xi^\ast K^\ast \Xi K)^{-1} \, \Xi^\ast 
               \, S^{\ast -\eta} V_0 \, , \quad
 p = \epsilon \, U_0 \, \Xi^\ast K^\ast \, (I - \epsilon \, \Xi K \Xi^\ast K^\ast)^{-1} 
               \, \Xi \, S^{-\eta} V_0  \, ,            \label{NLS_cc_q,p_reparam}
\ee  
with
\begin{center}
\renewcommand{\arraystretch}{1.5}
\begin{tabular}{c|c|c|c} 
 NLS  & $\eta$ & Sylvester equation & $\Xi$   \\
\hline
 continuous  & $0$  & $S K + K S^\ast = V_0 U_0$ 
             & $C \, e^{-x \, S - \imag t \, S^2}$  \\
\hline 
 semi-discrete & $1$ & $K - S K S^\ast = V_0 U_0$ 
             & $C \, S^{x+1} \, e^{- \imag \, t \, (S+S^{-1}-2I)}$ \\
\hline
 fully discrete & $1$ & $K - S K S^\ast = V_0 U_0$ 
             & $ C \, S^{x+1} \, [( I + \imag \, (I-S^{-1}) )( I - \imag \, (I-S))^{-1}]^t$ 
\end{tabular}
\end{center}
where $I = I_n$ and the matrices $A,B$ only appear through the combination 
\be
     C = B A \; . 
\ee

With the restriction to the \emph{focusing} case, i.e. $\epsilon = -1$, 
we now consider the special case where $S$ is \emph{diagonal}, hence
\be
   S = \mbox{diag}(s_1, \ldots, s_n) \, , \qquad
  U_0 = V_0^\intercal = (1,\ldots,1) \; . 
\ee
In the following we assume the spectrum condition (\ref{cc_Sylvester_spec_cond}), 
i.e. $s_i + s_j^\ast \neq 0$ in the continuous case, and 
$s_i s_j^\ast \neq 1$ in the discrete cases, for all $i,j=1,\ldots,n$. 
According to proposition~\ref{prop:cc_sup_sim_data}, with no restriction of generality
we can then assume that the eigenvalues $s_i$ are pairwise different.
As a consequence, $A$ and $B$ are diagonal, hence $C$ is diagonal. Writing 
\be
    \Xi = \mbox{diag}(\xi_1, \ldots, \xi_n) \, , \qquad  
      C = \mbox{diag}(c_1, \ldots, c_n) \, , \qquad K = (k_{ij}) \, , 
\ee
we obtain 
\begin{center}
\renewcommand{\arraystretch}{1.5}
\begin{tabular}{c|c|c} 
 NLS  & $\xi_i$ & $k_{ij}$   \\
\hline
 continuous         & $c_i \, e^{-x s_i - \imag \, t s_i^2}$  
                    &  $1/(s_i + s_j^\ast)$ \\
\hline 
 semi-discrete (AL) & $c_i \, s_i^{x+1} \, e^{-\imag \, t \, (s_i+s_i^{-1} -2)}$ 
                    & $1/(1-s_i s_j^\ast)$ \\
\hline
 fully discrete     & $c_i \, s_i^{x+1} \, [(1+ \imag \, (1-s_i^{-1}))(1- \imag \, (1 - s_i))^{-1}]^t$ 
                    & $1/(1-s_i s_j^\ast)$ 
\end{tabular}
\end{center}
$K$ is a Cauchy-like matrix and Hermitian.\footnote{We actually arranged the 
form of the Sylvester equations in order to achieve this (at the price of introducing 
$\eta$). }

\begin{lemma}
Let $S$ be diagonal with pairwise different eigenvalues $s_i$, subject to (\ref{cc_Sylvester_spec_cond}). \\
(i) The matrix $K$ of the continuous NLS case is positive definite if 
$\Re(s_i)>0$ for $i=1,\ldots,n$ (which means that $S$ is positive stable). \\
(ii) The matrix $K$ of the discrete NLS cases is positive definite if all 
$s_i$ lie in the open unit disk (which means that $S$ is stable with respect to the unit circle).
\end{lemma}
\noindent
\textit{Proof:} In case (i), the matrix $K$ is a positive semidefinite Cauchy matrix, see e.g. \cite{Bhat+Elsn01,Bhat07}. If the $s_i$ are pairwise different, Cauchy's determinant 
formula (see e.g. \cite{Bhat07}) shows that $K$ is invertible. Hence $K$ is positive definite. In case (ii),
we can write $K = D \cC D^\dagger$ with the Cauchy matrix $\cC = (1/(\la_i + \la_j^\ast))$ 
and $D = \mathrm{diag}((\la_1 +1)/\sqrt{2},\ldots,(\la_n +1)/\sqrt{2})$, where 
$s_i = (\la_i -1)/(\la_i +1)$ defines a bijection between the open unit disk and the open 
right half plane. Since the $s_i$ are assumed to be pairwise different, it follows that 
also the $\la_i$ are pairwise different. Hence $\cC$ is positive definite, which then 
also holds for $K$. 
\hfill $\square$

\begin{remark}
\label{rem:symmetric2}
Since a diagonal matrix $S$ is \emph{symmetric}, the Sylvester equation in the 
\emph{continuous} NLS case is the Lyapunov equation (\ref{Lyapunov}). 
The result formulated in part (i) of the preceding lemma can thus be deduced from a 
more general result in \cite{Snyd+Zakai70}, which we recalled in remark~\ref{rem:symmetric}. 
\hfill $\square$
\end{remark}

\begin{proposition}
\label{prop:cc_scalar_regularity}
Let $S = \mathrm{diag}(s_1, \ldots, s_n)$ with pairwise different $s_i$, and 
$\Re(s_i) > 0$ in the continuous NLS case, $|s_i| < 1$ in the discrete NLS cases.  
Then the associated solution (\ref{NLS_cc_q,p_reparam}) of the (continuous, semi-discrete, 
respectively fully discrete) focusing NLS equations is regular (for all $x,t \in \bbR$). 
\end{proposition}
\noindent
\textit{Proof:} Using a diagonal square root $\Xi^{1/2}$ of the diagonal matrix $\Xi$, 
we introduce
\bez
       L = \Xi^{1/2} \, K \, (\Xi^{1/2})^\dagger \; .
\eez
Since $K$ is Hermitian and positive definite by the preceding lemma, also $L$ is 
Hermitian, i.e. $L^\dagger = L$, and positive definite. Hence $L$ possesses a positive 
definite Hermitian square root $L^{1/2} = (L^{1/2})^\dagger$, and we can write 
\bez
   (\Xi K)^\ast (\Xi K) 
 &=& (\Xi^{1/2})^\ast \, L^\ast L \, (\Xi^{1/2})^{\ast -1}  \\
 &=& (\Xi^{1/2} \, L^{1/2})^\ast \, \left( [(L^{1/2})^\ast L^{1/2}] \, 
     [(L^{1/2})^\ast L^{1/2}]^\dagger \right) \, (\Xi^{1/2} \, L^{1/2})^{\ast -1} \; .
\eez
For the expression in (\ref{NLS_cc_q,p_reparam}) that has to be inverted, this implies 
\be
    I + (\Xi K)^\ast (\Xi K) 
  = (\Xi^{1/2} \, L^{1/2})^\ast \, \left( I + [(L^{1/2})^\ast L^{1/2}] \, 
     [(L^{1/2})^\ast L^{1/2}]^\dagger \right) \, (\Xi^{1/2} \, L^{1/2})^{\ast -1} \; .
\ee
Since any matrix of the form $I + A A^\dagger$ is invertible, this proves our assertion. 
\hfill $\square$
\vskip.2cm

Because of the following reason we can relax the assumptions in proposition~\ref{prop:cc_scalar_regularity} to the requirements that the eigenvalues 
$s_i$ of the diagonal matrix $S$ are pairwise different, and 
$s_i + s_j^\ast \neq 0$ in the continuous case, $s_i s_j^\ast \neq 1$ in the discrete cases. 
If we have matrix data $(S,C,U_0,V_0)$, where $S$ has an eigenvalue with negative real part 
(respectively which lies outside the unit circle), then we can apply a reflection symmetry 
and obtain new matrix data $(S',U',V')$ that determine the same solution, 
and $S'$ is given by $S$ where the respective eigenvalue now has positive real 
part, respectively lies inside the unit circle. 
Furthermore, we can choose the data such that $U' = U_0 B'^\ast$ and $V' = A' V_0$ 
with invertible matrices $A',B'$ that commute with $S'$. A reparametrisation 
transformation then achieves the form (\ref{NLS_cc_q,p_reparam}) with matrix data $(S',C',U_0,V_0)$. 
Hence the assumption in the proposition that the eigenvalues of $S$ have positive real parts, 
respectively lie in the open unit disk, is no restriction.

\subsection{Solutions of the matrix NLS equations}
\label{subsec:ccred_matrixsolitons}
By redefinitions of $K$ and $V$ in the discrete NLS cases, the solutions determined by 
proposition~\ref{prop:NLS_cc} can be expressed as\footnote{If $q$ and $p$ solves the respective 
NLS system, then also $-q$ and $p$.}
\be
 q = U \, Z^{-1} \, S^{\ast -\eta} V^\ast \, , \quad
 p = \epsilon \, U \, \Xi^\ast K^\ast \, Z^{\ast -1} \, S^{-\eta} V  \, ,
                \label{NLS_cc_q,p_eta}
\ee  
where
\be
     Z = \Xi^{\ast -1} - \epsilon \, K^\ast \Xi K
\ee
and 
\begin{center}
\renewcommand{\arraystretch}{1.5}
\begin{tabular}{c|c|c|c} 
 NLS  & $\eta$ & Sylvester equation & $\Xi$   \\
\hline
 continuous  & $0$  & $S K + K S^\ast = V U$ 
             & $e^{-x \, S - \imag t \, S^2}$  \\
\hline 
 semi-discrete & $1$ & $K - S K S^\ast = V U$ 
               & $S^{x+1} \, e^{- \imag \, t \, (S+S^{-1}-2I)}$ \\
\hline
 fully discrete & $1$ & $K - S K S^\ast = V U$ 
                & $S^{x+1} \, [( I + \imag \, (I-S^{-1}) )( I - \imag \, (I-S))^{-1}]^t$ 
\end{tabular}
\end{center}
This and what follows includes the case $m=1$, i.e. solutions of the scalar NLS equations. 
For $m>1$, the results are new.

\subsubsection{Rank one single solitons} 
Choosing $n=1$, $U$ is a column and $V$ a row vector (with 
$m$ components). Writing
\be
    S = s \, , \quad
    \Xi = \xi  \, , \quad
    U = \boldsymbol{u} \, , \quad
    V = \boldsymbol{v}^\intercal \, , \quad
    K = \kappa \, \boldsymbol{v}^\intercal \boldsymbol{u} \, , 
\ee
with ($m$-component) column vectors $\boldsymbol{u},\boldsymbol{v}$, 
we obtain\footnote{Note that 
$\cP = \boldsymbol{u} \boldsymbol{v}^\intercal/(\boldsymbol{v}^\intercal \boldsymbol{u})$ 
is a projector. Furthermore, $\tilde{\cP} = \boldsymbol{u} \boldsymbol{v}^\dagger /
(\boldsymbol{v}^\intercal \boldsymbol{u})^\ast$ satisfies 
$\tilde{\cP} \tilde{\cP}^\ast = \cP$.} 
\be
    q = \frac{\kappa^{-1} \, \txi^\ast}{1 -\epsilon \, |\txi|^2} \, s^{\ast -\eta} \, 
        \frac{\boldsymbol{u} \boldsymbol{v}^\dagger}{(\boldsymbol{v}^\intercal \boldsymbol{u})^\ast}
             \, , \qquad
    p = \frac{\kappa^{-1} }{1 -\epsilon \, |\txi|^2} 
        \, s^{-\eta} \, \frac{\boldsymbol{u} \boldsymbol{v}^\intercal}{
         \boldsymbol{v}^\intercal \boldsymbol{u} } + p_0 \, , 
        \label{ccred_single_matrix-soliton}
\ee
where $p_0 = - s^{-\eta} \kappa^{-1} \boldsymbol{u} \boldsymbol{v}^\intercal/(\boldsymbol{v}^\intercal \boldsymbol{u})$ and 
\begin{center}
\renewcommand{\arraystretch}{1.5}
\begin{tabular}{c|c|c|c} 
        NLS  & $\eta$ & $\kappa$ & $\txi$   \\
\hline
 continuous  & $0$  & $\frac{1}{s + s^\ast}$  
             & $\kappa \, \boldsymbol{v}^\intercal \boldsymbol{u} \, e^{- s \, x - \imag \, s^2 t }$  \\
\hline 
 semi-discrete & $1$ & $\frac{1}{1-s s^\ast}$ 
               & $\kappa \,\boldsymbol{v}^\intercal \boldsymbol{u} \, s^{x+1} \, 
                 e^{- \imag \, (s+s^{-1}-2) \, t }$ \\
\hline
 fully discrete & $1$ & $\frac{1}{1-s s^\ast}$ 
                & $\kappa \, \boldsymbol{v}^\intercal \boldsymbol{u} \, s^{x+1} \, 
                  \Big( \frac{1+\imag \, (1-s^{-1}) }{1-\imag \, (1-s)}\Big)^t$ 
\end{tabular}
\end{center}
Of course, we have to assume $\Re(s) \neq 0$, respectively $|s| \neq 1$. 
In the \emph{focusing} case, i.e. $\epsilon = -1$, these solutions are regular and describe 
a single soliton with an attached ``polarisation'' determined by the vectors $\boldsymbol{u}$ and $\boldsymbol{v}$. The solutions can be expressed as follows,
\be
 q &=& c \, \mathrm{sech}(\alpha \, (x - \rho \, t) - \delta)
       \, e^{\imag \, (\beta \, x - \nu  \, t - \theta)} \,  
       \frac{\boldsymbol{u} \boldsymbol{v}^\dagger}{(\boldsymbol{v}^\intercal \boldsymbol{u})^\ast}
               \, , \nonumber \\
 p &=& \frac{2 \, c}{1 + e^{-2 \, (\alpha \, (x 
       - \rho \, t) - \delta)} } \, \frac{\boldsymbol{u} \boldsymbol{v}^\intercal}{
         \boldsymbol{v}^\intercal \boldsymbol{u} }  + p_0 \, ,
\ee
where 
\be
   s = \alpha + \imag \, \beta \, , \quad 
   c = \alpha \, , \quad 
   \boldsymbol{v}^\intercal \boldsymbol{u} = \kappa^{-1} \, e^{\delta + \imag \, \theta} \, , \quad
   \rho = 2 \beta \, , \quad
   \nu = \beta^2 - \alpha^2     \label{ccred_matrix-soliton_NLS_param}
\ee
in the continuous NLS case, with $\alpha,\beta,\delta,\theta \in \bbR$. 
In the discrete NLS cases, 
\be
   s = e^{-\alpha - \imag \, \beta} \, , \quad
   c = \sinh(\alpha) \, e^{\imag \, \beta} \, , \quad
   \boldsymbol{v}^\intercal \boldsymbol{u} 
       = \kappa^{-1} \, e^{\alpha + \delta + \imag \, (\theta - \beta)} 
         \, ,  \label{ccred_matrix-soliton_dNLS_param}
\ee
and  
\be
   \rho = \frac{2}{\alpha} \, \sinh(\alpha) \, \sin(\beta) \, , \qquad
   \nu = 2 \, (1 - \cosh(\alpha) \, \cos(\beta))    \label{ccred_matrix-soliton_sdNLS_param}
\ee
in the semi-discrete case, whereas in the fully discrete case $\rho$ and $\nu$ are given 
in terms of $\alpha,\beta$ via
\be
    e^{\alpha \, \rho + \imag \, \nu} 
  = \frac{1 + \imag \, (1 - e^{\alpha + \imag \, \beta})}{1 - \imag \, (1 - e^{-\alpha - \imag \, \beta})}
    \; .                 \label{ccred_matrix-soliton_fdNLS_param}
\ee
Since here $q$ and $p$ are rank one matrices, these are the most elementary matrix soliton solutions. 
Such a soliton has velocity $\rho$. Whereas a scalar (i.e. $m=1$) NLS soliton carries a phase 
factor that can change via scattering with other solitons, here the phase is extended to the 
polarisation matrix. We shall see below how the latter changes in a scattering process.

\subsubsection{Asymptotics of superpositions}
The following is a preparation for the elaboration of the asymptotics of multi-soliton 
solutions in section~\ref{subsubsec:n-soliton}. 
We note that (\ref{cc_Z_superpos}) can be written as 
$Z = A_2^\ast \check{\Xi}^{\ast -1} B_2 + A_1^\ast \check{\Xi} B_1$ with 
$\check{\Xi} = \mbox{block-diagonal}( \Xi_1, \Xi_2^{\ast -1} )$ and 
\be
 A_1 = -\left( \begin{array}{cc} \epsilon K_1 & 0 \\ \epsilon K_{21} & I \end{array} \right) , \;
 B_1 = \left( \begin{array}{cc} K_1 & K_{12} \\ 0 & -I \end{array} \right) , \;
 A_2 = \left( \begin{array}{cc} I & K_{12} \\ 0 & K_2 \end{array} \right) , \; 
 B_2 = \left( \begin{array}{cc} I & 0 \\ -\epsilon K_{21} & -\epsilon K_2 \end{array} \right) 
\ee
(where $I$ stands for the identity matrix of the respective size). 
Assuming that $K_1$ and $K_2$ are invertible, this implies 
\be
  Z^{-1} = B_1^{-1} \, \check{\Xi}^{-1} \, \Big( I -\epsilon \, \tilde{K}^{\ast -1} 
           \, \check{\Xi}^{\ast -1} \, \tilde{K}^{-1} \, \check{\Xi}^{-1} 
              \Big)^{-1} \, A_1^{\ast -1} \, ,
\ee
where 
\be
     \tilde{K} = -\epsilon \, A_2^{-1} A_1 = B_1 B_2^{-1} 
  = \left( \begin{array}{cc} L_1 & - \epsilon \, K_{12} \, K_2^{-1} \\ 
     K_2^{-1} K_{21} & \epsilon \, K_2^{-1} \end{array} \right) 
\ee
(cf. (\ref{cc_reflection_K})), with the Schur complement of $K_2$ (with respect to $K$),
\be
    L_1 = K_1 - K_{12} K_2^{-1} K_{21} \; .
\ee
In a limit in which $\Xi_1^{-1} \to 0$ and $\Xi_2 \sim \hat{\Xi}_2$ (asymptotic form), we obtain 
\be
 Z^{-1} \sim B_1^{-1} \, 
    \left( \begin{array}{cc} 0 & 0 \\ 
     0 & \hat{\Xi}_2^\ast \, \left( I - \epsilon \, L_2^\ast \, 
         \hat{\Xi}_2 \, L_2 \, \hat{\Xi}_2^\ast \right)^{-1} 
           \end{array} \right) \, A_1^{\ast -1} \, ,   \label{cc_superpos_limit1a}
\ee
with the Schur complement of $K_1$ (with respect to $K$),
\be
    L_2 = K_2 - K_{21} K_1^{-1} K_{12} \; .
\ee
This allows us to compute the asymptotic form of $q$, given by (\ref{NLS_cc_q,p_eta}), in this limit. 
In order to compute the asymptotic form of $p$, we write
\be
   \Xi^\ast K^\ast Z^{\ast -1} 
 = B_1^{-1} \, ( B_1 \Xi^\ast K^\ast B_1^{\ast -1} \check{\Xi}^{\ast -1} ) \, 
   \Big( I -\epsilon \, \tilde{K}^{-1} \, \check{\Xi}^{-1} \, 
          \tilde{K}^{\ast -1} \, \check{\Xi}^{\ast -1} \Big)^{-1} \, A_1^{-1} \; . 
\ee
Since
\be
   B_1 \Xi^\ast K^\ast B_1^{\ast -1} \check{\Xi}^{\ast -1} 
 = \left( \begin{array}{cc} K_1 + K_{12} \Xi_2^\ast K_{21}^\ast K_1^{\ast -1} \Xi_1^{\ast -1} & 
         - K_{12} \Xi_2^\ast L_2^\ast \Xi_2 \\ 
   \Xi_2^\ast K_{21}^\ast K_1^{\ast -1} \Xi_1^{\ast -1} & 
         - \Xi_2^\ast L_2^\ast \Xi_2 \end{array} \right) \, ,  
\ee
we find
\be
  \Xi^\ast K^\ast Z^{\ast -1} \sim 
   B_1^{-1} \, 
    \left( \begin{array}{cc} 0 & 0 \\ 
     0 & \epsilon \, L_2^{-1} 
         \left( I - \epsilon \, L_2 \, 
         \hat{\Xi}_2^\ast \, L_2^\ast \, \hat{\Xi}_2 \right)^{-1} 
           \end{array} \right) \, A_1^{-1} \; .  \label{cc_superpos_limit1b}
\ee
In the same way, in a limit in which $\Xi_2^{-1} \to 0$ and $\Xi_1 \sim \hat{\Xi}_1$, we obtain
\be
 Z^{-1} &\sim& B_2^{-1} \, 
    \left( \begin{array}{cc} \hat{\Xi}_1^\ast \, \left( I - \epsilon \, 
      L_1^\ast \, \hat{\Xi}_1 \, L_1 
      \, \hat{\Xi}_1^\ast \right)^{-1} & 0 \\ 
     0 & 0
     \end{array} \right) \, A_2^{\ast -1} \, ,  \nonumber \\
\Xi^\ast K^\ast Z^{\ast -1} &\sim& 
   B_2^{-1} \, 
    \left( \begin{array}{cc} \epsilon \, L_1^{-1} 
         \left( I - \epsilon \, L_1 \, 
         \hat{\Xi}_1^\ast \, L_1^\ast \, \hat{\Xi}_1 \right)^{-1} & 0 \\ 
        0 & 0 
           \end{array} \right) \, A_2^{-1} \; .  \label{cc_superpos_limit2}
\ee

\subsubsection{Superposition of rank one solitons} 
\label{subsubsec:n-soliton}
Now we consider the \emph{focusing} case, i.e. $\epsilon = -1$. 
We choose $S = \mathrm{diag}(s_1,\ldots,s_n)$ and write
\be
  U = \left( \begin{array}{ccc} \boldsymbol{u}_1 & \cdots & \boldsymbol{u}_n \end{array} \right) 
      \, , \quad
  V = \left( \begin{array}{c} \boldsymbol{v}_1^\intercal \\ \vdots \\ \boldsymbol{v}_n^\intercal 
       \end{array} \right)  \, ,
\ee
with $m$-component column vectors $\boldsymbol{u}_i, \boldsymbol{v}_i$, $i=1,\ldots,n$, and 
$\Xi = \mathrm{diag}(\xi_1,\ldots,\xi_n)$. 
The corresponding Sylvester equation is solved by 
\be
   K = (k_{ij}) \qquad \mbox{where} \quad 
   k_{ij} = \left\{ \begin{array}{c@{\qquad}l} 
            \frac{\boldsymbol{v}_i^\intercal \boldsymbol{u}_j}{s_i + s_j^\ast} 
                   & \mbox{continuous NLS} \\
                        \frac{\boldsymbol{v}_i^\intercal \boldsymbol{u}_j}{1 - s_i s_j^\ast} 
                   & \mbox{(semi- and fully) discrete NLS} 
                   \end{array}  \right.  \; .   \label{cc_nsoliton_kij}
\ee
The NLS solution can now be expressed as
\be
   q = \sum_{i,j=1}^n \xi_i^\ast \, [(Z \Xi^\ast)^{-1}]_{ij} \, s_j^{\ast -\eta} \, 
        \boldsymbol{u}_i \boldsymbol{v}_j^\dagger \, , \qquad
   p = \sum_{i,k,j=1}^n (K^{-1})_{ik} \, [(Z^\ast \Xi)^{-1}]_{kj} \, s_j^{-\eta} \, 
        \boldsymbol{u}_i \boldsymbol{v}_j^\intercal + p_0 \, ,
\ee
with the constant term $p_0 = -U K^{-1} S^{-\eta} V$, dropped in the following, and 
\be
    (Z \Xi^\ast)_{ij} = \delta_{ij} + \sum_{k=1}^n k_{ik}^\ast \, k_{kj} \, \xi_k \xi_j^\ast \; .
\ee
For each $s_j$ we introduce real constants $\alpha_j, \beta_j, \rho_j, \nu_j$ as in 
(\ref{ccred_matrix-soliton_NLS_param}), (\ref{ccred_matrix-soliton_dNLS_param}), 
(\ref{ccred_matrix-soliton_sdNLS_param}), (\ref{ccred_matrix-soliton_fdNLS_param}), and
\be
  \kappa_i = \left\{ \begin{array}{l@{\qquad}l} (s_i + s_i^\ast)^{-1} & \mbox{continuous NLS} \\
             (1-s_i \, s_i^\ast)^{-1} & \mbox{discrete NLS} \end{array} \right. \, .
\ee
Furthermore, we choose\footnote{This is not really a restriction since for $\alpha_i \neq 0$ 
it can be achieved generically by a reflection symmetry.} 
$\alpha_i >0$, $i=1,\ldots,n$, and\footnote{For simplicity, we exclude the case where 
some of the $\rho_i$ are equal. } 
$\rho_1 < \rho_2 < \cdots < \rho_n$. 
Asymptotically, as $t \to -\infty$, 
the corresponding incoming solitons thus appear in decreasing numerical order along the $x$-axis. 
The outgoing solitons (as $t \to \infty$) then appear in increasing numerical order. 
To extract these asymptotic solitons from the solution in the limits $t \to \pm \infty$, 
we introduce a comoving coordinate $\bar{x}_i = x - \rho_i t$, with respect to which the 
$i$-th soliton is stationary up to a phase factor. We consider the limits $t \to \pm \infty$ 
while keeping $\bar{x}_i$ constant, for some \emph{fixed} $i$. Then 
$|\xi_j| = e^{-\alpha_j (x- \rho_j t) + \delta_j}
         = e^{-\alpha_j [\bar{x}_i - (\rho_j - \rho_i) t] + \delta_j}$. It follows that 
\be
    \lim_{t \to -\infty} \xi_j = 0 \qquad \mbox{if} \quad j > i \, , \qquad
    \lim_{t \to -\infty} \xi_j^{-1} = 0 \qquad \mbox{if} \quad j < i \, , 
\ee
and 
\be
    \lim_{t \to +\infty} \xi_j^{-1} = 0 \qquad \mbox{if} \quad j > i \, , \qquad
    \lim_{t \to +\infty} \xi_j = 0 \qquad \mbox{if} \quad j < i \; .
\ee
In order to find the corresponding asymptotic forms of $q$ and $p$, we express the solution 
as a superposition (\ref{cc_superpos}) with 
$S_1 = \mathrm{diag}(s_1,\ldots,s_{i-1})$ and $S_2 = \mathrm{diag}(s_i,\ldots,s_n)$. 
Let $U_1,U_2$ and $V_1,V_2$ be the corresponding parts of $U$ and $V$, respectively. 
Let $K_1,K_2$ and $K_{12},K_{21}$ be the corresponding diagonal and off-diagonal blocks 
of $K$. Using (\ref{cc_superpos_limit1a}), (\ref{cc_superpos_limit1b}), (\ref{cc_superpos_limit2}), 
with $(\hat{\Xi}_2)_{kl} = \xi_i \, \delta_{ik} \delta_{il}$, we obtain
\be
  q \sim \frac{\kappa_i^{-1}\, (\gamma_i^{\tminus}\, \txi_i)^\ast}{1+|\gamma_i^{\tminus}\, \txi_i|^2 } 
    \, s_i^{\ast -\eta} \, \frac{\boldsymbol{u}_i^{\tminus} \, 
        \boldsymbol{v}_i^{\tminus \dagger}}{(\gamma_i^{\tminus} \boldsymbol{v}_i^\intercal 
        \boldsymbol{u}_i)^\ast }       \, , \quad 
  p \sim \frac{\kappa_i^{-1}}{1+ |\gamma_i^{\tminus} \txi_i|^2} \, s_i^{-\eta}
   \, \frac{\boldsymbol{u}_i^{\tminus} \, \boldsymbol{v}_i^{\tminus \intercal} }{\gamma_i^{\tminus}
      \boldsymbol{v}_i^\intercal \boldsymbol{u}_i }
   \quad \mbox{as} \;  
  \begin{array}{l} t \to -\infty \\ \bar{x}_i = \mathrm{const.} \end{array}  \;
           \label{cc_incoming_q,p}
\ee
where we introduced 
\be
     \txi_i = k_{ii} \, \xi_i = \kappa_i \, \boldsymbol{v}_i^\intercal 
              \boldsymbol{u}_i \, \xi_i 
\ee 
and 
\be
    \gamma_i^{\tminus} = \frac{(L_2)_{ii}}{k_{ii}} \, , \quad
    \boldsymbol{u}_i^{\tminus} = (U_2 - U_1 K_1^{-1} K_{12})_i \, , \quad
    \boldsymbol{v}_i^{\tminus} = s_i^\eta \, [(S_2^{-\eta} V_2 
          - K_{21} K_1^{-1} S_1^{-\eta} V_1)_i]^\intercal \; .
\ee
The latter vectors can also be expressed as follows,
\be
  \boldsymbol{u}_i^{\tminus} 
 \!\! &=& \!\! \boldsymbol{u}_i - \sum_{j,k=1}^{i-1} \boldsymbol{u}_j \, (K_1^{-1})_{jk} k_{ki}  
   = \frac{1}{\det(K_1)} \, \det\left( \begin{array}{ccc} k_{11} & \cdots & k_{1i} \\
          \vdots & \ddots & \vdots \\
          k_{i-1,1} & \cdots & k_{i-1,i} \\
          \boldsymbol{u}_1 & \cdots & \boldsymbol{u}_i 
         \end{array} \right)  \, ,  \nonumber \\
  \boldsymbol{v}_i^{\tminus}
 \!\! &=& \!\! s_i^{\eta} \Big( s_i^{-\eta} \boldsymbol{v}_i 
           - \sum_{j,k=1}^{i-1} k_{ij} (K_1^{-1})_{jk} s_k^{-\eta} \, \boldsymbol{v}_k \Big)  
  = \frac{s_i^{\eta}}{\det(K_1)} \, \det\left( \begin{array}{cccc} 
          k_{11} & \cdots & k_{1,i-1} & s_1^{-\eta} \boldsymbol{v}_1 \\
          \vdots & \ddots & \vdots & \vdots \\
          k_{i1} & \cdots & k_{i,i-1} & s_i^{-\eta} \boldsymbol{v}_i 
         \end{array} \right) . \qquad        \label{cc_incoming_det}
\ee
The first equation e.g. can be verified by Laplace expansion of the determinant of the big matrix 
on the right hand side with respect to the last row, 
and subsequent Laplace expansion of the cofactors with respect to the last column. 
The expressions in (\ref{cc_incoming_det}) are examples of quasideterminants (see e.g. \cite{GGRW05}). 
Note also that 
\be
   (L_2)_{ii} = (K_2 - K_{21} K_1^{-1} K_{12})_{ii} = \frac{1}{\det(K_1)} 
        \det\left( \begin{array}{ccc} k_{11} & \cdots & k_{1i} \\
          \vdots & \ddots & \vdots \\
          k_{i1} & \cdots & k_{ii} 
         \end{array} \right) \, ,
\ee
so that all constituents of the formulae in (\ref{cc_incoming_q,p}) have nice expressions in terms 
of the vectors $\boldsymbol{u}_i$ and $\boldsymbol{v}_i$, and $k_{jl}$ with $j,l =1,\ldots,i$. 
The expressions in (\ref{cc_incoming_q,p}) are arranged in the form of the single soliton solution 
(\ref{ccred_single_matrix-soliton}). This is more clearly seen by noting the following result.

\begin{proposition}
\label{prop:cc_asympt_norm} 
\be
    \boldsymbol{v}_i^{\tminus \intercal} \boldsymbol{u}_i^{\tminus} 
  = \gamma_i^{\tminus} \, \boldsymbol{v}_i^\intercal \, \boldsymbol{u}_i \; .
               \label{cc_asympt_normalisation-}
\ee
\end{proposition}
\noindent
\textit{Proof:} In the continuous NLS case, we have
$\boldsymbol{v}_i^\intercal \boldsymbol{u}_j = (s_i + s_j^\ast) \, k_{ij}$. 
Hence
\bez
  &&    \boldsymbol{v}_i^{\tminus \intercal} \, \boldsymbol{u}_i^{\tminus} 
  = \Big( \boldsymbol{v}_i - \sum_{j,k=1}^{i-1} k_{ij} (K_1^{-1})_{jk} 
       \, \boldsymbol{v}_k \Big)^\intercal \, 
      \Big( \boldsymbol{u}_i - \sum_{l,r=1}^{i-1} \boldsymbol{u}_l \, 
         (K_1^{-1})_{lr} k_{ri} \Big) 
                 \nonumber \\
  &=& \boldsymbol{v}_i^\intercal \boldsymbol{u}_i
      - \sum_{l,r=1}^{i-1} \boldsymbol{v}_i^\intercal \boldsymbol{u}_l (K_1^{-1})_{lr} k_{ri}
      - \sum_{j,k=1}^{i-1} \boldsymbol{v}_k^\intercal \boldsymbol{u}_i k_{ij} (K_1^{-1})_{jk} 
      + \sum_{j,k,l,r=1}^{i-1} \boldsymbol{v}_k^\intercal \boldsymbol{u}_l k_{ij} 
        (K_1^{-1})_{jk} (K_1^{-1})_{lr} k_{ri} 
                        \nonumber \\ 
  &=& (s_i + s_i^\ast) \, \Big( k_{ii} 
         - \sum_{j,k=1}^{i-1} k_{ij} (K_1^{-1})_{jk} k_{ki} \Big) 
       - \sum_{j,k=1}^{i-1} k_{ij} (K_1^{-1})_{jk} \, s_k \, 
         \Big( k_{ki} - \sum_{l,r=1}^{i-1} k_{kl} (K_1^{-1})_{lr} k_{ri} \Big) \nonumber \\
   &&  -\sum_{l,r=1}^{i-1} \Big(k_{il} -\sum_{j,k=1}^{i-1} k_{ij} (K_1^{-1})_{jk} k_{kl} \Big) s_l^\ast 
        \, (K_1^{-1})_{lr} k_{ri} \; .
\eez
Now we observe that the last two sums vanish as a consequence of the identities
\bez
    k_{ri} - \sum_{j,k=1}^{i-1} k_{rj} (K_1^{-1})_{jk} k_{ki} = 0 \, , \quad
    k_{ir} - \sum_{j,k=1}^{i-1} k_{ij} (K_1^{-1})_{jk} k_{kr} = 0 
    \quad \mbox{for} \quad r<i \; . 
\eez
The first identity is e.g. obtained from the first identity in (\ref{cc_incoming_det}) by 
replacing $\boldsymbol{u}_j$ by $k_{rj}$, $j=1,\ldots,i$. 
Hence
\bez
    \boldsymbol{v}_i^{\tminus \intercal} \, \boldsymbol{u}_i^{\tminus} 
  = \frac{(L_2)_{ii}}{k_{ii}} \, \boldsymbol{v}_i^\intercal \boldsymbol{u}_i 
  = \gamma_i^{\tminus} \, \boldsymbol{v}_i^\intercal \boldsymbol{u}_i \; .
\eez
In the discrete NLS case, using 
$\boldsymbol{v}_i^\intercal \boldsymbol{u}_j = (1 - s_i s_j^\ast) \, k_{ij}$, 
we obtain 
\bez
      \boldsymbol{v}_i^{\tminus \intercal} \, \boldsymbol{u}_i^{\tminus} 
  &=& (1- s_i s_i^\ast) \, (L_2)_{ii}
      - s_i \sum_{j,k=1}^{i-1} k_{ij} (K_1^{-1})_{jk} \, s_k^{-1}  \, 
         \Big( k_{ki} - \sum_{l,r=1}^{i-1} k_{kl} (K_1^{-1})_{lr} k_{ri} \Big) \\
  && - s_i \sum_{l,r=1}^{i-1} \Big(k_{il} -\sum_{j,k=1}^{i-1} k_{ij} (K_1^{-1})_{jk} k_{kl} \Big) 
       s_l^\ast \, (K_1^{-1})_{lr} k_{ri} 
   = \gamma_i^{\tminus} \, \boldsymbol{v}_i^\intercal \boldsymbol{u}_i \; .
\eez
\hfill $\square$
\vskip.2cm

\noindent
Now we can rewrite (\ref{cc_incoming_q,p}) as follows,
\be
  q \sim \frac{\kappa_i^{-1}\, (\gamma_i^{\tminus}\, \txi_i)^\ast}{1+|\gamma_i^{\tminus}\, \txi_i|^2 } 
    \, s_i^{\ast -\eta} \, \frac{\boldsymbol{u}_i^{\tminus} \, 
        \boldsymbol{v}_i^{\tminus \dagger}}{(\boldsymbol{v}_i^{\tminus \intercal} 
        \boldsymbol{u}_i^{\tminus})^\ast }       \, , \quad
  p \sim \frac{\kappa_i^{-1}}{1+ |\gamma_i^{\tminus} \txi_i|^2} \, s_i^{-\eta}
   \, \frac{\boldsymbol{u}_i^{\tminus} \, \boldsymbol{v}_i^{\tminus \intercal} }{
      \boldsymbol{v}_i^{\tminus \intercal} \boldsymbol{u}_i^{\tminus} }
   \quad \mbox{as} \;  
  \begin{array}{l} t \to -\infty \\ \bar{x}_i = \mathrm{const.} \end{array}  \;
           \label{cc_incoming_q,p_2}
\ee

In the same way we obtain
\be
  q \sim \frac{\kappa_i^{-1} \, (\gamma_i^{\tplus} \, \txi_i)^\ast}{1+ |\gamma_i^{\tplus} \txi_i|^2} 
   \, s_i^{\ast -\eta} \, \frac{\boldsymbol{u}_i^{\tplus} \, \boldsymbol{v}_i^{\tplus \dagger}}{
     (\boldsymbol{v}_i^{\tplus \intercal} \boldsymbol{u}_i^{\tplus})^\ast }  \, , \quad
  p \sim \frac{\kappa_i^{-1}}{1+ |\gamma_i^{\tplus} \txi_i|^2} \, s_i^{-\eta} \, 
   \frac{\boldsymbol{u}_i^{\tplus} \, \boldsymbol{v}_i^{\tplus \intercal}}{
    \boldsymbol{v}_i^{\tplus \intercal} \boldsymbol{u}_i^{\tplus} }
  \quad \mbox{as} \;  
  \begin{array}{l} t \to +\infty \\ \bar{x}_i = \mathrm{const.} \end{array} ,
         \label{cc_outgoing_q,p}
\ee
where
\be
    \gamma_i^{\tplus} = \frac{(L_1)_{ii}}{k_{ii}} \, , \quad
    \boldsymbol{u}_i^{\tplus} = (U_1 - U_2 K_2^{-1} K_{21})_i \, , \quad
    \boldsymbol{v}_i^{\tplus} 
  = s_i^\eta \, [(S_1^{-\eta} V_1 - K_{12} K_2^{-1} S_2^{-\eta} V_2)_i]^\intercal \; . 
\ee
The vectors can also be expressed as follows,
\be
     \boldsymbol{u}_i^{\tplus}
  &=& \frac{1}{\det(K_2)} \, \det\left( \begin{array}{ccc} 
         \boldsymbol{u}_i & \cdots & \boldsymbol{u}_n \\ 
         k_{i+1,i} & \cdots & k_{i+1,n} \\
          \vdots & \ddots & \vdots \\
          k_{ni} & \cdots & k_{nn} 
         \end{array} \right)  \, , \nonumber \\ 
     \boldsymbol{v}_i^{\tplus} 
  &=& \frac{s_i^\eta}{\det(K_2)} \, \det\left( \begin{array}{cccc} 
          s_i^{-\eta} \boldsymbol{v}_i & k_{i,i+1} & \cdots & k_{in}  \\
          \vdots & \vdots & \ddots & \vdots  \\
          s_n^{-\eta} \boldsymbol{v}_n & k_{n,i+1}  & \cdots & k_{nn}  
         \end{array} \right) , \quad
\ee
and satisfy
\be
    \boldsymbol{v}_i^{\tplus \intercal} \boldsymbol{u}_i^{\tplus} 
  = \gamma_i^{\tplus} \, \boldsymbol{v}_i^\intercal \, \boldsymbol{u}_i \, ,
\ee
which is proved in the same way as proposition~\ref{prop:cc_asympt_norm}. Furthermore,
we have 
\be
   (L_1)_{ii} = (K_1 - K_{12} K_2^{-1} K_{21})_{ii} = \frac{1}{\det(K_2)} 
        \det\left( \begin{array}{ccc} k_{ii} & \cdots & k_{in} \\
          \vdots & \ddots & \vdots \\
          k_{n1} & \cdots & k_{nn} 
         \end{array} \right) \; .
\ee
All this nicely expresses the constituents of the formulae (\ref{cc_outgoing_q,p}) in terms of 
the vectors $\boldsymbol{u}_i$ and $\boldsymbol{v}_i$, and $k_{jl}$ with $j,l = i,\ldots,n$. 

Clearly, if $i=1$ respectively $i=n$, we have 
\be
 &&  q \sim \frac{\kappa_1^{-1} \, \txi_1^\ast}{ 1 + |\txi_1|^2} \, s_1^{\ast -\eta}
          \, \frac{\boldsymbol{u}_1 \, \boldsymbol{v}_1^\dagger}{( \boldsymbol{v}_1^\intercal 
          \boldsymbol{u}_1 )^\ast }
       \, , \quad
   p \sim \frac{\kappa_1^{-1}}{ 1 + |\txi_1|^2} \, s_1^{-\eta}
          \, \frac{\boldsymbol{u}_1 \, \boldsymbol{v}_1^\intercal}{ 
               \boldsymbol{v}_1^\intercal \boldsymbol{u}_1 }
      \qquad \mbox{as} \quad 
      \begin{array}{l} t \to -\infty \\ \bar{x}_1 = \mbox{const.} \end{array}
      \nonumber \\
 &&  q \sim \frac{\kappa_n^{-1} \, \txi_n^\ast}{ 1 + |\txi_n|^2} \, s_i^{-\eta}
       \, \frac{\boldsymbol{u}_n \, \boldsymbol{v}_n^\dagger}{(\boldsymbol{v}_n^\intercal 
          \boldsymbol{u}_n)^\ast }  \, , \quad
   p \sim \frac{\kappa_n^{-1}}{ 1 + |\txi_n|^2} \, s_n^{-\eta}
          \, \frac{\boldsymbol{u}_n \, \boldsymbol{v}_n^\intercal}{
             \boldsymbol{v}_n^\intercal \boldsymbol{u}_n }
      \qquad \mbox{as} \quad 
      \begin{array}{l} t \to +\infty \\ \bar{x}_n = \mbox{const.} \end{array}
\ee

By comparison of the asymptotic expressions (\ref{cc_incoming_q,p_2}) and 
(\ref{cc_outgoing_q,p}), the solitons experience a position shift and a change of polarisation 
through the scattering process.

\subsubsection{2-soliton solutions}
If $n=2$, we have
\be
    S = \left( \begin{array}{cc} s_1 & 0 \\ 0 & s_2 \end{array} \right) \, , \quad
    U = \left( \begin{array}{cc} \boldsymbol{u}_1 & \boldsymbol{u}_2 \end{array} \right) 
           \, , \quad
    V = \left( \begin{array}{c} \boldsymbol{v}_1^\intercal \\ \boldsymbol{v}_2^\intercal 
               \end{array} \right)   \; . 
\ee
Using a reparametrisation transformation (with diagonal transformation matrices) 
and a transformation (\ref{ext_transf}), we can achieve that
\be
    (\boldsymbol{v}_i^\intercal \boldsymbol{u}_j)^\ast = \boldsymbol{v}_j^\intercal \boldsymbol{u}_i 
    \, ,    \label{cc_Hermitian_K_cond}
\ee 
so that $K$ is Hermitian. But now we have to admit in the expression for $\xi_j$ a
factor $c_j \in \bbC$. The matrix $K$, given by (\ref{cc_nsoliton_kij}) in terms of the 
eigenvalues $s_j$ and the vectors $\boldsymbol{u}_j, \boldsymbol{v}_j$, $j=1,2$, has the form 
\be
    K = \left( \begin{array}{cc} k_1 & k \\ k^\ast & k_2 \end{array} \right) 
\ee 
(where the $k_i$ are real), and we obtain the following explicit solution,
\be
  q &=& \frac{1}{D} \Big[ \left( 1 + k_2^2 \, |\xi_2|^2 + k^2 \xi_1 \xi_2^\ast \right) 
           \, \xi_1^\ast s_1^{\ast -\eta} \boldsymbol{u}_1 \boldsymbol{v}_1^\dagger
         + \left( 1 + k_1^2 \, |\xi_1|^2 + k^{\ast 2} \xi_1^\ast \xi_2 \right) 
            \, \xi_2^\ast s_2^{\ast -\eta} \boldsymbol{u}_2 \boldsymbol{v}_2^\dagger      
                                  \nonumber  \\
     &&  - (k \, k_1 \, \xi_1 + k^\ast \, k_2 \, \xi_2) \, \xi_1^\ast \xi_2^\ast \, 
           (s_2^{\ast-\eta} \boldsymbol{u}_1 \boldsymbol{v}_2^\dagger 
            + s_1^{\ast-\eta} \boldsymbol{u}_2 \boldsymbol{v}_1^\dagger) \Big] \, , 
                                  \nonumber \\
 p &=& - \frac{1}{D} \Big[  
         \left( k_1 + \det(K) \, k_2 \, |\xi_2|^2 \right) \,
          k_1 |\xi_1|^2 s_1^{-\eta} \boldsymbol{u}_1 \boldsymbol{v}_1^\intercal
       + \left( k_2 + \det(K) \, k_1 \, |\xi_1|^2 \right) \, 
          k_2 |\xi_2|^2 s_2^{-\eta} \boldsymbol{u}_2 \boldsymbol{v}_2^\intercal  
                                  \nonumber \\
   &&  + \left( k^\ast - k \, \det(K) \, \xi_1 \xi_2^\ast \right) \, 
      \xi_1^\ast \xi_2 s_2^{-\eta} \boldsymbol{u}_1 \boldsymbol{v}_2^\intercal
       + \left( k - k^\ast \, \det(K) \, \xi_1^\ast \xi_2 \right) \, 
      \xi_1 \xi_2^\ast s_1^{-\eta} \boldsymbol{u}_2 \boldsymbol{v}_1^\intercal
               \Big] \, , 
\ee
where 
\be
   D = \Big| 1 - \frac{k}{k^\ast} \, \det(K) \, \xi_1 \xi_2^\ast \Big|^2 
    + \Big| k_1 \xi_1 + \frac{k^\ast}{k} k_2 \xi_2 \Big|^2 \; .
\ee

\begin{remark}
If $S=s I_n$ with $n \leq m$, we have a ``rank $n$ soliton''. 
It follows from proposition~\ref{prop:cc_sup_sim_data} that we can choose 
all $m$-component vectors composing $U$, respectively $V$, linearly independent. 
In particular, this means that there are at most $m^2$ different such solitons. 
If $n=m$, then with a reparametrisation transformation we can achieve that 
$U=V=I_m$, while replacing $\Xi$ by $C \Xi$ with a constant 
$m \times m$ matrix $C$. Imposing the condition (\ref{cc_Hermitian_K_cond}) 
on the vectors $\boldsymbol{u}_i, \boldsymbol{v}_j$, so that $K$ is Hermitian, and 
choosing $C$ symmetric, the resulting NLS solution is regular. In the continuous 
NLS case, $q$ is then symmetric. Superposing such rank $m$ solitons (with symmetric $C_i$), 
we obtain again continuous $m \times m$ matrix NLS solutions with symmetric $q$. 
This may be of interest for the aforementioned spin 1 Bose-Einstein condensate model.
\hfill $\square$
\end{remark}

\subsubsection{Regularity condition for matrix solitons in the focusing NLS case} 
We address the question whether a superposition (in the above sense of ``superposing'' the 
corresponding matrix data) of regular solutions again leads to a 
regular solution. Hence we should check the determinant of (\ref{cc_Z_superpos}) for the 
occurence of zeros. As an intermediate step, we compute the Schur complement of the 
first block-diagonal entry of $Z$, which is 
\be
   \mathfrak{D} := Z_2 + K_{21}^\ast \Xi_1 K_{12} 
 - (K_2^\ast \Xi_2 K_{21} + K_{21}^\ast \Xi_1 K_1) (Z_1 + K_{12}^\ast \Xi_2 K_{21})^{-1} 
   (K_1^\ast \Xi_1 K_{12} + K_{12}^\ast \Xi_2 K_2 ) \, . \;
\ee
Here we assume that, for the superposed data, a matrix $K$ as in (\ref{cc_superpos}) exists such that 
(\ref{cc_superpos_K12,K21_cont}), respectively (\ref{cc_superpos_K12,K21_discr}) holds. 
A sufficient condition is (\ref{cc_Sylvester_spec_cond}). 
We further assume that $Z_1$ and $Z_2$ are (for all $x,t$) invertible, so that 
the constituents of the superposition are regular, and also that
$\Xi_2^{-1} + K_{21} Z_1^{-1} K_{12}^\ast$ is invertible. By the 
matrix inversion lemma (see e.g. \cite{Bern09}, Corollary 2.8.8), the latter condition 
guarantees the existence of 
\be
   (Z_1 + K_{12}^\ast \Xi_2 K_{21})^{-1} 
 = Z_1^{-1} - Z_1^{-1} K_{12}^\ast (\Xi_2^{-1} + K_{21} Z_1^{-1} K_{12}^\ast)^{-1} 
   K_{21} Z_1^{-1} \; .
\ee
Using this formula in the preceding equation, and noting that
\be
   (\Xi_2^{-1} + K_{21} Z_1^{-1} K_{12}^\ast)^{-1} \, K_{21} Z_1^{-1} K_{12}^\ast \Xi_2 
 &=& \Xi_2 - (\Xi_2^{-1} + K_{21} Z_1^{-1} K_{12}^\ast)^{-1} \nonumber \\
 &=& \Xi_2 K_{21} Z_1^{-1} K_{12}^\ast \, (\Xi_2^{-1} + K_{21} Z_1^{-1} K_{12}^\ast)^{-1} \, , 
\ee
we obtain
\be
   \mathfrak{D} &=& Z_2 + K_{21}^\ast \Xi_1 K_{12} 
   - K_{21}^\ast \Xi_1 K_1 Z_1^{-1} K_1^\ast \Xi_1 K_{12} \nonumber \\
 && + K_{21}^\ast \Xi_1 K_1 Z_1^{-1} K_{12}^\ast (\Xi_2^{-1} + K_{21} Z_1^{-1} K_{12}^\ast)^{-1}
  K_{21} Z_1^{-1} K_1^\ast \Xi_1 K_{12}  \nonumber \\
 && - K_2^\ast (\Xi_2^{-1} + K_{21}^\ast Z_1^{-1} K_{12}^\ast)^{-1} K_{21} Z_1^{-1} K_1^\ast \Xi_1 K_{12}
       \nonumber \\
 && - K_{21}^\ast \Xi_1 K_1 Z_1^{-1} K_{12}^\ast (\Xi_2^{-1} + K_{21} Z_1^{-1} K_{12}^\ast)^{-1} K_2
            \nonumber \\
 && - K_2^\ast \Xi_2 K_2 + K_2^\ast (\Xi_2^{-1} + K_{21} Z_1^{-1} K_{12}^\ast)^{-1} K_2 \; .
\ee 
Using the identity $\Xi_1 K_1 Z_1^{-1} K_1^\ast = I - Z_1^{\ast -1} \Xi_1^{-1}$ in the third summand, 
this can rewritten as follows, 
\be
   \mathfrak{D} &=&  ( K_{21}^\ast \Xi_1 K_1 Z_1^{-1} K_{12}^\ast - K_2^\ast) 
                     (\Xi_2^{-1} + K_{21} Z_1^{-1} K_{12}^\ast)^{-1}
                     (K_{21} Z_1^{-1} K_1^\ast \Xi_1 K_{12} - K_2) \nonumber \\
 && + \Xi_2^{\ast -1} + K_{21}^\ast Z_1^{\ast -1} K_{12} \; .
\ee 
With the help of the identity $\Xi_1 K_1 Z_1^{-1} = (Z_1^{-1} K_1^\ast \Xi_1)^\ast$, we obtain
\be
   \mathfrak{D} &=&  ( K_{21} Z_1^{-1} K_1^\ast \Xi_1 K_{12} - K_2)^\ast 
                     (\Xi_2^{-1} + K_{21} Z_1^{-1} K_{12}^\ast)^{-1}
                     (K_{21} Z_1^{-1} K_1^\ast \Xi_1 K_{12} - K_2) \nonumber\\
 && + (\Xi_2^{-1} + K_{21} Z_1^{-1} K_{12}^\ast)^\ast \; .
\ee 

If the data  $(S_2,U_2,V_2)$ belong to a 1-soliton solution, writing 
$S_2 = s$, $\Xi_2 = \xi$, $U_2 = \boldsymbol{u}$, $V_2 = \boldsymbol{v}^\intercal$, 
$K_2 = k$, $K_{12} = \boldsymbol{k}_1$, and $K_{21} = \boldsymbol{k}_2^\intercal$, 
we have
\be
  Z = \left( \begin{array}{cc} Z_1 + \xi \, \boldsymbol{k}_1^\ast \boldsymbol{k}_2^\intercal & 
         K_1^\ast \Xi_1 \boldsymbol{k}_1 + \xi \, k \, \boldsymbol{k}_1^\ast \\ 
         \xi \, k^\ast \boldsymbol{k}_2^\intercal + \boldsymbol{k}_2^{\intercal \ast} \Xi_1 K_1 &
         Z_2 + \boldsymbol{k}_2^{\intercal \ast} \Xi_1 \boldsymbol{k}_1 \end{array} \right) \, ,
\ee
where $Z_2 = \xi^{\ast -1} + |k|^2 \, \xi$. 
Using determinant identities (see e.g. \cite{Bern09}, Fact 2.14.2), we find 
\be
  \det(Z) &=& \det(Z_1 + \xi \, \boldsymbol{k}_1^\ast \boldsymbol{k}_2^\intercal) \, \mathfrak{D} 
 = \det(Z_1) \, (1 + \xi \, \boldsymbol{k}_2^\intercal Z_1^{-1} \boldsymbol{k}_1^\ast) \,
   \mathfrak{D}          \nonumber \\
&=& \xi \, \det(Z_1) \, (\xi^{-1} + \boldsymbol{k}_2^\intercal Z_1^{-1} \boldsymbol{k}_1^\ast) \, 
   \Big( (\xi^{-1} + \boldsymbol{k}_2^\intercal Z_1^{-1} \boldsymbol{k}_1^\ast)^\ast \nonumber \\
&& + ( \boldsymbol{k}_2^\intercal Z_1^{-1} K_1^\ast \Xi_1 \boldsymbol{k}_1 - k)^\ast \, 
   ( \xi^{-1} + \boldsymbol{k}_2^\intercal Z_1^{-1} \boldsymbol{k}_1^\ast)^{-1} 
    ( \boldsymbol{k}_2^\intercal Z_1^{-1} K_1^\ast \Xi_1 \boldsymbol{k}_1 - k) \Big) \nonumber \\
&=& \xi \, \det(Z_1) \, \Big( |\xi^{-1} + \boldsymbol{k}_2^\intercal Z_1^{-1} \boldsymbol{k}_1^\ast|^2 
            + | \boldsymbol{k}_2^\intercal Z_1^{-1} K_1^\ast \Xi_1 \boldsymbol{k}_1 - k |^2 \Big)
    \; .
\ee
This can only vanish if 
$k = \boldsymbol{k}_2^\intercal Z_1^{-1} K_1^\ast \Xi_1 \boldsymbol{k}_1$ 
and simultaneously
$\xi^{-1} = -\boldsymbol{k}_2^\intercal Z_1^{-1} \boldsymbol{k}_1^\ast$ 
at some value of $x$ and $t$.
Generically it is not possible to satisfy both equations at any value of $x,t$. (Note that these  
equations consist of \emph{four} real equations.) 
In particular, this shows that the matrix $n$-soliton solutions are generically regular.

\subsubsection{Cases in which the Sylvester equation has not a unique solution}
\label{subsubsec:cc_exceptional}
Several of our results so far in this section rest upon the assumption that the condition 
(\ref{cc_Sylvester_spec_cond}) is satisfied, in which case the Sylvester equation in
proposition~\ref{prop:NLS_cc} possesses a unique solution. Now we address those cases 
in which this condition is violated, confining ourselves to the \emph{continuous} 
matrix NLS equation (\ref{NLS_cc}). 
The simplest such cases occur if the matrix $S$ consists of a single Jordan block with a purely 
imaginary eigenvalue. By application of a Galilean transformation (see remark~\ref{rem:cc_S-transf}), 
this case can be reduced to that where the eigenvalue is zero.

\begin{example}
(1) Let $n=1$ and $S = 0$. Writing $U=\boldsymbol{u}$ and $V=\boldsymbol{v}^\intercal$ 
with $m$-component vectors $\boldsymbol{u}, \boldsymbol{v}$, the Sylvester equation 
requires $\boldsymbol{v}^\intercal \boldsymbol{u} =0$ and leaves $K=k$ completely arbitrary. 
If $m=1$ (the scalar NLS case), the corresponding NLS solution vanishes identically. 
But if $m>1$ (the matrix NLS case), this is not so\footnote{This is in contrast to the 
Hermitian reduction case treated in section~\ref{sec:HermRed}, where  
the corresponding Sylvester equation also leads to the trivial solution if $m>1$. See 
remark~\ref{rem:Hred_no_except}. }
and we obtain the \emph{constant} rank one matrix NLS solution
\be
   q = \frac{1}{1- \epsilon \, |k|^2} \, \boldsymbol{u} \, \boldsymbol{v}^\dagger 
       \qquad \mbox{where} \quad \boldsymbol{v}^\intercal \boldsymbol{u} =0  \; .
\ee 
A Galilean transformation generalises it to
\be
    q = \frac{1}{1-\epsilon \, |k|^2} \, e^{\imag \, \beta \, (x - \beta \, t)} \, 
        \boldsymbol{u} \, \boldsymbol{v}^\dagger \, ,
        \label{cc_except_n=1_q}
\ee 
with $\beta \in \bbR$. This is what we would have obtained had we started above with 
$S = \imag \, \beta$, see remark~\ref{rem:cc_S-transf}. 
As a consequence of $\boldsymbol{v}^\intercal \boldsymbol{u} =0$, we have 
$q q^\ast = 0$, so that $q$ solves the linear and the nonlinear parts of the 
NLS equation separately. 
\\
(2) Let $n=2$ and $S$ a Jordan block with zero eigenvalue, hence
\be
 S = \left( \begin{array}{cc} 0 & 1 \\ 0 & 0 \end{array} \right) \, , \quad
 U = \left( \begin{array}{cc} \boldsymbol{a}_1 & \boldsymbol{a}_2 \end{array} \right) \, , \quad
 V = \left( \begin{array}{c} \boldsymbol{b}_1 \\ \boldsymbol{b}_2 \end{array} \right)
   \; .    \label{cc_except-sol_2x2Jordan}
\ee
The Sylvester equation only has a solution if $\boldsymbol{b}_2^\intercal \, \boldsymbol{a}_1 = 0$ 
and $\boldsymbol{b}_1^\intercal \, \boldsymbol{a}_1 = \boldsymbol{b}_2^\intercal \, \boldsymbol{a}_2$.
Let us more concretely consider the case $m=2$. The solution $K$ of the Sylvester equation 
depends on two arbitrary complex constants (expressing the non-uniqueness 
of $K$). The resulting components of $q$ are quotients of a linear polynomial in $x$ 
by a quadratic polynomial, hence $q$ vanishes as $|x| \to \infty$.
Applying a Galilean transformation (see remark~\ref{rem:cc_S-transf}), these solutions 
become $t$-dependent.\footnote{The absolute values of the components are still rational functions,  
hence these solutions resemble the ``rational'' soliton solutions known in the scalar NLS case 
\cite{Pere83}.} 
A special regular member from this family is
\be
    q = \frac{e^{\imag \, \beta \, (x-\beta \, t)}}{1+(x-2 \beta \, t)^2} \; \left(
        \begin{array}{cc} 1 & -(x-2 \beta \, t) \\ x-2 \beta \, t & 1 \end{array} 
        \right)  \qquad \qquad \beta \in \bbR  \; .
\ee
(3) For $n>2$, with $S$ still consisting of a Jordan block with 
eigenvalue zero, one obtains $t$-\emph{dependent rational} 
solutions. This includes regular solutions with a breather-like behaviour. 
For $n=4$, a special solution of the $2 \times 2$ matrix focusing NLS equation is determined by
\be
   S = \left( \begin{array}{rrrr} 0 & 1 & 0 & 0 \\
                                  0 & 0 & 1 & 0 \\
                                  0 & 0 & 0 & 1 \\
                                  0 & 0 & 0 & 0
       \end{array} \right) , \;
   U = \left( \begin{array}{rrrr} 0 & 0 & 1 & -1 \\ 1 & 0 & 0 & 0 \end{array} \right) , \;
   V = \left( \begin{array}{rr} 0 & -1 \\ 0 & -1 \\ 0 & 0 \\ 1 & 0 \end{array} \right) , \;
   K = \left( \begin{array}{rrrr} 0 & 0 & 0 & 0 \\
                                 -1 & 0 & 0 & 0 \\
                                 -1 & 1 & 0 & 0 \\
                                  0 & 1 &-1 & 0
       \end{array} \right) .  \nonumber \\   \label{cc2x2NLS_n=4_data}
\ee
This rational solution is regular, vanishes as $|x| \to \infty$, 
and shows a breather-like behaviour, see Fig.~\ref{fig:cc2x2NLS_n=4}. 
\begin{figure}[t] 
\begin{center} 
\resizebox{14.cm}{!}{
\includegraphics{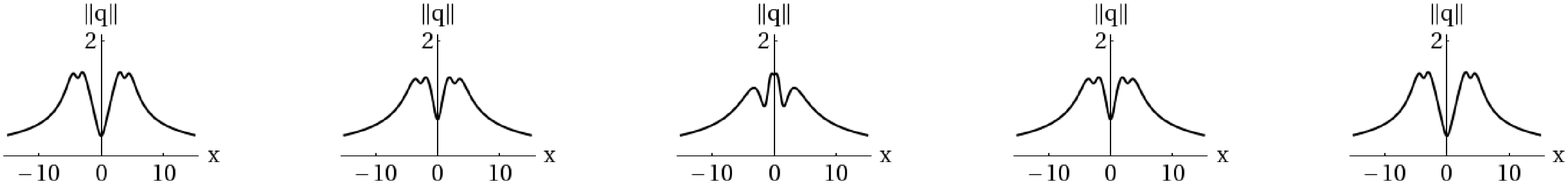}
}
\end{center} 
\caption{The norm $\|q\|$ of the $2 \times 2$ matrix focusing NLS solution, determined by 
the data (\ref{cc2x2NLS_n=4_data}), at $t=-4,-2,0,2,4$. This is a rational ``matrix breather''. 
\label{fig:cc2x2NLS_n=4}  }
\end{figure} 
\hfill $\square$
\end{example}

Of course, we can superpose any number of solutions of the kind described in the 
last example, see the next example, and also compose such data with (e.g. solitonic) data satisfying
(\ref{cc_Sylvester_spec_cond}), as in the next but one example. In doing so, we have 
to take the parameters $\beta_i$ into account, since a Galilean transformation applied 
to the composite solution can only eliminate or create one of these parameters. 
In the special case where $S= \mbox{diag}(\imag \, \beta_1, \ldots, \imag \, \beta_n)$, 
with real $\beta_i \neq 0$, the resulting solutions are periodic in $x$ and $t$, 
since they are rational expressions in trigonometric functions. 

\begin{example}
\label{ex:cc_ratsol_collision}
The following superposition of data of the kind considered in the preceding example 
determines a regular solution that vanishes as $|x| \to \infty$, see Fig.~\ref{fig:ratsolcollision}.
\bez
  S = \left( \begin{array}{cccc} \imag \, \beta & 1 & 0 & 0 \\ 0 & \imag \, \beta & 0 & 0 \\
                                 0 & 0 & 0 & 1 \\ 0 & 0 & 0 & 0 
             \end{array} \right) , \,
  U = \left( \begin{array}{cccc} 1 & 0 & 1 & 0 \\ 
                                 0 & 1 & 0 & 1  
             \end{array} \right) , \,
  V = \left( \begin{array}{cr} 1 & 0 \\ 0 & 1 \\ 1 & -1 \\ 0 & 1 
             \end{array} \right) , \,
  K = \left( \begin{array}{ccrr} 0 & 0 & -\frac{\imag}{2} & \frac{1}{2} \\ 1 & 0 & 0 & -\frac{\imag}{2} \\
                                 \frac{\imag}{2} & \frac{1-\imag}{2} & 0 & 0 \\ 0 & \frac{\imag}{2} & 1 & -1 
             \end{array} \right) . 
\eez
\begin{figure}[t] 
\begin{center} 
\resizebox{11.cm}{!}{
\includegraphics{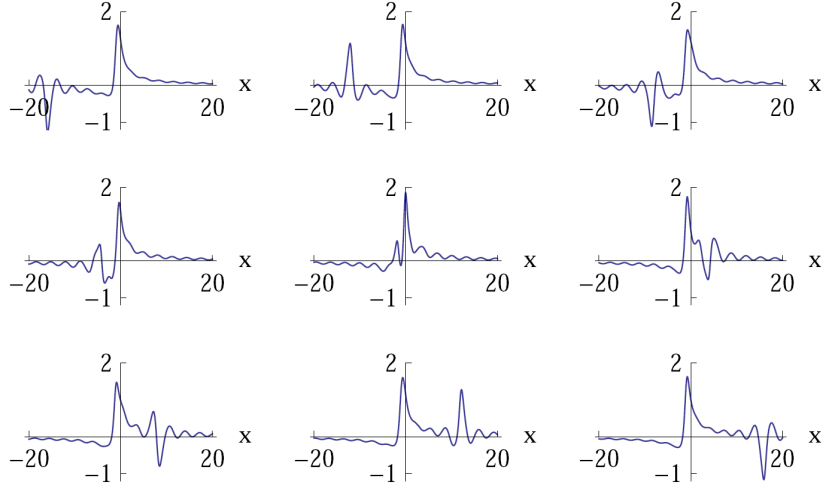}
}
\end{center} 
\caption{The real part of $q_{11}$ for the matrix NLS solution $q$ determined by the data in 
example~\ref{ex:cc_ratsol_collision} at $t=-4,\ldots,4$. 
\label{fig:ratsolcollision}  }
\end{figure} 
\hfill $\square$
\end{example}

\begin{example}
\label{ex:cc_degen+regular}
Let $S= \mathrm{diag}(s,0)$, $s \not\in \imag \, \bbR$, and $U,V$ as in (\ref{cc_except-sol_2x2Jordan}) 
with $m=2$.
The Sylvester equation then admits a solution $K$ iff 
$\boldsymbol{b}_2^\intercal \, \boldsymbol{a}_2 =0$. 
Writing $\boldsymbol{a}_i^\intercal = (a_{i1},a_{i2})$, $\boldsymbol{b}_i^\intercal = (b_{i1},b_{i2})$,
we solve this constraint by setting $b_{21} = c \, a_{22}$ and $b_{22} = -c \, a_{21}$ with 
$c \in \bbC$. Then $K$ takes the form
\be
   K = \left( \begin{array}{cc} \frac{a_{11} b_{11} + a_{12} b_{12}}{s+s^\ast} & 
        \frac{a_{21} b_{11} + a_{22} b_{12}}{s} \\  
       c \, \frac{a_{11} a_{22} - a_{12} a_{21}}{s^\ast} & k \end{array} \right)
       \, ,
\ee
where $k \in \bbC$ is another arbitrary constant. The resulting solution family of the  
$2 \times 2$ matrix focusing NLS equation depends on nine complex parameters and one real. Fig.~\ref{fig:degen+regular} shows a regular solution from this family. 
\begin{figure}[t] 
\begin{center} 
\resizebox{14.cm}{!}{
\includegraphics{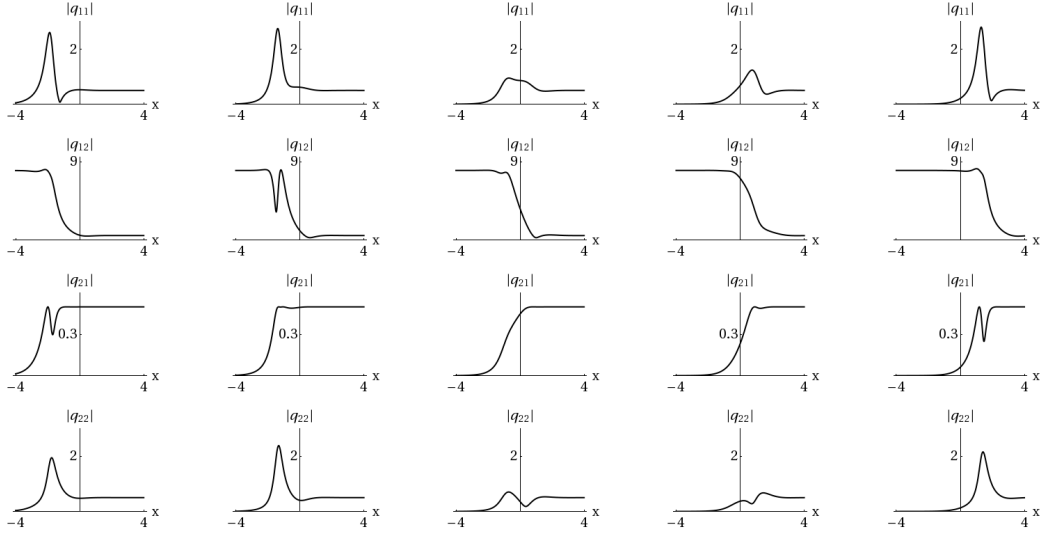}
}
\end{center} 
\caption{A sequence of snapshots of a solution from the family in 
example~\ref{ex:cc_degen+regular}. The plots show the absolute values of the components 
of the $2 \times 2$ matrix focusing NLS solution $q$ at times $t=-.2,0,.2,.4,.6$.
Here we chose the parameter values 
$s=2 \, (1+\imag)$, $c=k=a_{11}=a_{22}=1$ and $a_{21}=-1, b_{12}=\imag, a_{12}=b_{11}=0$. 
\label{fig:degen+regular}  }
\end{figure} 
\hfill $\square$
\end{example}

Recall that any NLS solution corresponding to matrix data $(S,U,V)$ consisting of several 
blocks can be regarded as a (nonlinear) superposition of solutions corresponding to these 
blocks, say $(S_1,U_1,V_1)$ and $(S_2,U_2,V_2)$. 
If $\sigma(S_1) \cap \sigma(-S_2^\ast) = \emptyset$ in the continuous NLS case 
(respectively $\sigma(S_1^\ast) \cap \sigma(S_2^{-1}) = \emptyset$ in the discrete cases), 
the solution is completely determined by the ``constituent solutions''. Soliton solutions 
are the examples par excellence. If $\sigma(S_1) \cap \sigma(-S_2^\ast) \neq \emptyset$ 
(respectively $\sigma(S_1^\ast) \cap \sigma(S_2^{-1}) \neq \emptyset$), 
there is no unique solution for the off-diagonal blocks of the matrix $K$ solving the 
Sylvester equation for the superposition. The parts then no longer determine the composite 
solution completely, since additional freedom enters through arbitrary parameters in $K$. 
But if the eigenvalue of $S_1$ has a non-zero real part in the continuous NLS case, 
respectively a modulus different from $1$ in the discrete NLS cases, then we can apply 
a reflection symmetry to $S_2$ (provided that the associated matrix $K_2$ is invertible). 
This leads to equivalent data involving Jordan blocks $S_1$ and $S_2'$, where 
now $\sigma(S_1) \cap \sigma(-S_2'^\ast) = \emptyset$. This means 
that such a NLS solution can be realised in different ways as a superposition, 
i.e. with different constituents.

\section{Hermitian conjugation reduction of the continuous NLS system}
\label{sec:HermRed}
\setcounter{equation}{0}
By decomposition, the Hermitian conjugation reduction condition (\ref{NLS_dag_red}) for the continuous 
matrix NLS system becomes
\be
     \bar{q} = \epsilon \, q^\dagger \, , \qquad  
     p = p^\dagger \, , \qquad  
     \bar{p} = \bar{p}^\dagger \, ,
\ee 
where $\epsilon = \pm 1$. It reduces the NLS system to the $m_1 \times m_2$ matrix \emph{NLS equation} 
(see e.g. \cite{APT04book}) 
\be
   \imag \, q_t + q_{xx} - 2 \, \epsilon \, q q^\dagger q = 0 \, ,  \label{matrix-NLS}
\ee
supplemented by $p_x = - \epsilon \, q q^\dagger$. 

If $m_2=1$ and $m_1>1$, (\ref{matrix-NLS}) is the \emph{vector NLS equation} (or \emph{coupled NLS equations}) 
\be
     \imag \, \vec{q}_t + \vec{q}_{xx} - 2 \, \epsilon \, |\vec{q}|^2 \, \vec{q} = 0 \, ,
\ee
first treated in \cite{Manak74}. 

If $m_1=m_2$ and if $q$ is \emph{symmetric} or \emph{skew-symmetric}, i.e. $q^\intercal = \pm q$, 
then (\ref{matrix-NLS}) coincides with (\ref{NLS_cc}). 
This includes the special case of a spin-1 Bose-Einstein condensate mentioned earlier. 
For the skew-symmetric case, see e.g. \cite{Nakk01}.

\begin{proposition}
\label{prop:NLS_eq_sol}
Let $S,V,\bar{V}$ be constant matrices of size $n \times n$, $n \times m_1$ and $n \times m_2$, 
respectively, and $K,\bar{K}$ constant Hermitian solutions of
\be
    S \, K + K \, S^\dagger = V V^\dagger 
    \, , \qquad
    S^\dagger \, \bar{K} + \bar{K} \, S = \bar{V} \bar{V}^\dagger \; .   \label{nlsHred_rank}
\ee
Then 
\be
  q = V^\dagger  \left( \Xi^{\dagger -1} - \epsilon \, \bar{K} \, \Xi \, K  \right)^{-1} 
      \, \bar{V}  
           \quad \mbox{where} \quad
  \Xi = e^{-x \, S - \imag t \, S^2}                   \label{nlsHred_q_sol} 
\ee
solves the $m_1 \times m_2$ matrix NLS equation (\ref{matrix-NLS}). 
\hfill $\square$
\end{proposition}
\noindent
\textit{Proof:} Setting $n_1 = n_2$, renamed to $n$, we recall from section~\ref{subsec:redsol} 
the reduction conditions to be imposed on the matrices entering the solution formulae. 
With the block structure expressed in (\ref{bJ_normalform}), (\ref{Hred_bGamma_choice}) 
and (\ref{S,U,V,K_blocks}), the statement then follows directly from 
proposition~\ref{prop:sol_NLSsystems}. 
\hfill  $\square$
\vskip.2cm

The class of matrix NLS solutions provided by proposition~\ref{prop:NLS_eq_sol} has also been 
obtained in \cite{ADM07,Demo+Mee08} via inverse scattering, including conditions under which the 
solutions are everywhere regular and exponentially localised (see also \cite{ABDM09sym}). 
Multiple soliton solutions of the matrix (including vector) NLS equation appeared previously also in 
\cite{Manak74,Radha+Laks95,RLH97,Tsuc+Wada98,Park+Shin00,Park+Shin02,Han+Shin04,KTA04,Tsuch04,APT04book,APT04DPDE,APT04IP,IUW07}, 
for example. 
\vskip.1cm

If $K$ and $\bar{K}$ solve the Lyapunov equations (\ref{nlsHred_rank}), then also $K^\dagger$ 
and $\bar{K}^\dagger$. As a consequence, there are then also Hermitian solutions of these 
equations. 
We further note that the equations (\ref{nlsHred_rank}) have unique solutions if
\be
      \sigma(S) \cap \sigma(-S^\dagger) = \emptyset \, ,  \label{Hred_spec_cond}
\ee
in which case $K$ and $\bar{K}$ are then necessarily Hermitian.

\begin{remark} 
\label{rem:Hred_no_except}
This concerns cases in which (\ref{Hred_spec_cond}) does not hold. 
If $S=0$, then (\ref{nlsHred_rank}) implies $V=\bar{V}=0$, leading  
to $q=0$. If $S$ is an $r \times r$ Jordan block with eigenvalue zero, i.e. $S_{ij} = \delta_{i,j-1}$, 
a more involved inspection of (\ref{nlsHred_rank}) shows that 
$K$ is left-upper-triangular, $\bar{K}$ right-lower-triangular, 
$V^\dagger = (\boldsymbol{v}_1,\ldots,\boldsymbol{v}_k,\boldsymbol{0},\ldots,\boldsymbol{0})$ 
and $\bar{V}^\dagger = (\boldsymbol{0},\ldots,\boldsymbol{0},\bar{\boldsymbol{v}}_l, \ldots,\bar{\boldsymbol{v}}_r)$, where $k=r/2, l=k+1$ if $r$ is even, and $k=(r-1)/2, l=k+2$ 
if $r$ is odd. (\ref{nlsHred_q_sol}) again leads to $q=0$. Corresponding solutions 
of the complex conjugation invariant matrix NLS equation, obtained in 
section~\ref{subsubsec:cc_exceptional}, have therefore no analogue in the Hermitian invariant 
NLS case.
\hfill $\square$
\end{remark}

\begin{remark} 
The first result noted in remark~\ref{rem:symmetric} for \emph{symmetric} $S$ 
can also be obtained from proposition~\ref{prop:NLS_eq_sol} by setting 
$\bar{V} = \pm V^\ast$ and $\bar{K} = K^\intercal$.
\hfill $\square$
\end{remark}

\begin{example}
Let $m_1=m_2=1$, hence we consider solutions of the \emph{scalar} NLS equation. 
Let $S = \mathrm{diag}(s_1,\ldots,s_{n})$ with pairwise different eigenvalues $s_i$. 
According to lemma~\ref{lemma:Toeplitz}, without restriction of generality we can set
\be
    V = A \left(\begin{array}{c} 1 \\ \vdots \\ 1 \end{array} \right) \, , \qquad
    \bar{V} = \bar{A} \left(\begin{array}{c} 1 \\ \vdots \\ 1 \end{array} \right) \, ,
\ee
with matrices $A,\bar{A}$ satisfying $[A,S] = 0 = [\bar{A}, S^\dagger]$. 
Using an argument similar to that in section~\ref{subsubsec:cc_Sylvester}, 
$A$ and $\bar{A}$ can be chosen invertible. By application of a reparametrisation 
transformation\footnote{The Hermitian conjugation reduction requires $B = \bar{A}^\dagger$
and $\bar{B} = A^\dagger$ in (\ref{decomp_repar_trans}). }
the solutions can be expressed as 
\be
  q = \left(\begin{array}{ccc} 1 & \cdots & 1 \end{array} \right) 
      \left( \tilde{\Xi}^{\dagger -1} - \epsilon \, \tilde{\Xi}^\dagger \, \bar{K} \, \tilde{\Xi} 
             \, K  \right)^{-1} \, 
      \left(\begin{array}{c} 1 \\ \vdots \\ 1 \end{array} \right) \, , 
\ee
where 
\be
   \tilde{\Xi} = C \, e^{-x \, S - \imag t \, S^2}  \quad \mbox{with} \quad
        C = \bar{A}^\dagger A  \, ,   \label{nls_tXi}
\ee
and $K,\bar{K}$ have to solve 
\be
       S \, K + K \, S^\dagger 
   = \left(\begin{array}{ccc} 1 & \cdots & 1 \\ 
                           \vdots & \ddots & \vdots \\ 
                                1 & \cdots & 1
           \end{array} \right) 
   = S^\dagger \, \bar{K} + \bar{K} \, S    \label{nlsI_rank_restr}
\ee
Since $S$ is \emph{symmetric} (i.e. $S^\intercal = S$) and $K = \bar{K}^\intercal$ Hermitian, 
the solutions coincide with those obtained via the complex conjugation reduction 
in section~\ref{sec:cc-reduction}, yielding the well-known soliton solutions of the 
scalar focusing NLS equation. 
\hfill $\square$
\end{example}

\begin{remark}
\label{rem:Hred_comm_constr}
The Hermitian conjugation reduction does \emph{not} 
work for the \emph{discrete} matrix NLS systems, except when $m_1=m_2$ and 
$q q^\dagger = q^\dagger q = |\chi|^2 I$ with a scalar $\chi$ 
\cite{TUW99,Tsuc00RMP,APT03,APT04DPDE,APT04book,APT06} (see also \cite{AOT99,Ohta00}). 
One is then led to the \emph{semi-discrete vector NLS equation} (or coupled NLS equations) 
\be
    \imag \, \dot{\vec{q}} + \vec{q}^{\,+} - 2 \, \vec{q} + \vec{q}^{\,-} 
        - \epsilon \, |\vec{q}|^2 \, ( \vec{q}^{\,+} + \vec{q}^{\,-} ) = 0 \, ,
        \label{cdNLSeqs}
\ee 
where the vector $\vec{q}$ is obtained via expansion of the matrix $q$ in terms of a 
basis of an appropriate Clifford algebra. A corresponding analysis in our framework 
will be postponed to a separate work. 
In appendix~D we present a \emph{non-local} modification of 
the Hermitian conjugation reduction which does work for the semi-discrete NLS system. 
\hfill $\square$
\end{remark}

\subsection{Solitons of the focusing matrix NLS equation}
\label{subsec:Hred_matrixsolitons}

\subsubsection{Rank one single solitons} 
Setting $n=1$, proposition~\ref{prop:NLS_eq_sol} determines rank one single soliton solutions of the 
focusing $m_1 \times m_2$ matrix NLS equation. Writing $V = \boldsymbol{v}^\dagger$ and $\bar{V} = \bar{\boldsymbol{v}}^\dagger$, 
with a reparametrisation we can achieve that the $m_1$-component column vector $\boldsymbol{v}$ and the 
$m_2$-component column vector $\bar{\boldsymbol{v}}$ satisfy 
$\| \boldsymbol{v} \|^2 := \boldsymbol{v}^\dagger \boldsymbol{v} = 1$ and 
$\| \bar{\boldsymbol{v}} \|^2 = 1$, 
at the price of having to replace $\Xi$ in (\ref{nlsHred_q_sol}) by 
\be
   \xi = c \, e^{-x \, s - \imag \, t \, s^2} 
  = 2 \alpha \, \txi \qquad \mbox{where} \quad
    \txi = e^{- \imag \, \left( \beta \, x + (\alpha^2-\beta^2) t \right)} 
      \, e^{-\alpha \, (x - 2 \beta \, t ) + \delta}   \, , 
\ee
with $c>0$ and $s= \alpha + \imag \, \beta$ with $\alpha,\beta \in \bbR$, 
$\alpha >0$.\footnote{$\alpha \neq 0$ is required in order to have a solution of the 
Sylvester equations (\ref{nlsHred_rank}). 
$\alpha >0$ can then always be achieved by a reflection symmetry.} 
Furthermore, we wrote $c = 2 \alpha \, e^{\delta}$ with $\delta \in \bbR$.
The solutions of the two equations (\ref{nlsHred_rank}) are then given by 
\be
   K = \bar{K} = \kappa  \qquad \quad \mbox{where} \qquad 
   \kappa = \frac{1}{s+s^\ast} = \frac{1}{2 \alpha} \, ,
\ee
and the matrix NLS solution (\ref{nlsHred_q_sol}), in the focusing case $\epsilon = -1$, 
takes the form
\be
  q = \frac{\kappa^{-1} \, \txi^\ast}{1+ |\txi|^2} \, \boldsymbol{v} \, \bar{\boldsymbol{v}}^\dagger 
    = \alpha \, \mathrm{sech}\left( \alpha \, (x - 2 \beta \, t) + \delta \right) 
        \, e^{\imag \, \left( \beta \, x + (\alpha^2-\beta^2) \, t \right)} \,
        \boldsymbol{v} \, \bar{\boldsymbol{v}}^\dagger  \; .
\ee
This describes a soliton with amplitude given by $\alpha$, 
with velocity $2 \beta$ and polarisation matrix $\boldsymbol{v} \bar{\boldsymbol{v}}^\dagger$.

\subsubsection{Superposing solutions} 
\label{subsubsec:Hred_super}
Given two data sets $(S_i,V_i,\bar{V}_i)$, $i=1,2$, that determine solutions according 
to proposition~\ref{prop:NLS_eq_sol}, with corresponding Hermitian solutions $K_i,\bar{K}_i$ of 
the corresponding Sylvester equations, we can superpose them, 
\be
    S = \left( \begin{array}{cc} S_1 & 0 \\ 0 & S_2 \end{array} \right) , \;
    V = \left( \begin{array}{c} V_1 \\ V_2 \end{array} \right) , \; 
    \bar{V} = \left( \begin{array}{c} \bar{V}_1 \\ \bar{V}_2 \end{array} \right) , \;
    K = \left( \begin{array}{cc} K_1 & K_{12} \\ K_{21} & K_2 \end{array} \right) , \; 
   \bar{K} = \left( \begin{array}{cc} \bar{K}_1 & \bar{K}_{12} \\ \bar{K}_{21} & \bar{K}_2 
        \end{array} \right) , \;        \label{Hred_superpos}
\ee
where $K_{12}, K_{21}$ and $\bar{K}_{12}, \bar{K}_{21}$ have to solve 
\be
  &&  S_1 K_{12} + K_{12} S_2^\dagger = V_1 V_2^\dagger \, , \quad 
      S_2 K_{21} + K_{21} S_1^\dagger = V_2 V_1^\dagger \, , \nonumber  \\
  &&  S_1^\dagger \bar{K}_{12} + \bar{K}_{12} S_2 = \bar{V}_1 \bar{V}_2^\dagger \, , \quad 
      S_2^\dagger \bar{K}_{21} + \bar{K}_{21} S_1 = \bar{V}_2 \bar{V}_1^\dagger 
       \; . 
\ee
This system reduces to only two equations by setting
\be
     K_{21} = K_{12}^\dagger \, , \qquad
     \bar{K}_{21} = \bar{K}_{12}^\dagger \; .  \label{Hred_K21=K12dag}
\ee
We have 
\be
  Z = \Xi^{\dagger -1} - \epsilon \, \bar{K} \, \Xi \, K 
    = \left( \begin{array}{cc} Z_1 - \epsilon \, \bar{K}_{12} \Xi_2 K_{21} & 
                 -\epsilon \, \bar{K}_1 \Xi_1 K_{12} - \epsilon \, \bar{K}_{12} \Xi_2 K_2 \\ 
                -\epsilon \, \bar{K}_2 \Xi_2 K_{21} - \epsilon \, \bar{K}_{21} \Xi_1 K_1 &
                Z_2 - \epsilon \, \bar{K}_{21} \Xi_1 K_{12} \end{array} \right) \, ,
\ee
where $Z_i = \Xi_i^{\dagger -1} - \epsilon \, \bar{K}_i \Xi_i K_i$. 

The following paves the way towards the asymptotic structure of multi-soliton 
solutions, worked out in section~\ref{subsubsec:Hred_n-soliton}, in analogy to the analysis in 
the case of the complex conjugation reduction in section~\ref{subsec:ccred_matrixsolitons}. 
First we note that 
$Z = \bar{A} \check{\Xi}^{\dagger -1} A + \bar{B} \check{\Xi} B$ with 
$\check{\Xi} = \mbox{block-diag}( \Xi_1, \Xi_2^{\dagger -1} )$ and 
\be
 \bar{A} = \left( \begin{array}{cc} I & \bar{K}_{12} \\ 0 & \bar{K}_2 \end{array} \right) , \; 
 A = \left( \begin{array}{cc} I & 0 \\ -\epsilon \, K_{21} & -\epsilon K_2 \end{array} \right) , \;
 \bar{B} = - \left( \begin{array}{cc} \epsilon \, \bar{K}_1 & 0 \\ \epsilon \bar{K}_{21} & I 
           \end{array} \right) , \;
 B = \left( \begin{array}{cc} K_1 & K_{12} \\ 0 & -I \end{array} \right) .
\ee
Assuming that $K_1$ and $\bar{K}_1$ are invertible, 
in a limit in which $\Xi_1^{-1} \to 0$ and $\Xi_2 \sim \hat{\Xi}_2$, we obtain 
\be
 Z^{-1} \sim B^{-1} \, 
    \left( \begin{array}{cc} 0 & 0 \\ 
     0 & \hat{\Xi}_2^\dagger \, \left( I - \epsilon \, \bar{L}_2 \, 
         \hat{\Xi}_2 \, L_2 \, \hat{\Xi}_2^\dagger \right)^{-1} 
           \end{array} \right) \, \bar{B}^{-1} \, ,   \label{Hred_superpos_limit1}
\ee
with the Schur complements of $K_1$ (with respect to $K$) and $\bar{K}_1$ (with respect to $\bar{K}$),
\be
          L_2 = K_2 - K_{21} K_1^{-1} K_{12} \, , \qquad 
    \bar{L}_2 = \bar{K}_2 - \bar{K}_{21} \bar{K}_1^{-1} \bar{K}_{12} \, , \label{Hred_L2}
\ee
which are also Hermitian.

\subsubsection{Superposition of rank one solitons}
\label{subsubsec:Hred_n-soliton}
A solution of the $m_1 \times m_2$ matrix NLS equation composed of $n$ rank one solitons 
is obtained from proposition~\ref{prop:NLS_eq_sol} by choosing $S = \mathrm{diag}(s_1,\ldots,s_n)$ 
with $s_i = \alpha_i + \imag \, \beta_i$. We assume $\alpha_i > 0$ (which can be achieved 
generically by a reflection symmetry, see also \cite{ABDM09sym}). Let us write
\be
   V = \left( \begin{array}{c} \boldsymbol{v}_1^\dagger \\ \vdots \\ \boldsymbol{v}_n^\dagger 
              \end{array} \right)         \, , \qquad
   \bar{V} = \left( \begin{array}{c} \bar{\boldsymbol{v}}_1^\dagger \\ \vdots \\
                     \bar{\boldsymbol{v}}_n^\dagger \end{array} \right) \, , 
\ee
where $\boldsymbol{v}_i$ and $\bar{\boldsymbol{v}}_i$ are $m_1$- respectively $m_2$-component 
column vectors. Conditions (\ref{nlsHred_rank}) are then solved by
\be
    K = (k_{ij}) \, , \quad \bar{K} = (\bar{k}_{ij}) \qquad \mbox{where} \quad
    k_{ij} = \frac{\boldsymbol{v}_i^\dagger \boldsymbol{v}_j}{s_i+s_j^\ast} \, , \quad
    \bar{k}_{ij} = \frac{\bar{\boldsymbol{v}}_i^\dagger \bar{\boldsymbol{v}}_j}{s_i^\ast+s_j} \; .
          \label{Hred_nsol_kij}
\ee
$K$ and $\bar{K}$ are Hermitian. Using a reparametrisation transformation, we can achieve that 
the vectors $\boldsymbol{v}_i,\bar{\boldsymbol{v}}_i$ are normalised, i.e. 
\be
    \| \boldsymbol{v}_i \| = 1 = \| \boldsymbol{\bar{v}}_i \| \; .  \label{Hred_nsol_normalis}
\ee
Hence $k_{ii} = \bar{k}_{ii} = 1/(2 \, \alpha_i)$. Now $\Xi$ in (\ref{nlsHred_q_sol}) has to 
be replaced by $\Xi = \mathrm{diag}(\xi_1,\ldots,\xi_n)$ with 
\be
    \xi_i = c_i \, e^{-x \, s_i - \imag \, s_i^2 \, t} 
          = 2 \alpha_i \, \txi_i  \, ,  \qquad
    \txi_i = e^{- \imag \, \left( \beta_i x + (\alpha_i^2 - \beta_i^2) \, t \right)} 
                    \, e^{- \alpha_i \, (x-2 \beta_i t) + \delta_i}  \, ,
\ee
where we set $c_i = 2 \alpha_i \, e^{\delta_i}$. 

Next we choose $\beta_1 < \beta_2 < \cdots < \beta_n$.\footnote{Here we leave aside the 
cases where some of the $\beta_i$ coincide.} 
As $t \to -\infty$ the solitons are then arranged along the $x$-axis in decreasing 
numerical order (recall that the velocity of the $i$-th soliton, when isolated from its 
partners, is $2 \, \beta_i$). 
After the interaction, as $t \to +\infty$, the order is reversed. 
In terms of the $i$-th comoving coordinate $\bar{x}_i = x - 2 \, \beta_i \, t$, we have  
$|\xi_j| = 2 \alpha_j \, \exp(-\alpha_j \, \bar{x}_i + 2 (\beta_j-\beta_i) \, t] + \delta_j)$. 
Hence, as $t\to -\infty$ (keeping $\bar{x}_i$ constant), 
we have $\xi_j \to 0$ for $j > i$ and $\xi_j^{-1} \to 0$ for $j<i$. 
Furthermore, as $t\to \infty$, we have $\xi_j^{-1} \to 0$ for $j>i$ and $\xi_j \to 0$ for $j<i$.
In order to find the corresponding limit of $q$, for \emph{fixed} $i$, we express the solution 
as a superposition (\ref{Hred_superpos}) with 
$S_1 = \mathrm{diag}(s_1,\ldots,s_{i-1})$ and $S_2 = \mathrm{diag}(s_i,\ldots,s_n)$. 
Let $V_1,V_2$ and $\bar{V}_1,\bar{V}_2$ be the corresponding parts of $V$ and $\bar{V}$, 
respectively. Let $K_1,K_2$ and $K_{12},K_{21}$ be the corresponding diagonal and off-diagonal 
blocks of $K$, and correspondingly for $\bar{K}$. Then we can use the results of 
section~\ref{subsubsec:Hred_super}, in particular (\ref{Hred_superpos_limit1})  
with $(\hat{\Xi}_2)_{kl} = \xi_i \, \delta_{ik} \delta_{il}$, in order to work out the 
limits of $q$ given by (\ref{nlsHred_q_sol}) as $t\to -\infty$, respectively $t \to +\infty$. 
To formulate the results, we introduce new vectors via 
\be
       \eta_i^{\tminus} \, \boldsymbol{v}_i^{\tminus} 
   &=& (V_2^\dagger - V_1^\dagger K_1^{-1} K_{12})_i 
    = \frac{1}{\det(K_1)} \, \det\left( \begin{array}{ccc} k_{11} & \cdots & k_{1i} \\
          \vdots & \ddots & \vdots \\
          k_{i-1,1} & \cdots & k_{i-1,i} \\
          \boldsymbol{v}_1 & \cdots & \boldsymbol{v}_i 
         \end{array} \right)  \, , \nonumber \\
      \bar{\eta}_i^{\tminus} \, \boldsymbol{\bar{v}}_i^{\tminus} 
  &=& (\bar{V}_2 - \bar{K}_{21} \bar{K}_1^{-1} \bar{V}_1)_i^\dagger 
   = \frac{1}{\det(\bar{K}_1)^\ast} \, \det\left( \begin{array}{cccc} 
          \bar{k}_{11}^\ast & \cdots & \bar{k}_{1,i-1}^\ast & \boldsymbol{\bar{v}}_1 \\
          \vdots & \ddots & \vdots & \vdots \\
          \bar{k}_{i,1}^\ast & \cdots & \bar{k}_{i,i-1}^\ast & \boldsymbol{\bar{v}}_i 
         \end{array} \right)  \, , 
\ee
and 
\be
      \eta_i^{\tplus} \, \boldsymbol{v}_i^{\tplus} 
  &=& (V_1^\dagger - K_2^{-1} K_{21} V_2^\dagger)_i 
       = \frac{1}{\det(K_2)} \, \det\left( \begin{array}{ccc} 
          \boldsymbol{v}_i & \cdots & \boldsymbol{v}_n \\
          k_{i+1,1} & \cdots & k_{i+1,n} \\
          \vdots & \ddots & \vdots \\
          k_{n1} & \cdots & k_{nn} 
         \end{array} \right)  \, , \nonumber \\
      \bar{\eta}_i^{\tplus} \, \boldsymbol{\bar{v}}_i^{\tplus} 
  &=& (\bar{V}_1 - \bar{K}_{12} \bar{K}_2^{-1} \bar{V}_2)_i^\dagger 
   = \frac{1}{\det(\bar{K}_2)^\ast} \, \det\left( \begin{array}{cccc} 
         \boldsymbol{\bar{v}}_i & \bar{k}_{i,i+1}^\ast & \cdots & \bar{k}_{in}^\ast \\ 
          \vdots & \vdots & \ddots & \vdots \\
          \boldsymbol{\bar{v}}_n & \bar{k}_{n,i+1}^\ast & \cdots & \bar{k}_{nn}^\ast 
         \end{array} \right)  \, , 
\ee
where the constants 
$\eta_i^{\tminus}, \bar{\eta}_i^{\tminus},\eta_i^{\tplus}, \bar{\eta}_i^{\tplus}$ 
are chosen such that the following normalisation conditions hold,
\be
   \| \boldsymbol{v}_i^{\tminus} \| = \| \boldsymbol{\bar{v}}_i^{\tminus} \| 
   = \| \boldsymbol{v}_i^{\tplus} \| = \| \boldsymbol{\bar{v}}_i^{\tplus} \| = 1 \; .
         \label{Hred_nsol_asympt_normalis}
\ee

\begin{proposition}
\be
   | \eta_i^{\tminus} |^2 = k_{ii}^{-1} \, (L_2)_{ii} \, , \qquad
   | \eta_i^{\tplus} |^2 = k_{ii}^{-1} \, (L_1)_{ii} \, , \nonumber \\
   | \bar{\eta}_i^{\tminus} |^2 = k_{ii}^{-1} \, (\bar{L}_2)_{ii} \, , \qquad
   | \bar{\eta}_i^{\tplus} |^2 = k_{ii}^{-1} \, (\bar{L}_1)_{ii} \, ,
\ee
with the Schur complements $L_1 = K_1 - K_{12} K_2^{-1} K_{21}$, 
$\bar{L}_1 = \bar{K}_1 - \bar{K}_{12} \bar{K}_2^{-1} \bar{K}_{21}$, and 
$L_2, \bar{L}_2$ given by (\ref{Hred_L2}).\footnote{Recall that $K_i,\bar{K}_i$ are Hermitian 
and (\ref{Hred_K21=K12dag}) holds, hence $L_i, \bar{L}_i$ are Hermitian and the diagonal 
entries are therefore real.} 
\end{proposition}
\noindent
\textit{Proof:} Based on (\ref{Hred_nsol_kij}) and (\ref{Hred_nsol_normalis}), the proof is 
similar to that of proposition~\ref{prop:cc_asympt_norm}. 
\hfill $\square$
\vskip.2cm

In particular, since we chose $\alpha_i >0$ and thus $k_{ii}>0$, this implies 
$(L_a)_{ii}>0, (\bar{L}_a)_{ii}>0$, $a=1,2$. Let
\be
    \gamma_i^{\tminus} = k_{ii}^{-1} \, \sqrt{ (L_2)_{ii} (\bar{L}_2)_{ii} } \, , \qquad
    \gamma_i^{\tplus} = k_{ii}^{-1} \, \sqrt{ (L_1)_{ii} (\bar{L}_1)_{ii} } \; .
\ee
Then the following asymptotic relations hold,\footnote{Clearly, if $i=1$ respectively $i=n$, 
keeping $\bar{x}_1$ respectively $\bar{x}_n$ constant, then we have 
$$ 
   q \sim \frac{k_{11}^{-1} \, \txi_1^\ast}{1 + |\txi_1|^2} 
               \, \boldsymbol{v}_1 \, \boldsymbol{\bar{v}}_1 
      \quad \mbox{as} \quad t \to -\infty
       \, , \quad
     q \sim \frac{k_{nn}^{-1} \, \txi_n^\ast}{1 
         + |\txi_n|^2} 
         \, \boldsymbol{v}_n \, \boldsymbol{\bar{v}}_n^{\dagger} 
        \quad \mbox{as} \quad t \to +\infty \, .
$$
}
\be
  q &\sim& \frac{k_{ii}^{-1} (\gamma_i^{\tminus} \, \txi_i)^\ast}{1 + |\gamma_i^{\tminus} \txi_i|^2} 
               \, \boldsymbol{v}_i^{\tminus} \, \boldsymbol{\bar{v}}_i^{\tminus\dagger}  
   \qquad \mbox{as} \quad t \to -\infty \, , \; \bar{x}_i = \mathrm{const.} \, , \\
  q &\sim& \frac{k_{ii}^{-1} (\gamma_i^{\tplus} \, \txi_i)^\ast}{1 
         + |\gamma_i^{\tplus} \txi_i|^2} 
         \, \boldsymbol{v}_i^{\tplus} \, \boldsymbol{\bar{v}}_i^{\tplus\dagger}  
   \qquad \mbox{as} \quad t \to +\infty \, , \; \bar{x}_i = \mathrm{const.} \; .
\ee
These asymptotic expressions determine the changes in relative positions and polarisations 
of the solitons due to the collision. For the vector NLS equation, such results have 
already been obtained in \cite{Tsuch04,APT04IP,APT04book} in a different way.

\subsubsection{2-soliton solutions}
For $n=2$ we can write
\be
   K = \left( \begin{array}{cc} \kappa_1 & k \\ k^\ast & \kappa_2 \end{array} \right) \, , \quad
   \bar{K} = \left( \begin{array}{cc} \kappa_1 & \bar{k}^\ast \\ \bar{k} & \kappa_2 \end{array} \right) \, ,
\ee
where
\be
   \kappa_i = \frac{1}{s_i + s_i^\ast} = \frac{1}{2 \alpha_i}   \, , \quad
   k = \frac{\boldsymbol{v}_1^\dagger \boldsymbol{v}_2}{s_1 + s_2^\ast} \, , \quad 
  \bar{k} = \frac{\bar{\boldsymbol{v}}_2^\dagger \bar{\boldsymbol{v}}_1}{s_1 + s_2^\ast} \; .
      \label{Hred_2sol_defs}
\ee
The corresponding $m_1 \times m_2$ matrix focusing NLS solution
is then explicitly given by
\be
 q &=& \frac{1}{D} \Big[  
    ( \xi_1^\ast + \kappa_2^2 |\xi_2|^2 \, \xi_1^\ast + k \bar{k} \, |\xi_1|^2 \xi_2^\ast ) 
                    \, \boldsymbol{v}_1 \bar{\boldsymbol{v}}_1^\dagger  
  + ( \xi_2^\ast + \kappa_1^2 |\xi_1|^2 \, \xi_2^\ast + k^\ast \bar{k}^\ast |\xi_2|^2 \xi_1^\ast ) 
                    \, \boldsymbol{v}_2 \bar{\boldsymbol{v}}_2^\dagger  \nonumber \\
   && - ( k \kappa_1 |\xi_1|^2 \xi_2^\ast + \bar{k}^\ast \kappa_2 |\xi_2|^2 \xi_1^\ast ) 
            \, \boldsymbol{v}_1 \bar{\boldsymbol{v}}_2^\dagger 
      - \left( \bar{k} \kappa_1 |\xi_1|^2 \xi_2^\ast + k^\ast \kappa_2 |\xi_2|^2 \xi_1^\ast \right) 
                    \, \boldsymbol{v}_2 \bar{\boldsymbol{v}}_1^\dagger \Big] \, , 
\ee 
where
\be
  D = \Big( 1 + \frac{\kappa_1}{\kappa_2} (\kappa_1 \kappa_2 - |k|^2) \, |\xi_1|^2 \Big)
      \Big( 1 + \frac{\kappa_2}{\kappa_1} (\kappa_1 \kappa_2 - |\bar{k}|^2) \, |\xi_2|^2 \Big) 
    + \frac{1}{\kappa_1\kappa_2} \left| \kappa_1 k \xi_1 + \kappa_2 \bar{k}^\ast \xi_2 \right|^2  \, .
\ee
As a consequence of the definitions (\ref{Hred_2sol_defs}) and $\alpha_i >0$, we have 
$\kappa_1 \kappa_2 - |k|^2 \geq 0$ and $\kappa_1 \kappa_2 - |\bar{k}|^2 \geq 0$, so that 
the solution is regular. 
If the vectors $\boldsymbol{v}_1$ and $\boldsymbol{v}_2$ are orthogonal (with respect to the 
Hermitian inner product), and also $\bar{\boldsymbol{v}}_1$ and $\bar{\boldsymbol{v}}_2$, 
then $K$ and $\bar{K}$ are diagonal and the superposition is simply the sum $q = q_1 + q_2$ 
of the single soliton solutions. 

For the above solution we have the following asymptotic relations,
\be
   q &\sim& \frac{\kappa_1^{-1} \, \txi_1^\ast}{1 + |\txi_1|^2} 
               \, \boldsymbol{v}_1 \, \boldsymbol{\bar{v}}_1^\dagger  
      + \frac{\kappa_2^{-1} \, \gamma_2^{\tminus} \, \txi_2^\ast}{1 + |\gamma_2^{\tminus} \txi_2|^2} 
               \, \boldsymbol{v}_2^{\tminus} \, \boldsymbol{\bar{v}}_2^{\tminus \dagger} 
   \quad \mbox{as} \quad t \to -\infty \, , \\
  q &\sim& \frac{\kappa_1^{-1} \, \gamma_1^{\tplus} \, \txi_1^\ast}{1 + |\gamma_1^{\tplus} \txi_1|^2} 
               \, \boldsymbol{v}_1^{\tplus} \, \boldsymbol{\bar{v}}_1^{\tplus \dagger} 
     + \frac{\kappa_2^{-1} \, \txi_2^\ast}{1 + |\txi_2|^2} 
               \, \boldsymbol{v}_2 \, \boldsymbol{\bar{v}}_2^\dagger
   \quad \mbox{as} \quad t \to +\infty \, ,
\ee 
where $\txi_i = \kappa_i \, \xi_i$, $\gamma_2^{\tminus} = \gamma_1^{\tplus} 
= \sqrt{\det(K) \det(\bar{K})}/(\kappa_1 \, \kappa_2)$, and 
\be
 && \boldsymbol{v}_2^{\tminus} = \sqrt{\frac{\kappa_1 \, \kappa_2}{\det(K)}} (\boldsymbol{v}_2
       - \frac{k}{\kappa_1} \, \boldsymbol{v}_1 ) \, , \quad
   \boldsymbol{\bar{v}}_2^{\tminus} = \sqrt{\frac{\kappa_1 \, \kappa_2}{\det(\bar{K})}} 
   (\boldsymbol{\bar{v}}_2 - \frac{\bar{k}}{\kappa_1} \, \boldsymbol{\bar{v}}_1 ) \, , \nonumber \\
 &&  \boldsymbol{v}_1^{\tplus} = \sqrt{\frac{\kappa_1 \, \kappa_2}{\det(K)}} (\boldsymbol{v}_1
       - \frac{k^\ast}{\kappa_2} \, \boldsymbol{v}_2 ) \, , \quad
   \boldsymbol{\bar{v}}_2^{\tplus} = \sqrt{\frac{\kappa_1 \, \kappa_2}{\det(\bar{K})}}
   (\boldsymbol{\bar{v}}_1 - \frac{\bar{k}^\ast}{\kappa_2} \, \boldsymbol{\bar{v}}_2 ) \; .
\ee

\section{Conclusions}
\label{sec:concl} 
\setcounter{equation}{0}
An appealing aspect of the bidifferential calculus approach is the simplicity of the 
basic structure one starts with, as compared with the typical complexity of a Lax pair. 
Our work demonstrates this in case of NLS systems and their discretisations (see 
also appendix~E). 
The step from the continuum NLS bidifferential calculus to its discretisations 
is quite evident and the general solution generating method of section~\ref{sec:framework} 
is easily adapted. 
In this work we concentrated on the application of this method, which seems to be the 
most direct and simplest method to generate a large class of relevant exact solutions. 
In \cite{DMH08bidiff} we have shown how other methods, like B\"acklund and Darboux 
transformations, universally arise in the bidifferential calculus framework. 
The basic general result formulated in theorem~\ref{theorem:main} may actually be 
regarded as a shortcut to the binary Darboux transformation method \cite{DMH08bidiff}. 

In this work we presented infinite families of exact solutions of matrix continuous, semi- and 
fully discrete NLS systems and some of their reductions. The solutions are parametrised 
by constant matrices, and the only task left is to solve a Sylvester equation 
(which appeared as a ``rank condition'' in the literature). In the focusing case, they 
include the soliton solutions. But there are more and a complete 
understanding of the relations between matrix data and the behaviour of solutions has 
not yet been reached. But the expression of solutions in terms of matrix data turned out 
to be very convenient in this respect. The regularity proofs and the asymptotic analysis 
for some classes of solutions in sections~\ref{sec:cc-reduction} and \ref{sec:HermRed} 
provide good examples. 

Besides a comparatively short treatment of the previously (with different methods) 
studied matrix continuous NLS equation with a Hermitian conjugation invariance, 
we considered in great detail another matrix NLS equation with a complex conjugation 
invariance. They coincide in the scalar case. 
But in the \emph{square} matrix case, the latter has \emph{discrete} counterparts, whereas 
the Hermitian conjugation invariant matrix NLS equation seems to admit a discretisation 
only if additional constraints are satisfied \cite{TUW99,Tsuc00RMP,APT03,APT04book,APT06} 
(but see also appendix~D for a corresponding non-local version). We therefore concentrated on 
an exploration of continuous and discrete matrix NLS solutions of the complex 
conjugation invariant NLS equations. We would also like to generate solutions 
of \emph{vector discrete} NLS equations \cite{TUW99,Tsuc00RMP,APT03,APT04book,APT06}, using 
the framework presented here, and plan to elaborate on this in more detail in a separate work. 
Of further interest is the possibility to extend the present work in order to 
obtain solutions of $(2+1)$-dimensional generalisations of matrix NLS equations. 

In the defocusing NLS case we have not yet reached, via the method of section~\ref{sec:framework}, 
the important class of dark soliton solutions (see in particular 
\cite{Zakh+Shab73,Blow+Dora85,AEK87,IWY91,Radha+Laks95,Kivs+Luth98,Park+Shin00,Kivs+Agra03,CLH05,Maru+Ohta06}) 
and also rational solutions as e.g. obtained in \cite{Hone97}.
A generalisation of this method seems to be required.

\vskip.2cm
\noindent
\textbf{Acknowledgment.} F M-H thanks Nils Kanning for discussions and comments. The authors are also 
grateful to Takayuki Tsuchida for some helpful comments on the original manuscript. 

\renewcommand{\theequation} {\Alph{section}.\arabic{equation}}
\renewcommand{\thesection} {\Alph{section}}
\newtheorem{lemmaA}{Lemma}[section]

\section*{Appendix A: Solutions of the matrix KdV equation}
\addcontentsline{toc}{section}{\numberline{}Appendix A: Solutions of the matrix KdV equation}
\setcounter{section}{1}
\setcounter{equation}{0}
Let $\cB_0$ be the algebra of smooth complex functions on $\bbR^2$, and $\cB$ its extension by 
adjoining an operator $\pa_x$ that satisfies $\pa_x f = f_x + f \, \pa_x$ (where 
$f_x$ denotes the partial derivative of $f$ with respect to $x$). 
We define linear maps 
$\d, \bar{\d} \, : \, \cB \rightarrow \cB \otimes \bigwedge^1(\bbC^2)$ via 
\be
    \d f = [\pa_x, f] \, \zeta_1 + \frac{1}{2} [\pa_x^2,f] \, \zeta_2 \, ,  \qquad
   \bar{\d} f = -\frac{1}{2} [\pa_x^2,f] \, \zeta_1 + \frac{1}{3} [\pa_t-\pa_x^3,f] \, \zeta_2 \, ,
\ee
where $\zeta_1,\zeta_2$ is a basis of $\bigwedge^1(\bbC^2)$. 
In particular, we have $\d(\pa_x) = \bar{\d}(\pa_x) =0$. 
The maps $\d,\bar{\d}$ extend to linear maps 
$\cB \otimes \bigwedge(\bbC^2) \rightarrow \cB \otimes \bigwedge(\bbC^2)$ 
satisfying (\ref{bidiff}) and the graded Leibniz rule. Moreover, they extend to 
$\mathrm{Mat}_1(\cB) \otimes \bigwedge(\bbC^2)$. 
For $\phi \in \mathrm{Mat}(m,m,\cB_0)$, we have
\be
    \d \phi = \phi_x \, \zeta_1 + \frac{1}{2} (\phi_{xx} + 2 \, \phi_x \pa_x) \, \zeta_2 \, ,
\ee
and thus
\be
    \d \phi \wedge \d\phi = -\frac{1}{2} (\phi_x{}^2)_x \, \zeta_1 \wedge \zeta_2 \, , \qquad
    \bar{\d} \d \phi = \left(-\frac{1}{3} \phi_{x t} 
        + \frac{1}{12} \phi_{xxxx} \right) \, \zeta_1 \wedge \zeta_2 \, , 
\ee
so that (\ref{univeq}) turns out to be equivalent to the $m \times m$ \emph{matrix KdV equation}
\be
    u_t - \frac{1}{4} u_{xxx} - \frac{3}{2} (u^2)_x = 0 \qquad 
    \mbox{where} \quad u = \phi_x  \; .  \label{matKdV}
\ee

In the following we apply results of section~\ref{sec:framework} 
in order to generate exact solutions of (\ref{matKdV}). 
Inspection of the linear equation $\bar{\d} \bX = (\d \bX) \, \bP$ 
suggests setting $\bP = \bS - \bI \pa_x$, where $\d$- and $\bar{\d}$-constancy requires that 
$\bS$ does not depend on $x$ and $t$. The linear equation is then equivalent to
\be
    \bX_{xx} = -2 \, \bX_x \bS \, , \qquad  \bX_t = \bX_x \bS^2 \; .
\ee
These equations make sense for $\bS \in \mathrm{Mat}(n,n,\bbC)$ and 
$\bX \in \mathrm{Mat}(n,n,\cB_0)$. They are solved by
\be
    \bX = \bA_0 + \bA_1 \, \hat{\bXi} \, , \qquad
    \hat{\bXi} = e^{- 2 (x \bS + t \bS^3)} \, ,
\ee
where $\bA_i$ are constant complex matrices. Assuming $\bA_0$ invertible, 
application of theorem~\ref{theorem:main} reduces to an application of 
corollary~\ref{cor:univ_sol}. 
(\ref{constr_univ}) takes the form
\be
    \bS \hat{\bK} + \hat{\bK} \bS = \bV \hat{\bU} \, ,     \label{KdV-rank}
\ee
and (\ref{phi_univ}) becomes
\be
    \phi = \hat{\bU} \, \hat{\bXi} \, ( \bI - \hat{\bK} \, \hat{\bXi} )^{-1} \, \bV \; .
           \label{KdV_sol}
\ee
This solves the $m \times m$ matrix KdV equation for any 
matrix data $(\bS,\hat{\bU},\bV)$ for which (\ref{KdV-rank}) admits a solution $\hat{\bK}$. 
Under the additional assumption that $\bS$ is positive stable, so that 
$e^{-y \bS} \to 0$ as $y \to \infty$, the condition\footnote{For 
$m=1$, (\ref{KdV-rank}) is a well-known \emph{rank one condition} in the context of the 
scalar KdV equation. 
See \cite{GHS06,GHS09} for a corresponding general analysis in the framework of bounded linear operators on 
a Banach space. } 
(\ref{KdV-rank}) is solved by \cite{Heinz51,Bhat+Rose97}
\be
    \hat{\bK} = \int_0^\infty e^{-y\bS} \bV \hat{\bU} e^{-y \bS} \, dy \; .  \label{KdV_K}
\ee

In the scalar case, i.e. $m=1$, we can apply more directly the formulae of theorem~\ref{theorem:main} 
to obtain
\be
      \phi 
  &=& \tbU \bY \bX^{-1} \tbV 
   = \mathrm{tr}(\tbV \tbU \bY \bX^{-1})
   = \mathrm{tr}((\bX \bS +\bX_x-\tbR \bX) \bX^{-1}) \nonumber  \\
  &=& \mathrm{tr}(\bS-\tbR) + \mathrm{tr}(\bX_x \bX^{-1}) \, ,
\ee
where $\tbR = \bR + \pa_x$.  
The second step requires that the matrices do not contain operator terms (involving $\pa_x$). 
Dropping the irrelevant constant summand, this becomes 
\be
    \phi = (\log \tau)_x \, , 
\ee
where
\be
    \tau = \det \left( \bI - e^{-x \bS} \hat{\bK} e^{-x \bS - 2 t \bS^3} \right) 
         = \det \left( \bI - \int_x^\infty e^{-y \bS} \bV \hat{\bU} e^{-y \bS} e^{-2t \bS^3} \, dy  \right)
           \; .
\ee
In the last step we inserted (\ref{KdV_K}), assuming positive stable $\bS$.
This determines the family of (here in general complex) KdV solutions derived in \cite{Akto+Mee06} 
(see (4.1) therein). 

Alternatively, using lemma~\ref{lemma:Toeplitz}, (\ref{KdV_sol}) can be expressed in the scalar 
case ($m=1$) as 
$\phi = \bV_0^\intercal \, \bC \, \hat{\bXi} \, ( \bI - \hat{\bK} \, \bC \, \hat{\bXi} )^{-1} \, \bV_0$,  
with $\bV_0^\intercal = (1, \ldots, 1)$. Here $\hat{\bK}$ now has to solve 
$\bS \hat{\bK} + \hat{\bK} \bS = \bV_0 \bV_0^\intercal$, and $\bC$ is any $n \times n$ matrix over $\bbC$ 
that commutes with $\bS$. If $\bS$ is diagonal and if the eigenvalues satisfy $s_i + s_j \neq 0$ 
for all $i,j = 1,\ldots,n$, then $\hat{\bK}$ is the Cauchy matrix with entries $1/(s_i+s_j)$.

\section*{Appendix B: A relation between the semi-discrete NLS and modified KdV equations}
\addcontentsline{toc}{section}{\numberline{}Appendix B: A relation between the semi-discrete NLS and modified KdV equations}
\setcounter{section}{2}
\setcounter{equation}{0}
Let us recall from (\ref{sd_NLS_sys}) the $m \times m$ matrix equation for $\cU$ 
in the semi-discrete NLS (AL) case, 
 \be
  \imag \, J \, \dot{\cU} - 2 \, \cU + \cU^+ \, (I - \cU^2)  + (I - \cU^2) \, \cU^- = 0 \; .
       \label{AL_U_eq}
\ee
Substituting 
\be
    \cU = \imag \, (\imag J)^x \, e^{-2 \imag \, J \, t} \, \cV \qquad \quad 
            x \in \bbZ \, ,         \label{cU->cW}
\ee
with a new matrix variable $\cV$, we find\footnote{Note that $\cV$ anticommutes with $J$, 
since $\cU$ does, and we have $\cU^2 = - \cV^2$. } 
\be
  \dot{\cV} + \cV^+ \, (I + \cV^2) - (I + \cV^2) \, \cV^- = 0 \, , \label{MdmKdV}
\ee 
which is the $m \times m$ matrix (semi-) \emph{discrete mKdV} equation. As a complex 
matrix equation, the latter is thus an equivalent form of (\ref{AL_U_eq}).

\begin{proposition}
Let $\bS,\bU,\bV$ be complex matrices of size $n \times n$, $m \times n$ and $n \times m$, respectively, 
and $\bK$ a solution of 
\be
     \bS^{-1} \bK - \bK \bS = \bV \bU \; .   \label{mKdV_rank_cond}
\ee 
Then 
\be
  \cV = - \bU \, (\bI + (\bXi \bK)^2 )^{-1} \, \bXi \, \bV  \quad \mbox{with} \quad
  \bXi = \bS^x \, e^{-t (\bS-\bS^{-1})}  \label{mKdV_sol}
\ee
solves the $m \times m$ discrete modified KdV equation (\ref{MdmKdV}). 
\end{proposition}
\noindent 
\textit{Proof:} From proposition~\ref{prop:NLSsol} we recall that 
\bez
   \cU = \bU \, (\bXi^{-1} - \bK \, \bXi \, \bK)^{-1} \, \bV \, , \quad \mbox{with} \quad
   \bXi = \bS^x \, e^{ -\imag \, t \, (\bS+\bS^{-1}-2\bI) \bJ } 
\eez
solves (\ref{AL_U_eq}) if $\bK$ is a solution of 
$\bS^{-1} \bK - \bK \bS = \bV \bU$. Via (\ref{cU->cW}), with the redefinitions
\bez
   \bS \mapsto - \imag \, \bS \, \bJ \, , \qquad
   \bV \mapsto \imag \, \bV \, , \qquad 
   \bK \mapsto \bJ \, \bK  \, , 
\eez
(and keeping $\bU$) and using (\ref{JSUVK_rels}), this is translated into (\ref{mKdV_rank_cond}) 
and (\ref{mKdV_sol}). 
\hfill $\square$
\vskip.2cm 

The proof moreover gives a concrete map between a class of solutions of 
(\ref{AL_U_eq}) and a class of solutions of (\ref{MdmKdV}).

\section*{Appendix C: Equivalence of the fully discrete NLS equation with the corresponding 
equation of Ablowitz and Ladik in the scalar case}
\addcontentsline{toc}{section}{\numberline{}Appendix C: Equivalence of the fully discrete NLS equation 
with the corresponding equation of Ablowitz and Ladik in the scalar case}
\setcounter{section}{3}
\setcounter{equation}{0}
After elimination of the terms cubic in $q$ in favour of $p$, (\ref{fdNLS_cc}) reads
\be
  &&  \imag \, (q_+ - q) + (q^+ - q)_+ - (q - q^-) + (p_+^+ - p) \, q_+
        + q \, (p_+ - p^- )^\ast = 0 \, , \nonumber \\
  &&  p^+ - p = - \epsilon \, q^+ q^\ast \; .              \label{fdNLS_cc_2}
\ee
We can solve the second equation with
\be
    p(x,t) = - \epsilon \sum_{k=-\infty}^x q(k,t) \, q(k-1,t)^\ast  
    \qquad \quad     x,t \in \bbZ \, , 
\ee
assuming the existence of this infinite series. Inserting this in the first equation, leads to
\be
  && \imag \, [q(x,t+1) - q(x,t)] + q(x+1,t+1) - q(x,t+1) - [q(x,t) - q(x-1,t)] \nonumber \\
  && - \epsilon \, q(x,t) \sum_{k=-\infty}^x \Big( q(k,t+1)^\ast \, q(k-1,t+1)
           - q(k-1,t)^\ast \, q(k-2,t) \Big)  \nonumber \\ 
  && - \epsilon \sum_{k=-\infty}^{x+1} \Big( q(k,t+1) \, q(k-1,t+1)^\ast - q(k-1,t) \, q(k-2,t)^\ast
      \Big) \, q(x,t+1) = 0 \; . 
\ee 
Expressing this in terms of a new variable $r$ related to $q$ by $q(x,t) = r(x,-t+1)$, 
and substituting $t$ by $-t$ in the resulting equation, in the scalar case we end up with 
equation (2.17) in \cite{Ablo+Ladi76SAM} (with an obvious change in notation).
The infinite sum in the last equation appears to be an awkward non-locality. But see 
\cite{Tsuch09} for how to achieve locality with the help of conservation laws.

\section*{Appendix D: A non-local Hermitian reduction of the semi-dicrete NLS system}
\addcontentsline{toc}{section}{\numberline{}Appendix D: A non-local Hermitian reduction of the 
semi-dicrete NLS system}
\setcounter{section}{4}
\setcounter{equation}{0}
Imposing the condition (see also \cite{Gerd+Ivan82})
\be
    \bar{q}(x,t) = \epsilon \, q(-x,t)^\dagger \qquad \mbox{with} \quad \epsilon = \pm 1 \, 
\ee
reduces the two equations (\ref{ALsystem}) to the single equation
\be
  && \imag \, \dot{q}(x,t) + q(x+1,t) - 2 q(x,t) + q(x-1,t) 
      - \epsilon \, q(x+1,t) \, q(-x,t)^\dagger q(x,t)   \nonumber \\
  && - \epsilon \, q(x,t) \, q(-x,t)^\dagger q(x-1,t) = 0 \; .  \label{nonlocal_sdNLS}
\ee
This is a non-local equation unless one imposes the restriction to functions $q(x,t)$ that 
are even in $x$. Setting
\be
   U = V^\dagger \, , \quad
   \bar{U} = \epsilon \, \bar{V}^\dagger \, , \quad
   \bar{S} = (S^{-1})^\dagger \, , \quad
   K^\dagger = - K \, , \quad
   \bar{K}^\dagger = - \bar{K} \, , 
\ee
then $\bar{\Xi}$ is given by $\Xi^\dagger$ with $x$ replaced by $-x$, and 
proposition~\ref{prop:sol_NLSsystems} implies that
\be
    q = V^\dagger \, \Big( (S^\dagger)^x \, e^{-\imag \, t \, \omega(S^\dagger) } 
         - \epsilon \, \bar{K} \, S^x \, e^{-\imag \, t \, \omega(S) } K \Big)^{-1} \, \bar{V} \, , 
\ee
with $\omega(S)$ defined in (\ref{omega}), solves (\ref{nonlocal_sdNLS}) if $K, \bar{K}$ solve
\be
   S^{-1} K - K (S^{-1})^\dagger = V V^\dagger \, , \qquad
   S^\dagger \bar{K} - \bar{K} S = \bar{V} \bar{V}^\dagger \; .
\ee

\section*{Appendix E: Lax pairs for the NLS systems}
\addcontentsline{toc}{section}{\numberline{}Appendix E: Lax pairs for the NLS systems}
\setcounter{section}{5}
\setcounter{equation}{0}
(\ref{univeq}) is the integrability condition of the linear equation
\be
    \bar{\d} \Psi = (\d \phi) \, \Psi + (\d \Psi) \, \Delta \, ,  \label{linsys}
\ee
if $\Delta$ solves (cf. \cite{DMH08bidiff})
\be
    \bar{\d}  \Delta = (\d \Delta) \, \Delta \; . \label{Delta-eq}
\ee
In the following we show how Lax pairs for the NLS systems originate from these equations, 
by using the respective bidifferential calculus. (\ref{Delta-eq}) is obviously satisfied 
if $\d \Delta = \bar{\d} \Delta = 0$. These linear equations allow $\Delta$ to carry an 
arbitrary constant factor which turns out to play the role of a ``spectral parameter'' 
in the Lax pair. Although (\ref{linsys}) is linear in this parameter, the resulting 
Lax pair exhibits a nonlinear parameter dependence. 

\paragraph{Continuous NLS system.}
(\ref{Delta-eq}) is trivially satisfied if $\bar{\d} \Delta = \d \Delta =0$. 
Using the bidifferential calculus that led to the continuous NLS system, this means 
$\Delta_x = \Delta_t = [J,\Delta] =0$. A particular solution is 
$\Delta = 2 \, \lambda \, I$ with a constant $\lambda$. Then (\ref{linsys}) becomes
\be
   \Psi_x = \frac{1}{2} J [J, \varphi] \, \Psi + \lambda \, [J, \Psi] \, , \qquad
   - \imag \, \Psi_t = J \varphi_x \Psi + 2 \lambda  \Psi_x 
\ee
(where $\phi = J \varphi$). Writing
\be
    \Psi = \psi \, e^{-x \, \lambda \, J - 2 \imag \, t \, \lambda^2 J} \, , 
\ee
after some straight computations we obtain the linear system in the canonical form
\be
   \psi_x = \mathcal{L} \, \psi  \, , \qquad  \psi_t = \mathcal{M} \, \psi \, , 
\ee
where
\be
   && \mathcal{L} = \lambda \, J + \cU 
     = \left( \begin{array}{cc} \lambda \, I_{m_1} & q \\ \bar{q} & -\lambda \, I_{m_2} \end{array} \right)
                  \, , \nonumber \\
   && \mathcal{M} = \imag \, J \, ( \cU_x - \cU^2 ) + 2 \imag \, \lambda \, \mathcal{L}
                  = \imag \, \left( \begin{array}{cc} 2 \lambda^2 \, I_{m_1} - q \bar{q} & 2 \lambda q + q_x \\
                    2 \lambda \bar{q} - \bar{q}_x & - 2 \lambda^2 \, I_{m_2} + \bar{q} q \end{array} \right) 
                    \; .
\ee
This is a familar Lax pair for the matrix NLS system \cite{APT04book}. 
The integrability condition (zero curvature condition) 
\be
    \mathcal{L}_t - \mathcal{M}_x + [\mathcal{L},\mathcal{M}] = 0 
\ee 
indeed reproduces the NLS system (\ref{NLSsystem}). 
Whereas our linear system (\ref{linsys}) is \emph{linear} in the (spectral) 
parameter $\lambda$, the above Lax pair has a nonlinear $\lambda$-dependence. 

\paragraph{Semi-discrete NLS system.} 
(\ref{Delta-eq}) is trivially satisfied with $\Delta = 2 \, \alpha \, \bbS^{-1}$, 
where $\alpha$ is a constant. Then (\ref{linsys}) becomes
\be
  \Psi - \Psi^- &=& \cU \Psi^- + \alpha \, [ J , \Psi ] \, ,  \\
   -\imag \, \dot{\Psi} &=& J \, (\varphi^+ - \varphi) \, \Psi + 2 \, \alpha \, (\Psi^+ - \Psi) \nonumber \\
                &=& J \, (\cU^+ - \cU - \cU^+ \cU) \, \Psi + 2 \, \alpha \, (\Psi^+ - \Psi) \, , 
\ee
where we used the second of equations (\ref{sd_NLS_sys}) in the last step. 
Multiplying from the right by $(1/2)(I+J)$,  
and introducing $\psi = \frac{1}{2} \Psi^- \, (I+J)$, we obtain
\be
    \psi^+ = \mathcal{L} \psi \, , \quad
    \dot{\psi} = \mathcal{M} \psi \, , 
\ee
where
\be
   \mathcal{L} = [(1+\alpha) I - \alpha J ]^{-1} (I+\cU) \, , \qquad
   \mathcal{M} = \imag \, J (\cU - \cU^- - \cU \, \cU^-) + 2 \imag \, \alpha \, (\mathcal{L}-I) \; . 
\ee
The integrability condition of this linear system, i.e. 
$\dot{\mathcal{L}}= \mathcal{M}^+ \mathcal{L} - \mathcal{L} \mathcal{M}$, is invariant under the
transformations $\mathcal{L}' = \mathcal{S}^+ \mathcal{L} \mathcal{S}^{-1}$, 
$\mathcal{M}' = \mathcal{S} \mathcal{M} \mathcal{S}^{-1} + \dot{\mathcal{S}} \mathcal{S}^{-1}$. 
Setting
\be
    \mathcal{S} = \la^x \, e^{\imag \, t \, (\la^2-1)}
    \left(\begin{array}{cc} I_{m_1} & 0 \\ 0 & \la \, I_{m_2} \end{array} \right) \, , \qquad
    \alpha = \frac{1}{2}(\la^2-1) 
\ee
(where $x$ is the discrete variable), we recover the usual Lax pair of the AL system, 
\be
  \mathcal{L}' = \left(\begin{array}{cc} \la \, I_{m_1} & q \\ \bar{q} & \la^{-1} \, I_{m_2} \end{array} \right) 
                 \, , \qquad
    \mathcal{M}' = \imag \left(\begin{array}{cc} (\la^2-1) \, I_{m_1} - q \bar{q}^- & \la q - \la^{-1}q^- \\
        \la \bar{q}^- -\la^{-1} \bar{q} & (1 - \la^{-2}) \, I_{m_2} + \bar{q} q^- \end{array} \right) \; .
           \label{sdNLS_Lax-pair}
\ee

\paragraph{Fully discrete NLS system.}
Choosing again $\Delta = 2 \alpha \, \bbS^{-1}$, (\ref{linsys}) yields
\be
  \Psi - \Psi^- = \cU \, \Psi^- + \alpha \, [J , \Psi ]  \, , \qquad
   -\imag \, (\Psi_+ - \Psi) = J (\varphi^+_+ - \varphi) \, \Psi_+ + 2\alpha \, (\Psi^+_+ - \Psi) 
   \; .
\ee
Introducing again $\psi = \frac{1}{2} \Psi^- (I+J)$ leads to
\be
    \psi^+ = \mathcal{L} \psi \, , \qquad
    \psi-\psi_- = \mathcal{M} \psi \, ,
\ee
with
\be
    \mathcal{L} = [(1+\alpha) I - \alpha J]^{-1} (I+\cU) \, , \qquad
    \mathcal{M} = -\frac{1}{\imag+2\alpha} [J (\varphi-\varphi^-_-) + 2 \, \alpha \, (\mathcal{L}-I)] \; .
\ee
The integrability condition 
$\mathcal{L} - \mathcal{L}_- = \mathcal{M}^+ \mathcal{L} - \mathcal{L}_- \mathcal{M}$ 
is invariant under the transformation
$\mathcal{L}' = \mathcal{S}^+ \mathcal{L} \mathcal{S}^{-1}$, 
$\mathcal{M}' = I - \mathcal{S}_- (I-\mathcal{M}) \mathcal{S}^{-1}$.
Choosing 
\be
  \mathcal{S} = \la^x \, (1-\imag -\la^2)^{-t} 
     \left(\begin{array}{cc} I_{m_1} & 0 \\ 0 & \la \, I_{m_2} \end{array}\right) \, , \qquad
    \alpha = \frac{1}{2} (\la^2-1) 
\ee
(where $x,t \in \bbZ$), we obtain a Lax pair close to (\ref{sdNLS_Lax-pair}), 
\be
  \mathcal{L}' = \left(\begin{array}{cc} \la \, I_{m_1} & q \\ \bar{q} & \la^{-1} \, I_{m_2} \end{array}\right) 
                 \, , \; 
  \mathcal{M}' = \left(\begin{array}{cc} (\la^2 + \imag) \, I_{m_1} + p - p^-_- & \la q - \la^{-1} q^-_- \\ 
          \la \bar{q}^-_- - \la^{-1} \bar{q} & (2 + \imag -\la^{-2}) \, I_{m_2} -\bar{p} + \bar{p}^-_-
                       \end{array}\right) \, . \;
\ee

\section*{Appendix F: Continuum limit}
\addcontentsline{toc}{section}{\numberline{}Appendix F: Continuum limit}
\setcounter{section}{6}
\setcounter{equation}{0}
Let us introduce lattice spacing parameters $a,b$ in the basic relations that determine 
the bidifferential calculus for the fully discrete NLS system, 
\be
 \d f = \frac{1}{a} \, [ \bbT \bbS , f ] \, \zeta_1 + \frac{1}{2} [J,f] \, \zeta_2 
                 \, , \qquad
 \bar{\d} f = - \frac{\imag}{b} \, [\bbT,f] \, \zeta_1 - \frac{1}{a} \, [\bbS^{-1},f] \, \zeta_2 \, , 
\ee
where now $x \in a \bbZ$ and $t \in b \bbZ$. 
$\frac{1}{a} [ \bbT \bbS , f ]$ taken at a lattice point $(x,t)$ reads 
$\frac{1}{a} [f(x+a,t+b)-f(x,t)] \, \bbT \bbS$, which shows that an approach to the continuum 
can only work if we let first $b \to 0$ and only afterwards $a \to 0$. The first limit 
yields the bidifferential calculus for the semi-discrete NLS system and then the second 
that for the continuum NLS system (provided that $f$ is differentiable). 
In proposition~\ref{prop:NLSsol} we only have to replace the expression for $\bXi$ by
\be
   \bXi &=& \bS^{x/a} \, \Big( \bI - \imag \, \frac{b}{a^2} (\bS^{-1} - \bI) \bJ \Big)^{t/b} 
          \Big( \bI - \imag \, \frac{b}{a^2} (\bI - \bS) \bJ]^{-1} \Big)^{-t/b} \\
        && \stackrel{b \to 0}{\longrightarrow} \quad
           \bS^{x/a} \, e^{- \imag \, t \, (\bS^{-1} - \bI) \bJ/a^2} \, 
            e^{\imag \, t \, (\bI - \bS) \bJ/a^2} 
            = \bS^{x/a} \, e^{- \imag \, t \, (\bS + \bS^{-1} - 2 \bI) \bJ/a^2 } \nonumber  \\
        && \stackrel{a \to 0}{\longrightarrow} \quad
              e^{-x \tbP} e^{- \imag \, t \, \tbP^2} \; .  \nonumber 
\ee
In the last step we expressed $\bS$ in terms of $\tbP$ via $\bS = (\bI + a \, \tbP)^{-1}$, 
which replaces (\ref{bS<->tbP}). Renaming $\tbP$ to $\bS$, yields the expression for 
$\bXi$ in the continuous NLS case in proposition~\ref{prop:NLSsol}.

\small
\addcontentsline{toc}{section}{\numberline{}References}

\end{document}